\documentclass[11pt]{article}
\usepackage{xypic}
\usepackage{mathrsfs}
\usepackage{amssymb}
\usepackage{amsfonts}
\usepackage{amsmath}

\def\ben{\begin{enumerate}}
\def\een{\end{enumerate}}

\def\bit{\begin{itemize}}
\def\eit{\end{itemize}}

\def\0{\leqno}

\input xy
\xyoption{all}
\input{tcilatex}
\begin{document}

\begin{center}
{\Large {HAMILTONIAN MECHANICS}\ \ \bigskip }

{\Large {ON DUALS OF GENERALIZED LIE ALGEBROIDS}\bigskip }

%{\Large \bigskip }\textbf{\bigskip }

\textbf{by }

\textbf{CONSTANTIN M. ARCU\c{S} }
\end{center}

%\bigskip

%\bigskip

%\bigskip

%\ \ \ \textit{In memory of my uncles}

%\ \ \ \textit{Prof. Dr. Gheorghe RADU and Acad. Dr. Doc. Cornelius RADU}

%\medskip
%
%\ \ \ \textit{Dedicated to} \textit{Acad. Prof. Dr. Doc. Radu MIRON at his 83%
%$^{th}$ anniversary}

\begin{abstract}
%In some previous papers, a geometric description of Hamiltonian Mechanics on
%dual of Lie algebroids has been developed.
A new description, different by the classical theory of Hamiltonian
Mechanics, in the general framework of generalized Lie algebroids is
presented. In the particular case of Lie algebroids, new and important
results are obtained. We present the \emph{dual mechanical systems} called
by use, \emph{dual mechanical }$\left( \rho ,\eta \right) $\emph{-systems,
Hamilton mechanical }$\left( \rho ,\eta \right) $\emph{-systems} or \emph{%
Cartan mechanical }$\left( \rho ,\eta \right) $\emph{-systems.}
%and we develop their geometries.
We obtain the canonical $\left( \rho ,\eta \right) $\emph{-}semi(spray)
associated to a dual mechanical $\left( \rho ,\eta \right) $-system. The
Hamilton mechanical $(\rho ,\eta)$-systems are the spaces necessary to
develop a Hamiltonian formalism. We obtain the $(\rho ,\eta)$-semispray
associated to a regular Hamiltonian $H$ and external force $F_{e}$ and we
derive the equations of Hamilton-Jacobi type. \ \ \bigskip\newline
\end{abstract}

\tableofcontents

\section{Introduction}

The concept of Hamilton space, introduced in $\left[ 15\right] $, was
intensively studied in $\left[ 6,7,8,11,14\right] $, and it has been
successful, as a geometric theory of the Hamiltonian function. In the
classical sense, a regular Hamiltonian on $T^{\ast }M$ is a smooth function $%
\begin{array}[b]{ccc}
T^{\ast }M & ^{\underrightarrow{~H\ }} & \mathbb{R}%
\end{array}%
$ such that the Hessian matrix with entries%
\begin{equation*}
\begin{array}[b]{c}
g^{ij}\left( x,p\right) =\frac{1}{2}\frac{\partial ^{2}H\left( x,p\right) }{%
\partial p_{i}\partial p_{j}}%
\end{array}%
\end{equation*}%
is everywhere nondegenerate on $T^{\ast }M$ (or on a domain of $T^{\ast }M$)
and a Hamilton space is a pair $H^{n}=\left( M,H\right) ,$ where $H$ is a
regular Hamiltonian. (see $\left[ 16\right] $) The case when $H$ is square
of a function on $T^{\ast }M,$ positively, $1$-homogeneous with respect to
the momentum $p_{i},$ provides an important class of Hamilton spaces called
Cartan spaces. The modern formulation of the geometry of Cartan spaces was
given by R. Miron $\left[ 12,13\right] $ although some results where
obtained by \'{E}. Cartan $\left[ 5\right] $ and A. Kawaguchi $\left[ 9%
\right] .$

The geometry of $T^{\ast }M$ is from one point of view different from that
of $TM,$ because not exists a natural tangent structure and a semispray can
not be introduced as usual for the tangent bundle. Two geometrical
ingredients are of great importance on $T^{\ast }M$: the canonical $1$-form $%
p_{i}dx^{i}$ and its exterior derivative $dp_{i}\wedge dx^{i}$ (the
canonical symplectic structure of $T^{\ast }M$). They are systematicaly used
to defined new useful tools in the classical theory.

A Hamiltonian description of Mechanics on duals of Lie algebroids was
presented in $\left[ 10\right] .$ (see also $\left[ 17,18,19,20,21\right] $)
The role of cotangent bundle of the configuration manifold was played by the
prolongation $\mathcal{L}^{\tau ^{\ast }}E$ of $E$ along the projection $%
\begin{array}[b]{ccc}
E^{\ast } & ^{\underrightarrow{~\tau ^{\ast }\ }} & M.%
\end{array}%
$ The Lie algebroid version of the classical results concerning the
universality of the standard Liouville $1$-form on cotangent bundles is
presented in \emph{Theorem 3.4} and \emph{Corollary 3.6.} Given a
Hamiltonian function $%
\begin{array}[b]{ccc}
E^{\ast } & ^{\underrightarrow{~H\ }} & \mathbb{R}%
\end{array}%
$ and the symplectic form $\Omega _{E}$ on $E^{\ast },$ the dynamics are
obtained solving the equation%
\begin{equation*}
i_{\xi _{H}}\Omega _{E}=d^{\mathcal{L}^{\tau ^{\ast }}E}H
\end{equation*}%
with the usual notations. The solutions of $\xi _{H}$ (curves in $E^{\ast }$%
) are the ones of the Hamilton equations for $H.$

The purpose of the present paper is to find the answer to the following
question:

\begin{itemize}
\item \textit{Could we to give a Hamiltonian description of Mechanics on
duals of generalized Lie algebroids (see }$\left[ 1,2,3\right] $\textit{)
similar with the Lagrangian description of Mechanics on generalized Lie
algebroids presented in the paper }$\left[ 4\right] $\textit{\ without the
symplectic form?}
\end{itemize}

In Sections $3,4,5$ and $6$ we set up the basic notions and terminology. In
Section $7$ we present for the first time the \emph{dual mechanical systems}
called by use,\emph{\ dual mechanical }$\left( \rho ,\eta \right) $\emph{%
-systems, Hamilton mechanical }$\left( \rho ,\eta \right) $\emph{-systems}
or \emph{Cartan mechanical }$\left( \rho ,\eta \right) $\emph{-systems.}
%We develop their geometries as parts of Lagrangian Mechanics.

In Section $8$ we obtain the \emph{\ canonical }$\left( \rho ,\eta \right) $%
\emph{-semispray associated to the dual mechanical }$\left( \rho ,\eta
\right) $\emph{-system }$\left( \left( \overset{\ast }{E},\overset{\ast }{%
\pi },M\right) ,\overset{\ast }{F}_{e},\left( \rho ,\eta \right) \Gamma
\right) $\emph{\ and from locally invertible }$\mathbf{B}^{\mathbf{v}}$\emph{%
-morphism }$\left( g,h\right) .$ Also, we present \emph{the canonical }$%
\left( \rho ,\eta \right) $\emph{-spray associated to mechanical system }$%
\left( \left( \overset{\ast }{E},\overset{\ast }{\pi },M\right) ,\overset{%
\ast }{F}_{e},\left( \rho ,\eta \right) \Gamma \right) $\emph{\ and from
locally invertible }$\mathbf{B}^{\mathbf{v}}$\emph{-morphism }$(g,h).$

The Section $9$ is dedicated to study the geometry of Hamilton mechanical $%
\left( \rho ,\eta \right) $-systems. These mechanical systems are the spaces
necessary to obtain a Hamiltonian formalism\textbf{\ }in the general
framewok of generalized Lie algebroids. We determine and we study the $%
\left( \rho ,\eta \right) $-semispray associated to a regular Hamiltonian $H$
and external force $\overset{\ast }{F}_{e}$ which are applied on the dual of
the total space of a generalized Lie algebroid and we derive the equations
of Hamilton-Jacobi type.

Finally, we obtain that the\ integral curves of the canonical $\left( \rho
,\eta \right) $-semispray associated to Hamilton mechanical $\left( \rho
,\eta \right) $-system $\left( \left( \overset{\ast }{E},\overset{\ast }{\pi
},M\right) ,\overset{\ast }{F}_{e},\left( \rho ,\eta \right) \Gamma \right) $%
\ and from locally invertible\emph{\ }$\mathbf{B}^{\mathbf{v}}$-morphism $%
\left( g,h\right) $\ are the $\left( g,h\right) $-lifts solutions for the
equations of Hamilton-Jacobi type $\left( 9.10\right) $.

Our researches are very important because, if $h=Id_{M}=\eta ,$ then all
results presented in this paper become new results in the framework of Lie
algebroids.

\section{Preliminaries}

Let$\mathbf{~Vect},$ $\mathbf{Liealg},~\mathbf{Mod}$\textbf{,} $\mathbf{Man}$
and $\mathbf{B}^{\mathbf{v}}$ be the category of real vector spaces, Lie
algebras, modules, manifolds and vector bundles respectively.

We know that if $\left( E,\pi ,M\right) \in \left\vert \mathbf{B}^{\mathbf{v}%
}\right\vert $ so that $M$ is paracompact and if $A\subseteq M$ is closed,
then for any section $u$ over $A$ it exists $\tilde{u}\in $ $\Gamma \left(
E,\pi ,M\right) $ so that $\tilde{u}_{|A}=u.$ In the following, we consider
only vector bundles with paracompact base.

Aditionally, if $\left( E,\pi ,M\right) \in \left\vert \mathbf{B}^{\mathbf{v}%
}\right\vert ,$ $\Gamma \left( E,\pi ,M\right) =\left\{ u\in \mathbf{Man}%
\left( M,E\right) :u\circ \pi =Id_{M}\right\} $ and $\mathcal{F}\left(
M\right) =\mathbf{Man}\left( M,\mathbb{R}\right) ,$ then $\left( \Gamma
\left( E,\pi ,M\right) ,+,\cdot \right) $ is a $\mathcal{F}\left( M\right) $%
-module. If \ $\left( \varphi ,\varphi _{0}\right) \in \mathbf{B}^{\mathbf{v}%
}\left( \left( E,\pi ,M\right) ,\left( E^{\prime },\pi ^{\prime },M^{\prime
}\right) \right) $ such that $\varphi _{0}\in Iso_{\mathbf{Man}}\left(
M,M^{\prime }\right) ,$ then, using the operation
\begin{equation*}
\begin{array}{ccc}
\mathcal{F}\left( M\right) \times \Gamma \left( E^{\prime },\pi ^{\prime
},M^{\prime }\right) & ^{\underrightarrow{~\ \ \cdot ~\ \ }} & \Gamma \left(
E^{\prime },\pi ^{\prime },M^{\prime }\right) \\
\left( f,u^{\prime }\right) & \longmapsto & f\circ \varphi _{0}^{-1}\cdot
u^{\prime }%
\end{array}%
\end{equation*}%
it results that $\left( \Gamma \left( E^{\prime },\pi ^{\prime },M^{\prime
}\right) ,+,\cdot \right) $ is a $\mathcal{F}\left( M\right) $-module and we
obtain the $\mathbf{Mod}$-morphism%
\begin{equation*}
\begin{array}{ccc}
\Gamma \left( E,\pi ,M\right) & ^{\underrightarrow{~\ \ \Gamma \left(
\varphi ,\varphi _{0}\right) ~\ \ }} & \Gamma \left( E^{\prime },\pi
^{\prime },M^{\prime }\right) \\
u & \longmapsto & \Gamma \left( \varphi ,\varphi _{0}\right) u%
\end{array}%
\end{equation*}%
defined by
\begin{equation*}
\begin{array}{c}
\Gamma \left( \varphi ,\varphi _{0}\right) u\left( y\right) =\varphi \left(
u_{\varphi _{0}^{-1}\left( y\right) }\right) ,%
\end{array}%
\end{equation*}%
for any $y\in M^{\prime }.$

Let $M,N\in \left\vert \mathbf{Man}\right\vert ,$ $h\in Iso_{\mathbf{Man}%
}\left( M,N\right) $ and $\eta \in Iso_{\mathbf{Man}}\left( N,M\right) $.

We know (see $\left[ 2,3\right] $) that if $\left( F,\nu ,N\right) \in
\left\vert \mathbf{B}^{\mathbf{v}}\right\vert $ so that there exists
\begin{equation*}
\begin{array}{c}
\left( \rho ,\eta \right) \in \mathbf{B}^{\mathbf{v}}\left( \left( F,\nu
,N\right) ,\left( TM,\tau _{M},M\right) \right)%
\end{array}%
\end{equation*}%
and an operation
\begin{equation*}
\begin{array}{ccc}
\Gamma \left( F,\nu ,N\right) \times \Gamma \left( F,\nu ,N\right) & ^{%
\underrightarrow{\left[ ,\right] _{F,h}}} & \Gamma \left( F,\nu ,N\right) \\
\left( u,v\right) & \longmapsto & \left[ u,v\right] _{F,h}%
\end{array}%
\end{equation*}%
with the following properties:\bigskip

\noindent $\qquad GLA_{1}$. the equality holds good
\begin{equation*}
\begin{array}{c}
\left[ u,f\cdot v\right] _{F,h}=f\left[ u,v\right] _{F,h}+\Gamma \left(
Th\circ \rho ,h\circ \eta \right) \left( u\right) f\cdot v,%
\end{array}%
\end{equation*}%
\qquad \quad\ \ for all $u,v\in \Gamma \left( F,\nu ,N\right) $ and $f\in
\mathcal{F}\left( N\right) .$

\medskip $GLA_{2}$. the $4$-tuple $\left( \Gamma \left( F,\nu ,N\right)
,+,\cdot ,\left[ ,\right] _{F,h}\right) $ is a Lie $\mathcal{F}\left(
N\right) $-algebra,

$GLA_{3}$. the $\mathbf{Mod}$-morphism $\Gamma \left( Th\circ \rho ,h\circ
\eta \right) $ is a $\mathbf{LieAlg}$-morphism of
\begin{equation*}
\left( \Gamma \left( F,\nu ,N\right) ,+,\cdot ,\left[ ,\right] _{F,h}\right)
\end{equation*}%
source and
\begin{equation*}
\left( \Gamma \left( TN,\tau _{N},N\right) ,+,\cdot ,\left[ ,\right]
_{TN}\right)
\end{equation*}%
target, \medskip \noindent then the triple $\left( \left( F,\nu ,N\right) ,%
\left[ ,\right] _{F,h},\left( \rho ,\eta \right) \right) $ is called
generalized Lie algebroid.\emph{\ }

In particular, if $h=Id_{M}=\eta ,$ then we obtain the definition of the Lie
algebroid.

We can discuss about \emph{the category }$\mathbf{GLA}$\emph{\ of
generalized Lie algebroids.} (see $\left[ 3\right] $)

Examples of objects of this category are presented in the paper $\left[ 2%
\right] .$

Let $\left( \left( F,\nu ,N\right) ,\left[ ,\right] _{F,h},\left( \rho ,\eta
\right) \right) $ be an object of the category $\mathbf{GLA}$.

\begin{itemize}
\item Locally, for any $\alpha ,\beta \in \overline{1,p},$ we set $\left[
t_{\alpha },t_{\beta }\right] _{F,h}=L_{\alpha \beta }^{\gamma }t_{\gamma }.$
We easily obtain that $L_{\alpha \beta }^{\gamma }=-L_{\beta \alpha
}^{\gamma },~$for any $\alpha ,\beta ,\gamma \in \overline{1,p}.$
\end{itemize}

The real local functions $L_{\alpha \beta }^{\gamma },~\alpha ,\beta ,\gamma
\in \overline{1,p}$ will be called the \emph{structure functions of the
generalized Lie algebroid }$\left( \left( F,\nu ,N\right) ,\left[ ,\right]
_{F,h},\left( \rho ,\eta \right) \right) .$

\begin{itemize}
\item We assume the following diagrams:%
\begin{equation*}
\begin{array}[b]{ccccc}
F & ^{\underrightarrow{~\ \ \ \rho ~\ \ }} & TM & ^{\underrightarrow{~\ \ \
Th~\ \ }} & TN \\
~\downarrow \nu &  & ~\ \ \ \downarrow \tau _{M} &  & ~\ \ \ \downarrow \tau
_{N} \\
N & ^{\underrightarrow{~\ \ \ \eta ~\ \ }} & M & ^{\underrightarrow{~\ \ \
h~\ \ }} & N \\
&  &  &  &  \\
\left( \chi ^{\tilde{\imath}},z^{\alpha }\right) &  & \left(
x^{i},y^{i}\right) &  & \left( \chi ^{\tilde{\imath}},z^{\tilde{\imath}%
}\right)%
\end{array}%
\end{equation*}

where $i,\tilde{\imath}\in \overline{1,m}$ and $\alpha \in \overline{1,p}.$

If%
\begin{equation*}
\left( \chi ^{\tilde{\imath}},z^{\alpha }\right) \longrightarrow \left( \chi
^{\tilde{\imath}\prime }\left( \chi ^{\tilde{\imath}}\right) ,z^{\alpha
\prime }\left( \chi ^{\tilde{\imath}},z^{\alpha }\right) \right) ,
\end{equation*}%
\begin{equation*}
\left( x^{i},y^{i}\right) \longrightarrow \left( x^{i%
%TCIMACRO{\U{b4}}%
%BeginExpansion
{\acute{}}%
%EndExpansion
}\left( x^{i}\right) ,y^{i%
%TCIMACRO{\U{b4}}%
%BeginExpansion
{\acute{}}%
%EndExpansion
}\left( x^{i},y^{i}\right) \right)
\end{equation*}%
and
\begin{equation*}
\left( \chi ^{\tilde{\imath}},z^{\tilde{\imath}}\right) \longrightarrow
\left( \chi ^{\tilde{\imath}\prime }\left( \chi ^{\tilde{\imath}}\right) ,z^{%
\tilde{\imath}\prime }\left( \chi ^{\tilde{\imath}},z^{\tilde{\imath}%
}\right) \right) ,
\end{equation*}%
then
\begin{equation*}
\begin{array}[b]{c}
z^{\alpha
%TCIMACRO{\U{b4}}%
%BeginExpansion
{\acute{}}%
%EndExpansion
}=\Lambda _{\alpha }^{\alpha
%TCIMACRO{\U{b4}}%
%BeginExpansion
{\acute{}}%
%EndExpansion
}z^{\alpha }%
\end{array}%
,
\end{equation*}%
\begin{equation*}
\begin{array}[b]{c}
y^{i%
%TCIMACRO{\U{b4}}%
%BeginExpansion
{\acute{}}%
%EndExpansion
}=\frac{\partial x^{i%
%TCIMACRO{\U{b4}}%
%BeginExpansion
{\acute{}}%
%EndExpansion
}}{\partial x^{i}}y^{i}%
\end{array}%
\end{equation*}%
and
\begin{equation*}
\begin{array}{c}
z^{\tilde{\imath}\prime }=\frac{\partial \chi ^{\tilde{\imath}\prime }}{%
\partial \chi ^{\tilde{\imath}}}z^{\tilde{\imath}}.%
\end{array}%
\end{equation*}

\item We assume that $\left( \theta ,\mu \right) \overset{put}{=}\left(
Th\circ \rho ,h\circ \eta \right) $. If $z^{\alpha }t_{\alpha }\in \Gamma
\left( F,\nu ,N\right) $ is arbitrary, then
\begin{equation*}
\begin{array}[t]{l}
\displaystyle%
\begin{array}{c}
\Gamma \left( Th\circ \rho ,h\circ \eta \right) \left( z^{\alpha }t_{\alpha
}\right) f\left( h\circ \eta \left( \varkappa \right) \right) =\vspace*{1mm}
\\
=\left( \theta _{\alpha }^{\tilde{\imath}}z^{\alpha }\frac{\partial f}{%
\partial \varkappa ^{\tilde{\imath}}}\right) \left( h\circ \eta \left(
\varkappa \right) \right) =\left( \left( \rho _{\alpha }^{i}\circ h\right)
\left( z^{\alpha }\circ h\right) \frac{\partial f\circ h}{\partial x^{i}}%
\right) \left( \eta \left( \varkappa \right) \right) ,%
\end{array}%
\end{array}%
\leqno(2.1)
\end{equation*}%
for any $f\in \mathcal{F}\left( N\right) $ and $\varkappa \in N.$
\end{itemize}

The coefficients $\rho _{\alpha }^{i}$ respectively $\theta _{\alpha }^{%
\tilde{\imath}}$ change to $\rho _{\alpha
%TCIMACRO{\U{b4}}%
%BeginExpansion
{\acute{}}%
%EndExpansion
}^{i%
%TCIMACRO{\U{b4}}%
%BeginExpansion
{\acute{}}%
%EndExpansion
}$ respectively $\theta _{\alpha
%TCIMACRO{\U{b4}}%
%BeginExpansion
{\acute{}}%
%EndExpansion
}^{\tilde{\imath}%
%TCIMACRO{\U{b4}}%
%BeginExpansion
{\acute{}}%
%EndExpansion
}$ according to the rule:
\begin{equation*}
\begin{array}{c}
\rho _{\alpha
%TCIMACRO{\U{b4}}%
%BeginExpansion
{\acute{}}%
%EndExpansion
}^{i%
%TCIMACRO{\U{b4}}%
%BeginExpansion
{\acute{}}%
%EndExpansion
}=\Lambda _{\alpha
%TCIMACRO{\U{b4}}%
%BeginExpansion
{\acute{}}%
%EndExpansion
}^{\alpha }\rho _{\alpha }^{i}\displaystyle\frac{\partial x^{i%
%TCIMACRO{\U{b4}}%
%BeginExpansion
{\acute{}}%
%EndExpansion
}}{\partial x^{i}},%
\end{array}%
\leqno(2.2)
\end{equation*}%
respectively%
\begin{equation*}
\begin{array}{c}
\theta _{\alpha
%TCIMACRO{\U{b4}}%
%BeginExpansion
{\acute{}}%
%EndExpansion
}^{\tilde{\imath}%
%TCIMACRO{\U{b4}}%
%BeginExpansion
{\acute{}}%
%EndExpansion
}=\Lambda _{\alpha
%TCIMACRO{\U{b4}}%
%BeginExpansion
{\acute{}}%
%EndExpansion
}^{\alpha }\theta _{\alpha }^{\tilde{\imath}}\displaystyle\frac{\partial
\varkappa ^{\tilde{\imath}%
%TCIMACRO{\U{b4}}%
%BeginExpansion
{\acute{}}%
%EndExpansion
}}{\partial \varkappa ^{\tilde{\imath}}},%
\end{array}%
\leqno(2.3)
\end{equation*}%
where
\begin{equation*}
\left\Vert \Lambda _{\alpha
%TCIMACRO{\U{b4}}%
%BeginExpansion
{\acute{}}%
%EndExpansion
}^{\alpha }\right\Vert =\left\Vert \Lambda _{\alpha }^{\alpha
%TCIMACRO{\U{b4}}%
%BeginExpansion
{\acute{}}%
%EndExpansion
}\right\Vert ^{-1}.
\end{equation*}

\emph{Remark 2.1 } The following equalities hold good:%
\begin{equation*}
\begin{array}{c}
\displaystyle\rho _{\alpha }^{i}\circ h\frac{\partial f\circ h}{\partial
x^{i}}=\left( \theta _{\alpha }^{\tilde{\imath}}\frac{\partial f}{\partial
\varkappa ^{\tilde{\imath}}}\right) \circ h,\forall f\in \mathcal{F}\left(
N\right) .%
\end{array}%
\leqno(2.4)
\end{equation*}%
\emph{and }%
\begin{equation*}
\begin{array}{c}
\displaystyle\left( L_{\alpha \beta }^{\gamma }\circ h\right) \left( \rho
_{\gamma }^{k}\circ h\right) =\left( \rho _{\alpha }^{i}\circ h\right) \frac{%
\partial \left( \rho _{\beta }^{k}\circ h\right) }{\partial x^{i}}-\left(
\rho _{\beta }^{j}\circ h\right) \frac{\partial \left( \rho _{\alpha
}^{k}\circ h\right) }{\partial x^{j}}.%
\end{array}%
\leqno(2.5)
\end{equation*}

Let $\left( E,\pi ,M\right) \in \left\vert \mathbf{B}^{\mathbf{v}%
}\right\vert $ and $\left( \overset{\ast }{E},\overset{\ast }{\pi },M\right)
$ its dual. We have the $\mathbf{B}^{\mathbf{v}}$-morphism%
\begin{equation*}
\begin{array}{ccc}
~\ \ \ \ \ \ \ \ \ \ \ \ \ \overset{\ast }{\pi }^{\ast }\left( h^{\ast
}F\right) & \hookrightarrow & F \\
\overset{\ast }{\pi }^{\ast }\left( h^{\ast }\nu \right) \downarrow &  &
~\downarrow \nu \\
~\ \ \ \ \ \ \ \ \ \ \ \ M & ^{\underrightarrow{~\ \ h\circ \overset{\ast }{%
\pi }~\ \ }} & N%
\end{array}%
\leqno(2.6)
\end{equation*}

Let $\Big(\overset{\overset{\ast }{\pi }^{\ast }\left( h^{\ast }F\right) }{%
\rho },Id_{E}\Big)$ be the\emph{\ }$\mathbf{B}^{\mathbf{v}}$-morphism of%
\emph{\ }$\left( \overset{\ast }{\pi }^{\ast }\left( h^{\ast }F\right) ,%
\overset{\ast }{\pi }^{\ast }\left( h^{\ast }\nu \right) ,\overset{\ast }{E}%
\right) $\ source and $\left( T\overset{\ast }{E},\tau _{\overset{\ast }{E}},%
\overset{\ast }{E}\right) $\ target, where%
\begin{equation*}
\begin{array}{rcl}
\overset{\ast }{\pi }^{\ast }\left( h^{\ast }F\right) & ^{\underrightarrow{%
\overset{\overset{\ast }{\pi }^{\ast }\left( h^{\ast }F\right) }{\rho }}} & T%
\overset{\ast }{E} \\
\displaystyle Z^{\alpha }T_{\alpha }\left( \overset{\ast }{u}_{x}\right) &
\longmapsto & \displaystyle\left( Z^{\alpha }\cdot \rho _{\alpha }^{i}\circ
h\circ \overset{\ast }{\pi }\right) \frac{\partial }{\partial x^{i}}\left(
\overset{\ast }{u}_{x}\right)%
\end{array}%
\leqno(2.7)
\end{equation*}

Using the operation
\begin{equation*}
\begin{array}{ccc}
\Gamma \left( \overset{\ast }{\pi }^{\ast }\left( h^{\ast }F\right) ,\overset%
{\ast }{\pi }^{\ast }\left( h^{\ast }\nu \right) ,\overset{\ast }{E}\right)
^{2} & ^{\underrightarrow{~\ \ \left[ ,\right] _{\overset{\ast }{\pi }^{\ast
}\left( h^{\ast }F\right) }~\ \ }} & \Gamma \left( \overset{\ast }{\pi }%
^{\ast }\left( h^{\ast }F\right) ,\overset{\ast }{\pi }^{\ast }\left(
h^{\ast }\nu \right) ,\overset{\ast }{E}\right)%
\end{array}%
\end{equation*}%
defined by%
\begin{equation*}
\begin{array}{ll}
\left[ T_{\alpha },T_{\beta }\right] _{\overset{\ast }{\pi }^{\ast }\left(
h^{\ast }F\right) } & =\left( L_{\alpha \beta }^{\gamma }\circ h\circ
\overset{\ast }{\pi }\right) T_{\gamma },\vspace*{1mm} \\
\left[ T_{\alpha },fT_{\beta }\right] _{\overset{\ast }{\pi }^{\ast }\left(
h^{\ast }F\right) } & \displaystyle=f\left( L_{\alpha \beta }^{\gamma }\circ
h\circ \overset{\ast }{\pi }\right) T_{\gamma }+\left( \rho _{\alpha
}^{i}\circ h\circ \overset{\ast }{\pi }\right) \frac{\partial f}{\partial
x^{i}}T_{\beta },\vspace*{1mm} \\
\left[ fT_{\alpha },T_{\beta }\right] _{\overset{\ast }{\pi }^{\ast }\left(
h^{\ast }F\right) } & =-\left[ T_{\beta },fT_{\alpha }\right] _{\overset{%
\ast }{\pi }^{\ast }\left( h^{\ast }F\right) },%
\end{array}%
\leqno(2.8)
\end{equation*}%
for any $f\in \mathcal{F}\left( \overset{\ast }{E}\right) ,$ it results that
\begin{equation*}
\begin{array}{c}
\left( \left( \overset{\ast }{\pi }^{\ast }\left( h^{\ast }F\right) ,\overset%
{\ast }{\pi }^{\ast }\left( h^{\ast }\nu \right) ,\overset{\ast }{E}\right) ,%
\left[ ,\right] _{\overset{\ast }{\pi }^{\ast }\left( h^{\ast }F\right)
},\left( \overset{\overset{\ast }{\pi }^{\ast }\left( h^{\ast }F\right) }{%
\rho },Id_{\overset{\ast }{E}}\right) \right)%
\end{array}%
\end{equation*}%
is a Lie algebroid.

\section{Natural and adapted basis}

In the following we consider the following diagram:
\begin{equation*}
\begin{array}{c}
\xymatrix{\overset{\ast }{E}\ar[d]_{\overset{\ast }{\pi }}&\left( F,\left[
,\right] _{F,h},\left( \rho ,\eta \right) \right)\ar[d]^\nu\\ M\ar[r]^h&N}%
\end{array}%
\leqno(3.1)
\end{equation*}%
where $\left( E,\pi ,M\right) \in \left\vert \mathbf{B}^{\mathbf{v}%
}\right\vert $ and $\left( \left( F,\nu ,N\right) ,\left[ ,\right]
_{F,h},\left( \rho ,\eta \right) \right) $ is a generalized Lie algebroid.

Let $\left( \rho ,\eta \right) \Gamma $ be a $\left( \rho ,\eta \right) $%
-connection for the vector bundle $\left( \overset{\ast }{E},\overset{\ast }{%
\pi },M\right) .$

We take $\left( x^{i},p_{a}\right) $ as canonical local coordinates on $%
\left( \overset{\ast }{E},\overset{\ast }{\pi },M\right) ,$ where $i\in
\overline{1,m}$ and $a\in \overline{1,r}.$ Let
\begin{equation*}
\left( x^{i},p_{a}\right) \longrightarrow \left( x^{i%
%TCIMACRO{\U{b4}}%
%BeginExpansion
{\acute{}}%
%EndExpansion
}\left( x^{i}\right) ,p_{a^{\prime }}\left( x^{i},p_{a}\right) \right)
\end{equation*}%
be a change of coordinates on $\left( \overset{\ast }{E},\overset{\ast }{\pi
},M\right) $. Then the coordinates $p_{a}$ change to $p_{a^{\prime }}$ by
the rule:
\begin{equation*}
\begin{array}{c}
p_{a^{\prime }}=M_{a^{\prime }}^{a}p_{a}.%
\end{array}%
\leqno(3.2)
\end{equation*}

Let
\begin{equation*}
\begin{array}[b]{c}
\left( \frac{\partial }{\partial x^{i}},\frac{\partial }{\partial p_{a}}%
\right) \overset{put}{=}\left( \overset{\ast }{\partial }_{i},\overset{\cdot
}{\partial }^{a}\right)%
\end{array}%
\leqno(3.3)
\end{equation*}%
be the natural base of the dual tangent Lie algebroid $\left( \left( T%
\overset{\ast }{E},\tau _{\overset{\ast }{E}},\overset{\ast }{E}\right) ,%
\left[ ,\right] _{T\overset{\ast }{E}},\left( Id_{T\overset{\ast }{E}},Id_{%
\overset{\ast }{E}}\right) \right) .$

For any sections%
\begin{equation*}
\begin{array}{c}
Z^{\alpha }T_{\alpha }\in \Gamma \left( \overset{\ast }{\pi }^{\ast }\left(
h^{\ast }F\right) ,\overset{\ast }{\pi }^{\ast }\left( h^{\ast }F\right) ,%
\overset{\ast }{E}\right)%
\end{array}%
\end{equation*}%
and%
\begin{equation*}
\begin{array}{c}
Y_{a}\displaystyle\overset{\cdot }{\partial }^{a}\in \Gamma \left( VT\overset%
{\ast }{E},\tau _{\overset{\ast }{E}},\overset{\ast }{E}\right)%
\end{array}%
\end{equation*}%
we obtain the section%
\begin{equation*}
\begin{array}{c}
Z^{\alpha }\overset{\ast }{\tilde{\partial}}_{\alpha }+Y_{a}\overset{\cdot }{%
\tilde{\partial}}^{a}=:Z^{\alpha }\left( T_{\alpha }\oplus \left( \rho
_{\alpha }^{i}\circ h\circ \overset{\ast }{\pi }\right) \overset{\ast }{%
\partial }_{i}\right) +Y_{a}\left( 0_{\overset{\ast }{\pi }^{\ast }\left(
h^{\ast }F\right) }\oplus \overset{\cdot }{\partial }^{a}\right) \vspace*{1mm%
} \\
=Z^{\alpha }T_{\alpha }\oplus \left( Z^{\alpha }\left( \rho _{\alpha
}^{i}\circ h\circ \overset{\ast }{\pi }\right) \overset{\ast }{\partial }%
_{i}+Y_{a}\overset{\cdot }{\partial }^{a}\right) \in \Gamma \left( \overset{%
\ast }{\pi }^{\ast }\left( h^{\ast }F\right) \oplus T\overset{\ast }{E},%
\overset{\oplus }{\pi },\overset{\ast }{E}\right) .%
\end{array}%
\end{equation*}

Since we have
\begin{equation*}
\begin{array}{c}
Z^{\alpha }\overset{\ast }{\tilde{\partial}}_{\alpha }+Y_{a}\overset{\cdot }{%
\tilde{\partial}}^{a}=0 \\
\Updownarrow \\
Z^{\alpha }T_{\alpha }=0~\wedge Z^{\alpha }\left( \rho _{\alpha }^{i}\circ
h\circ \overset{\ast }{\pi }\right) \overset{\ast }{\partial }_{i}+Y_{a}%
\overset{\cdot }{\partial }^{a}=0,%
\end{array}%
\end{equation*}%
it implies $Z^{\alpha }=0,~\alpha \in \overline{1,p}$ and $Y_{a}=0,~a\in
\overline{1,r}.$

Therefore, the sections $\displaystyle\overset{\ast }{\tilde{\partial}}%
_{1},...,\overset{\ast }{\tilde{\partial}}_{p},\overset{\cdot }{\tilde{%
\partial}}^{1},...,\overset{\cdot }{\tilde{\partial}}^{r}$ are linearly
independent.\smallskip

We consider the vector subbundle $\left( \left( \rho ,\eta \right) T\overset{%
\ast }{E},\left( \rho ,\eta \right) \tau _{\overset{\ast }{E}},\overset{\ast
}{E}\right) $ of the vector bundle\break $\left( \overset{\ast }{\pi }^{\ast
}\left( h^{\ast }F\right) \oplus T\overset{\ast }{E},\overset{\oplus }{\pi },%
\overset{\ast }{E}\right) ,$ for which the $\mathcal{F}\left( \overset{\ast }%
{E}\right) $-module of sections is the $\mathcal{F}\left( \overset{\ast }{E}%
\right) $-submodule of $\left( \Gamma \left( \overset{\ast }{\pi }^{\ast
}\left( h^{\ast }F\right) \oplus T\overset{\ast }{E},\overset{\oplus }{\pi },%
\overset{\ast }{E}\right) ,+,\cdot \right) ,$ generated by the set of
sections $\left( \overset{\ast }{\tilde{\partial}}_{\alpha },\overset{\cdot }%
{\tilde{\partial}}^{a}\right) $ which is called the \emph{natural }$\left(
\rho ,\eta \right) $\emph{-base.}

The matrix of coordinate transformation on $\left( \left( \rho ,\eta \right)
T\overset{\ast }{E},\left( \rho ,\eta \right) \tau _{\overset{\ast }{E}},%
\overset{\ast }{E}\right) $ at a change of fibred charts is
\begin{equation*}
\left\Vert
\begin{array}{cc}
\Lambda _{\alpha }^{\alpha
%TCIMACRO{\U{b4}}%
%BeginExpansion
{\acute{}}%
%EndExpansion
}\circ h\circ \overset{\ast }{\pi } & 0\vspace*{1mm} \\
\left( \rho _{a}^{i}\circ h\circ \overset{\ast }{\pi }\right) \displaystyle%
\frac{\partial M_{b}^{a%
%TCIMACRO{\U{b4}}%
%BeginExpansion
{\acute{}}%
%EndExpansion
}\circ \overset{\ast }{\pi }}{\partial x_{i}}y^{b} & M_{a}^{a%
%TCIMACRO{\U{b4}}%
%BeginExpansion
{\acute{}}%
%EndExpansion
}\circ \overset{\ast }{\pi }%
\end{array}%
\right\Vert .\leqno(3.4)
\end{equation*}

We have the following

\textbf{Theorem 3.1 }\emph{Let} $\left( \overset{\ast }{\tilde{\rho}},Id_{%
\overset{\ast }{E}}\right) $\ \emph{be the} $\mathbf{B}^{\mathbf{v}}$\emph{%
-morphism of }$\left( \left( \rho ,\eta \right) T\overset{\ast }{E},\left(
\rho ,\eta \right) \tau _{\overset{\ast }{E}},\overset{\ast }{E}\right) $\
\emph{source and }$\left( T\overset{\ast }{E},\tau _{\overset{\ast }{E}},%
\overset{\ast }{E}\right) $\ \emph{target, where}
\begin{equation*}
\begin{array}{rcl}
\left( \rho ,\eta \right) T\overset{\ast }{E}\!\!\! & \!\!^{\underrightarrow{%
~\ \ \overset{\ast }{\tilde{\rho}}}~\ \ }\!\!\! & \!\!T\overset{\ast }{E}%
\vspace*{2mm} \\
\left( Z^{\alpha }\displaystyle\overset{\ast }{\tilde{\partial}}_{\alpha
}+Y_{a}\overset{\cdot }{\tilde{\partial}}^{a}\right) \!(\overset{\ast }{u}%
_{x})\!\!\!\! & \!\!\longmapsto \!\!\! & \!\!\left( \!Z^{\alpha }\!\left(
\rho _{\alpha }^{i}{\circ }h{\circ }\overset{\ast }{\pi }\!\right) \!\overset%
{\ast }{\partial }_{i}{+}Y_{a}\overset{\cdot }{\partial }^{a}\right) \!(%
\overset{\ast }{u}_{x})\!\!%
\end{array}%
\leqno(3.5)
\end{equation*}

\emph{Using the operation}
\begin{equation*}
\begin{array}{ccc}
\Gamma \left( \left( \rho ,\eta \right) T\overset{\ast }{E},\left( \rho
,\eta \right) \tau _{\overset{\ast }{E}},\overset{\ast }{E}\right) ^{2} & ^{%
\underrightarrow{~\ \ \left[ ,\right] _{\left( \rho ,\eta \right) T\overset{%
\ast }{E}}~\ \ }} & \Gamma \left( \left( \rho ,\eta \right) T\overset{\ast }{%
E},\left( \rho ,\eta \right) \tau _{\overset{\ast }{E}},\overset{\ast }{E}%
\right)%
\end{array}%
\end{equation*}%
\emph{defined by}%
\begin{equation*}
\begin{array}{l}
\left[ \left( Z_{1}^{\alpha }\displaystyle\overset{\ast }{\tilde{\partial}}%
_{\alpha }+Y_{a}^{1}\overset{\cdot }{\tilde{\partial}}^{a}\right) ,\left(
Z_{2}^{\beta }\displaystyle\overset{\ast }{\tilde{\partial}}_{\beta
}+Y_{b}^{2}\overset{\cdot }{\tilde{\partial}}^{b}\right) \right] _{\left(
\rho ,\eta \right) T\overset{\ast }{E}}\vspace*{1mm} \\
\displaystyle=\left[ Z_{1}^{\alpha }T_{a},Z_{2}^{\beta }T_{\beta }\right] _{%
\overset{\ast }{\pi }^{\ast }\left( h^{\ast }F\right) }\oplus \left[ \left(
\rho _{\alpha }^{i}\circ h\circ \overset{\ast }{\pi }\right) Z_{1}^{\alpha }%
\overset{\ast }{\partial }_{i}+Y_{a}^{1}\overset{\cdot }{\partial }%
^{a},\right. \vspace*{1mm} \\
\hfill \displaystyle\left. \left( \rho _{\beta }^{j}\circ h\circ \overset{%
\ast }{\pi }\right) Z_{2}^{\beta }\overset{\ast }{\partial }_{j}+Y_{b}^{2}%
\overset{\cdot }{\partial }^{b}\right] _{T\overset{\ast }{E}},%
\end{array}%
\leqno(3.6)
\end{equation*}%
\emph{for any} $\left( Z_{1}^{\alpha }\displaystyle\overset{\ast }{\tilde{%
\partial}}_{\alpha }+Y_{a}^{1}\overset{\cdot }{\tilde{\partial}}^{a}\right) $%
\emph{\ and }$\left( Z_{2}^{\beta }\displaystyle\overset{\ast }{\tilde{%
\partial}}_{\beta }+Y_{b}^{2}\overset{\cdot }{\tilde{\partial}}^{b}\right) ,$
\emph{we obtain that the couple }%
\begin{equation*}
\left( \left[ ,\right] _{\left( \rho ,\eta \right) T\overset{\ast }{E}%
},\left( \overset{\ast }{\tilde{\rho}},Id_{\overset{\ast }{E}}\right) \right)
\end{equation*}%
\emph{\ is a Lie algebroid structure for the vector bundle }$\left( \left(
\rho ,\eta \right) T\overset{\ast }{E},\left( \rho ,\eta \right) \tau _{%
\overset{\ast }{E}},\overset{\ast }{E}\right) .$

The Lie algebroid
\begin{equation*}
\begin{array}{c}
\left( \left( \left( \rho ,\eta \right) T\overset{\ast }{E},\left( \rho
,\eta \right) \tau _{\overset{\ast }{E}},\overset{\ast }{E}\right) ,\left[ ,%
\right] _{\left( \rho ,\eta \right) T\overset{\ast }{E}},\left( \overset{%
\ast }{\tilde{\rho}},Id_{\overset{\ast }{E}}\right) \right)%
\end{array}%
,
\end{equation*}%
is called the \emph{Lie algebroid generalized tangent bundle of dual vector
bundle }$\left( \overset{\ast }{E},\overset{\ast }{\pi },M\right) .$

\textit{Remark 3.1 }The following equalities hold good:%
\begin{equation*}
\begin{array}{c}
\begin{array}[b]{cl}
\left[ \overset{\ast }{\tilde{\partial}}_{\alpha },\overset{\ast }{\tilde{%
\partial}}_{\beta }\right] _{\left( \rho ,\eta \right) T\overset{\ast }{E}}
& =\left( L_{\alpha \beta }^{\gamma }\circ h\circ \overset{\ast }{\pi }%
\right) \overset{\ast }{\tilde{\partial}}_{\gamma } \\
\left[ \overset{\ast }{\tilde{\partial}}_{\alpha },\overset{\cdot }{\tilde{%
\partial}}^{b}\right] _{\left( \rho ,\eta \right) T\overset{\ast }{E}} &
=0_{\left( \rho ,\eta \right) T\overset{\ast }{E}} \\
\left[ \overset{\cdot }{\tilde{\partial}}^{a},\overset{\cdot }{\tilde{%
\partial}}^{b}\right] _{\left( \rho ,\eta \right) T\overset{\ast }{E}} &
=0_{\left( \rho ,\eta \right) T\overset{\ast }{E}}%
\end{array}%
\end{array}%
\leqno(3.7)
\end{equation*}

We consider the $\mathbf{B}^{\mathbf{v}}$-morphism $\left( \left( \rho ,\eta
\right) \overset{\ast }{\pi }!,Id_{\overset{\ast }{E}}\right) $ given by the
commutative diagram%
\begin{equation*}
\begin{array}{c}
\xymatrix{\left( \rho ,\eta \right) T\overset{\ast }{E}\ar[r]^{( \rho ,\eta
) \overset{\ast }{\pi!} }\ar[d]_{(\rho,\eta)\tau_{\overset{\ast }{E}}}&
\overset{\ast }{\pi }^{\ast }\left( h^{\ast }F\right) \ar[d]^{pr_1} \\
\overset{\ast }{E}\ar[r]^{id_{\overset{\ast }{E}}}& \overset{\ast }{E}}%
\end{array}%
\leqno(3.8)
\end{equation*}

Using the components, this is defined as:%
\begin{equation*}
\begin{array}[b]{c}
\left( \rho ,\eta \right) \overset{\ast }{\pi }!\left( \tilde{Z}^{\alpha }%
\overset{\ast }{\tilde{\partial}}_{\alpha }+Y_{a}\overset{\cdot }{\tilde{%
\partial}}^{a}\right) \left( \overset{\ast }{u}_{x}\right) =\left( \tilde{Z}%
^{\alpha }\tilde{T}_{\alpha }\right) \left( \overset{\ast }{u}_{x}\right) ,%
\end{array}%
\leqno(3.9)
\end{equation*}%
for any $\displaystyle\tilde{Z}^{\alpha }\overset{\ast }{\tilde{\partial}}%
_{\alpha }+Y_{a}\overset{\cdot }{\tilde{\partial}}^{a}\in \left( \left( \rho
,\eta \right) T\overset{\ast }{E},\left( \rho ,\eta \right) \tau _{\overset{%
\ast }{E}},\overset{\ast }{E}\right) .$\medskip

Using the $\mathbf{B}^{\mathbf{v}}$-morphism $\left( \left( \rho ,\eta
\right) \overset{\ast }{\pi }!,Id_{\overset{\ast }{E}}\right) $ and the $%
\mathbf{B}^{\mathbf{v}}$-morphism $\left( 2.7\right) $ we obtain the \emph{%
tangent }$\left( \rho ,\eta \right) $\emph{-application }$\left( \left( \rho
,\eta \right) T\overset{\ast }{\pi },h\circ \overset{\ast }{\pi }\right) $%
\emph{\ }of $\left( \left( \rho ,\eta \right) T\overset{\ast }{E},\left(
\rho ,\eta \right) \tau _{\overset{\ast }{E}},\overset{\ast }{E}\right) $
source and $\left( F,\nu ,N\right) $ target.

Using the $\mathbf{B}^{\mathbf{v}}$-morphisms $\left( 2.6\right) $ and $%
\left( 3.7\right) $ we obtain the \emph{tangent }$\left( \rho ,\eta \right) $%
\emph{-application }$\left( \left( \rho ,\eta \right) T\overset{\ast }{\pi }%
,h\circ \overset{\ast }{\pi }\right) $ of $\left( \left( \rho ,\eta \right) T%
\overset{\ast }{E},\left( \rho ,\eta \right) \tau _{\overset{\ast }{E}},%
\overset{\ast }{E}\right) $ source and $\left( F,\nu ,N\right) $ target.

\textbf{Definition 3.1} The kernel of the tangent $\left( \rho ,\eta \right)
$-application\ is written
\begin{equation*}
\left( V\left( \rho ,\eta \right) T\overset{\ast }{E},\left( \rho ,\eta
\right) \tau _{\overset{\ast }{E}},\overset{\ast }{E}\right)
\end{equation*}%
and is called \emph{the vertical subbundle}.\bigskip

We remark that the set $\left\{ \displaystyle\overset{\cdot }{\tilde{\partial%
}}^{a},~a\in \overline{1,r}\right\} $ is a base of the $\mathcal{F}\left(
\overset{\ast }{E}\right) $-module
\begin{equation*}
\left( \Gamma \left( V\left( \rho ,\eta \right) T\overset{\ast }{E},\left(
\rho ,\eta \right) \tau _{\overset{\ast }{E}},\overset{\ast }{E}\right)
,+,\cdot \right) .
\end{equation*}

\textbf{Proposition 3.1} \emph{The short sequence of vector bundles }%
\begin{equation*}
\begin{array}{c}
\xymatrix{0\ar@{^(->}[r]^i\ar[d]&V(\rho,\eta)T\overset{\ast
}{E}\ar[d]\ar@{^(->}[r]^i&(\rho,\eta)T\overset{\ast
}{E}\ar[r]^{(\rho,\eta)\overset{\ast }{\pi }!}\ar[d]&\overset{\ast }{\pi
}^{\ast }\left( h^{\ast }F\right)\ar[r]\ar[d]&0\ar[d]\\ \overset{\ast
}{E}\ar[r]^{Id_{\overset{\ast }{E}}}&\overset{\ast
}{E}\ar[r]^{Id_{\overset{\ast }{E}}}&\overset{\ast
}{E}\ar[r]^{Id_{\overset{\ast }{E}}}&\overset{\ast }
{E}\ar[r]^{Id_{\overset{\ast }{E}}}&\overset{\ast }{E}}%
\end{array}%
\leqno(3.10)
\end{equation*}%
\emph{is exact.}

Let $\left( \rho ,\eta \right) \Gamma $ be a $\left( \rho ,\eta \right) $%
-connection for the vector bundle$\left( \overset{\ast }{E},\overset{\ast }{%
\pi },M\right) ,$ i. e. \textit{a }$\mathbf{Man}$-morphism of $\left( \rho
,\eta \right) T\overset{\ast }{E}$ source and $V\left( \rho ,\eta \right) T%
\overset{\ast }{E}$ target defined by%
\begin{equation*}
\begin{array}[b]{c}
\left( \rho ,\eta \right) \Gamma \left( \tilde{Z}^{\alpha }\overset{\ast }{%
\tilde{\partial}}_{\alpha }+Y_{b}\overset{\cdot }{\tilde{\partial}}%
^{b}\right) \left( \overset{\ast }{u}_{x}\right) =\left( Y_{b}-\left( \rho
,\eta \right) \Gamma _{b\alpha }\tilde{Z}^{\alpha }\right) \overset{\cdot }{%
\tilde{\partial}}^{b}\left( \overset{\ast }{u}_{x}\right) ,%
\end{array}%
\leqno(3.11)
\end{equation*}%
such that the $\mathbf{B}^{\mathbf{v}}$-morphism $\left( \left( \rho ,\eta
\right) \Gamma ,Id_{\overset{\ast }{E}}\right) $ is a split to the left in
the previous exact sequence. Its components satisfy the law of transformation%
\emph{\ }%
\begin{equation*}
\begin{array}[b]{c}
\left( \rho ,\eta \right) \Gamma _{b%
%TCIMACRO{\U{b4}}%
%BeginExpansion
{\acute{}}%
%EndExpansion
\gamma
%TCIMACRO{\U{b4}}%
%BeginExpansion
{\acute{}}%
%EndExpansion
}=M_{b%
%TCIMACRO{\U{b4}}%
%BeginExpansion
{\acute{}}%
%EndExpansion
}^{b}{\circ }\overset{\ast }{\pi }\left[ -\left( \rho _{\gamma }^{i}\circ
h\circ \overset{\ast }{\pi }\right) \frac{\partial M_{b}^{a%
%TCIMACRO{\U{b4}}%
%BeginExpansion
{\acute{}}%
%EndExpansion
}{\circ }\overset{\ast }{\pi }}{\partial x^{i}}p_{a%
%TCIMACRO{\U{b4}}%
%BeginExpansion
{\acute{}}%
%EndExpansion
}+\left( \rho ,\eta \right) \Gamma _{b\gamma }\right] \left( \Lambda
_{\gamma
%TCIMACRO{\U{b4}}%
%BeginExpansion
{\acute{}}%
%EndExpansion
}^{\gamma }\circ h\circ \overset{\ast }{\pi }\right) .%
\end{array}%
\leqno(3.12)
\end{equation*}

The kernel of the $\mathbf{B}^{\mathbf{v}}$-morphism $\left( \left( \rho
,\eta \right) \Gamma ,Id_{\overset{\ast }{E}}\right) $\ is written $\left(
H\left( \rho ,\eta \right) T\overset{\ast }{E},\left( \rho ,\eta \right)
\tau _{\overset{\ast }{E}},\overset{\ast }{E}\right) $ and is called the
\emph{horizontal vector subbundle}.

We remark that the horizontal and the vertical vector subbundles are
interior differential systems of the Lie algebroid generalized tangent
bundle
\begin{equation*}
\begin{array}{c}
\left( \left( \left( \rho ,\eta \right) T\overset{\ast }{E},\left( \rho
,\eta \right) \tau _{\overset{\ast }{E}},\overset{\ast }{E}\right) ,\left[ ,%
\right] _{\left( \rho ,\eta \right) T\overset{\ast }{E}},\left( \overset{%
\ast }{\tilde{\rho}},Id_{\overset{\ast }{E}}\right) \right) .%
\end{array}%
\end{equation*}

We put the problem of finding a base for the $\mathcal{F}\left( \overset{%
\ast }{E}\right) $-module
\begin{equation*}
\left( \Gamma \left( H\left( \rho ,\eta \right) T\overset{\ast }{E},\left(
\rho ,\eta \right) \tau _{\overset{\ast }{E}},\overset{\ast }{E}\right)
,+,\cdot \right)
\end{equation*}%
of the type\textbf{\ }
\begin{equation*}
\begin{array}[t]{l}
\overset{\ast }{\tilde{\delta}}_{\alpha }=Z_{\alpha }^{\beta }\overset{\ast }%
{\tilde{\partial}}_{\beta }+Y_{a\alpha }\overset{\cdot }{\tilde{\partial}}%
^{a},\alpha \in \overline{1,r}%
\end{array}%
\end{equation*}%
which satisfies the following conditions:
\begin{equation*}
\begin{array}{rcl}
\displaystyle\Gamma \left( \left( \rho ,\eta \right) \overset{\ast }{\pi }%
!,Id_{\overset{\ast }{E}}\right) \left( \overset{\ast }{\tilde{\delta}}%
_{\alpha }\right) & = & T_{\alpha }\vspace*{2mm}, \\
\displaystyle\Gamma \left( \left( \rho ,\eta \right) \Gamma ,Id_{\overset{%
\ast }{E}}\right) \left( \overset{\ast }{\tilde{\delta}}_{\alpha }\right) & =
& 0.%
\end{array}%
\leqno(3.13)
\end{equation*}

Then we obtain the sections
\begin{equation*}
\begin{array}[t]{l}
\displaystyle\overset{\ast }{\tilde{\delta}}_{\alpha }=\overset{\ast }{%
\tilde{\partial}}_{\alpha }+\left( \rho ,\eta \right) \Gamma _{b\alpha }%
\overset{\cdot }{\tilde{\partial}}^{b}=T_{\alpha }\oplus \left( \left( \rho
_{\alpha }^{i}\circ h\circ \overset{\ast }{\pi }\right) \overset{\ast }{%
\partial }_{i}-\left( \rho ,\eta \right) \Gamma _{b\alpha }\overset{\cdot }{%
\partial }^{b}\right) .%
\end{array}%
\leqno(3.14)
\end{equation*}%
such that their law of change is a tensorial law under a change of vector
fiber charts.

The base $\left( \overset{\ast }{\tilde{\delta}}_{\alpha },\overset{\cdot }{%
\tilde{\partial}}^{a}\right) $ will be called the \emph{adapted }$\left(
\rho ,\eta \right) $\emph{-base.}

\textit{Remark 3.2 }The following equality holds good%
\begin{equation*}
\begin{array}{l}
\Gamma \left( \overset{\ast }{\tilde{\rho}},Id_{\overset{\ast }{E}}\right)
\left( \overset{\ast }{\tilde{\delta}}_{\alpha }\right) =\left( \rho
_{\alpha }^{i}\circ h\circ \overset{\ast }{\pi }\right) \overset{\ast }{%
\partial }_{i}-\left( \rho ,\eta \right) \Gamma _{b\alpha }\dot{\partial}%
^{b}.%
\end{array}%
\leqno(3.15)
\end{equation*}

Moreover, if $\left( \rho ,\eta \right) \Gamma $ is the $\left( \rho ,\eta
\right) $-connection associated to a connection $\Gamma $ (see $\left[ 1%
\right] $), then we obtain
\begin{equation*}
\begin{array}{l}
\Gamma \left( \overset{\ast }{\tilde{\rho}},Id_{\overset{\ast }{E}}\right)
\left( \overset{\ast }{\tilde{\delta}}_{\alpha }\right) =\left( \rho
_{\alpha }^{i}\circ h\circ \overset{\ast }{\pi }\right) \overset{\ast }{%
\delta }_{i},%
\end{array}%
\leqno(3.16)
\end{equation*}%
where $\left( \overset{\ast }{\delta }_{i},\dot{\partial}^{a}\right) $ is
the adapted base for the $\mathcal{F}\left( \overset{\ast }{E}\right) $%
-module $\left( \Gamma \left( T\overset{\ast }{E},\tau _{\overset{\ast }{E}},%
\overset{\ast }{E}\right) ,+,\cdot \right) .$

\textbf{Theorem 3.2 }\emph{The following equality holds good\ }%
\begin{equation*}
\begin{array}{c}
\left[ \overset{\ast }{\tilde{\delta}}_{\alpha },\overset{\ast }{\tilde{%
\delta}}_{\beta }\right] _{\left( \rho ,\eta \right) T\overset{\ast }{E}%
}=\left( L_{\alpha \beta }^{\gamma }\circ h\circ \overset{\ast }{\pi }%
\right) \overset{\ast }{\tilde{\delta}}_{\gamma }+\left( \rho ,\eta
,h\right) \mathbb{R}_{b\,\ \alpha \beta }\overset{\cdot }{\tilde{\partial}}%
^{b},%
\end{array}%
\leqno(3.17)
\end{equation*}%
\emph{where }%
\begin{equation*}
\begin{array}{ll}
\left( \rho ,\eta ,h\right) \mathbb{R}_{b\,\ \alpha \beta } & \displaystyle%
=\Gamma \left( \overset{\ast }{\tilde{\rho}},Id_{\overset{\ast }{E}}\right)
\left( \overset{\ast }{\tilde{\delta}}_{\beta }\right) \left( \left( \rho
,\eta \right) \Gamma _{b\alpha }\right) \vspace*{2mm} \\
& \displaystyle+\ \Gamma \left( \overset{\ast }{\tilde{\rho}},Id_{\overset{%
\ast }{E}}\right) \left( \overset{\ast }{\tilde{\delta}}_{\alpha }\right)
\left( \left( \rho ,\eta \right) \Gamma _{b\beta }\right) -\left( L_{\alpha
\beta }^{\gamma }\circ h\circ \overset{\ast }{\pi }\right) \left( \rho ,\eta
\right) \Gamma _{b\gamma },%
\end{array}%
\leqno(3.18)
\end{equation*}

\emph{Moreover, we have: }%
\begin{equation*}
\begin{array}{c}
\left[ \overset{\ast }{\tilde{\delta}}_{\alpha },\overset{\cdot }{\tilde{%
\partial}}^{a}\right] _{\left( \rho ,\eta \right) T\overset{\ast }{E}%
}=-\Gamma \left( \overset{\ast }{\tilde{\rho}},Id_{\overset{\ast }{E}%
}\right) \left( \overset{\cdot }{\tilde{\partial}}^{a}\right) \left( \left(
\rho ,\eta \right) \Gamma _{b\alpha }\right) \overset{\cdot }{\tilde{\partial%
}}^{b},%
\end{array}%
\leqno(3.19)
\end{equation*}%
\emph{and}
\begin{equation*}
\begin{array}{c}
\Gamma \left( \overset{\ast }{\tilde{\rho}},Id_{\overset{\ast }{E}}\right) %
\left[ \overset{\ast }{\tilde{\delta}}_{\alpha },\overset{\ast }{\tilde{%
\delta}}_{\beta }\right] _{\left( \rho ,\eta \right) T\overset{\ast }{E}}=%
\left[ \Gamma \left( \overset{\ast }{\tilde{\rho}},Id_{\overset{\ast }{E}%
}\right) \left( \overset{\ast }{\tilde{\delta}}_{\alpha }\right) ,\Gamma
\left( \overset{\ast }{\tilde{\rho}},Id_{\overset{\ast }{E}}\right) \left(
\overset{\ast }{\tilde{\delta}}_{\beta }\right) \right] _{T\overset{\ast }{E}%
}.%
\end{array}%
\leqno(3.20)
\end{equation*}

Let $\left( d\tilde{z}^{\alpha },d\tilde{p}_{a}\right) $ be the natural dual
$\left( \rho ,\eta \right) $-base of natural $\left( \rho ,\eta \right) $%
-base $\left( \displaystyle\overset{\ast }{\partial }_{\alpha },\displaystyle%
\overset{\cdot }{\tilde{\partial}}^{a}\right) .$

This is determined by the equations
\begin{equation*}
\begin{array}{c}
\left\{
\begin{array}{cc}
\displaystyle\left\langle d\tilde{z}^{\alpha },\overset{\ast }{\tilde{%
\partial}}_{\beta }\right\rangle =\delta _{\beta }^{\alpha }, & \displaystyle%
\left\langle d\tilde{z}^{\alpha },\overset{\cdot }{\tilde{\partial}}%
^{b}\right\rangle =0,\vspace*{2mm} \\
\displaystyle\left\langle d\tilde{p}_{a},\overset{\ast }{\tilde{\partial}}%
_{\beta }\right\rangle =0, & \displaystyle\left\langle d\tilde{p}_{a},%
\overset{\cdot }{\tilde{\partial}}^{b}\right\rangle =\delta _{a}^{b}.%
\end{array}%
\right.%
\end{array}%
\end{equation*}

We consider the problem of finding a base for the $\mathcal{F}\left( \overset%
{\ast }{E}\right) $-module
\begin{equation*}
\left( \Gamma \left( \left( V\left( \rho ,\eta \right) T\overset{\ast }{E}%
\right) ^{\ast },\left( \left( \rho ,\eta \right) \tau _{\overset{\ast }{E}%
}\right) ^{\ast },\overset{\ast }{E}\right) ,+,\cdot \right)
\end{equation*}%
of the type
\begin{equation*}
\begin{array}{c}
\delta \tilde{p}_{a}=\theta _{a\alpha }d\tilde{z}^{\alpha }+\omega _{a}^{b}d%
\tilde{p}_{b},~a\in \overline{1,r}%
\end{array}%
\end{equation*}%
which satisfies the following conditions:
\begin{equation*}
\begin{array}{c}
\left\langle \delta \tilde{p}_{a},\overset{\cdot }{\tilde{\partial}}%
^{b}\right\rangle =\delta _{a}^{b}\wedge \left\langle \delta \tilde{p}_{a},%
\overset{\ast }{\tilde{\delta}}_{\alpha }\right\rangle =0,%
\end{array}%
\leqno(3.21)
\end{equation*}

We obtain the sections
\begin{equation*}
\begin{array}{l}
\delta \tilde{p}_{a}=-\left( \rho ,\eta \right) \Gamma _{a\alpha }d\tilde{z}%
^{\alpha }+d\tilde{p}_{a},a\in \overline{1,r}.%
\end{array}%
\leqno(3.22)
\end{equation*}%
such that their changing rule is tensorial under a change of vector fiber
charts. The base $\left( d\tilde{z}^{\alpha },\delta \tilde{p}_{a}\right) $
will be called the \emph{adapted dual }$\left( \rho ,\eta \right) $\emph{%
-base.}

\section{The lift of a differentiable curve}

We consider the following diagram:
\begin{equation*}
\begin{array}{c}
\xymatrix{\overset{\ast }{E}\ar[d]_{\overset{\ast }{\pi }}&\left( F,\left[
,\right] _{F,h},\left( \rho ,\eta \right) \right)\ar[d]^\nu\\ M\ar[r]^h&N}%
\end{array}%
\leqno(4.1)
\end{equation*}%
where $\left( E,\pi ,M\right) \in \left\vert \mathbf{B}^{\mathbf{v}%
}\right\vert $ and $\left( \left( F,\nu ,N\right) ,\left[ ,\right]
_{F,h},\left( \rho ,\eta \right) \right) \in \left\vert \mathbf{GLA}%
\right\vert .$

We admit that $\left( \rho ,\eta \right) \Gamma $ is a $\left( \rho ,\eta
\right) $-connection for the vector bundle $\left( \overset{\ast }{E},%
\overset{\ast }{\pi },M\right) $ and $%
\begin{array}[b]{ccc}
I & ^{\underrightarrow{\ c\ }} & M%
\end{array}%
$ is a differentiable curve. We know that
\begin{equation*}
\begin{array}{c}
\left( \overset{\ast }{E}_{|\func{Im}\left( \eta \circ h\circ c\right) },%
\overset{\ast }{\pi }_{|\func{Im}\left( \eta \circ h\circ c\right) },\func{Im%
}\left( \eta \circ h\circ c\right) \right)%
\end{array}%
\end{equation*}%
is a vector subbundle of the vector bundle $\left( \overset{\ast }{E},%
\overset{\ast }{\pi },M\right) .$

\bigskip \textbf{Definition 4.1} If
\begin{equation*}
\begin{array}{ccl}
I & ^{\underrightarrow{\ \ \dot{c}\ \ }} & \overset{\ast }{E}_{|\func{Im}%
\left( \eta \circ h\circ c\right) }\vspace*{1mm} \\
t & \longmapsto & p_{a}\left( t\right) s^{a}\left( \eta \circ h\circ c\left(
t\right) \right)%
\end{array}%
\leqno(4.2)
\end{equation*}%
is a differentiable curve such that there exists $g\in \mathbf{Man}\left(
\overset{\ast }{E},F\right) $ such that the following conditions are
satisfied:

\begin{itemize}
\item[1.] $\left( g,h\right) \in \mathbf{B}^{v}\left( \left( \overset{\ast }{%
E},\overset{\ast }{\pi },M\right) ,\left( F,\nu ,N\right) \right) $ and

\item[2.] $\rho \circ g\circ \dot{c}\left( t\right) =\displaystyle\frac{%
d\left( \eta \circ h\circ c\right) ^{i}\left( t\right) }{dt}\frac{\partial }{%
\partial x^{i}}\left( \left( \eta \circ h\circ c\right) \left( t\right)
\right) ,$ for any $t\in I,$ \smallskip then we will say that $\dot{c}$\emph{%
\ is the }$\left( g,h\right) $\emph{-lift of the differentiable curve }$c.$
\end{itemize}

\emph{Remark 4.1 }The condition $2$ is equivalent with the following
affirmation:
\begin{equation*}
\begin{array}[b]{c}
\rho _{\alpha }^{i}\left( \eta \circ h\circ c\left( t\right) \right)
g^{\alpha a}\left( h\circ c\left( t\right) \right) p_{a}\left( t\right) =%
\frac{d\left( \eta \circ h\circ c\right) ^{i}\left( t\right) }{dt},~i\in
\overline{1,m}.%
\end{array}%
\leqno(4.3)
\end{equation*}

\textbf{Definition 4.2} If $%
\begin{array}{ccc}
I & ^{\underrightarrow{\ \ \dot{c}\ \ }} & \overset{\ast }{E}_{|\func{Im}%
\left( \eta \circ h\circ c\right) }%
\end{array}%
$ is a differentiable $\left( g,h\right) $-lift of the differentiable curve $%
c,$ then the section%
\begin{equation*}
\begin{array}{ccc}
\func{Im}\left( \eta \circ h\circ c\right) & ^{\underrightarrow{\overset{%
\ast }{u}\left( c,\dot{c}\right) }} & \overset{\ast }{E}_{|\func{Im}\left(
\eta \circ h\circ c\right) }\vspace*{1mm} \\
\eta \circ h\circ c\left( t\right) & \longmapsto & \dot{c}\left( t\right)%
\end{array}%
\leqno(4.4)
\end{equation*}%
will be called the\emph{\ canonical section associated to the couple } $%
\left( c,\dot{c}\right) .$

\textbf{Definition 4.3 }If $\left( g,h\right) \in \mathbf{B}^{\mathbf{v}%
}\left( \left( \overset{\ast }{E},\overset{\ast }{\pi },M\right) ,\left(
F,\nu ,N\right) \right) $ has the components
\begin{equation*}
\begin{array}{c}
g^{\alpha a};a\in \overline{1,r},~\alpha \in \overline{1,p}%
\end{array}%
\end{equation*}%
such that for any vector local $\left( n+p\right) $-chart $\left(
V,t_{V}\right) $ of $\left( F,\nu ,N\right) $ there exists the real
functions
\begin{equation*}
\begin{array}{ccc}
V & ^{\underrightarrow{~\ \ \ \tilde{g}_{a\alpha }~\ \ }} & \mathbb{R}%
\end{array}%
;~a\in \overline{1,r},~\alpha \in \overline{1,p}
\end{equation*}%
such that
\begin{equation*}
\begin{array}{c}
\tilde{g}_{a\alpha }\left( \varkappa \right) \cdot g^{\alpha b}\left(
\varkappa \right) =\delta _{a}^{b},~\forall \varkappa \in V,%
\end{array}%
\leqno(4.4)
\end{equation*}%
then we will say that \emph{the }$\mathbf{B}^{\mathbf{v}}$\emph{-morphism }$%
\left( g,h\right) $\emph{\ is locally invertible.}

\emph{Remark 4.2 }In particular, if $\left( Id_{TM},Id_{M},Id_{M}\right)
=\left( \rho ,\eta ,h\right) $ and the $\mathbf{B}^{\mathbf{v}}$ morphism $%
\left( g,Id_{M}\right) $ is locally invertible, then we have the
differentiable $\left( g,Id_{M}\right) $-lift
\begin{equation*}
\begin{array}{ccl}
I & ^{\underrightarrow{\ \ \dot{c}\ \ }} & \overset{\ast }{TM} \\
t & \longmapsto & \displaystyle\tilde{g}_{ji}\left( c\left( t\right) \right)
\frac{dc^{j}\left( t\right) }{dt}dx^{i}\left( c\left( t\right) \right)%
\end{array}%
.\leqno(4.6)
\end{equation*}

\textbf{Definition 4.4 }If $%
\begin{array}{ccl}
I & ^{\underrightarrow{\ \ \dot{c}\ \ }} & \overset{\ast }{E}_{|\func{Im}%
\left( \eta \circ h\circ c\right) }%
\end{array}%
$ is a differentiable $\left( g,h\right) $-lift for the curve $c$ such that
its components functions $\left( p_{b},~b\in \overline{1,r}\right) $ are
solutions for the differentiable system of equations:%
\begin{equation*}
\begin{array}[b]{c}
\frac{du_{b}}{dt}+\left( \rho ,\eta \right) \Gamma _{b\alpha }\circ \overset{%
\ast }{u}\left( c,\dot{c}\right) \circ \left( \eta \circ h\circ c\right)
\cdot g^{\alpha a}\circ h\circ c\cdot u_{a}=0,%
\end{array}%
\leqno(4.7)
\end{equation*}%
then we will say that \emph{the }$\left( g,h\right) $\emph{-lift }$\dot{c}$%
\emph{\ is parallel with respect to the }$\left( \rho ,\eta \right) $\emph{%
-connection }$\left( \rho ,\eta \right) \overset{\ast }{\Gamma }.$

\emph{Remark 4.3 } $\left( \tilde{g}_{ji}\circ c\cdot \displaystyle\frac{%
dc^{i}}{dt},~j\in \overline{1,m}\right) $ are solutions for the
differentiable system of equations%
\begin{equation*}
\begin{array}[b]{c}
\frac{du_{j}}{dt}+\Gamma _{jk}\circ \overset{\ast }{u}\left( c,\dot{c}%
\right) \circ c\cdot g^{kh}\circ c\cdot u_{h}=0,%
\end{array}%
\leqno(4.8)
\end{equation*}%
namely%
\begin{equation*}
\begin{array}{l}
\displaystyle\frac{d}{dt}\left( \tilde{g}_{ji}\circ c\left( t\right) \cdot
\frac{dc^{i}\left( t\right) }{dt}\right) \vspace*{1mm} \\
\qquad \displaystyle+\Gamma _{jk}\left( c\left( t\right) ,\left( \tilde{g}%
_{ji}\circ c\left( t\right) \cdot \frac{dc^{j}\left( t\right) }{dt}\right)
\cdot dx^{i}\left( c\left( t\right) \right) \right) \cdot \frac{dc^{k}\left(
t\right) }{dt}=0,%
\end{array}%
\leqno(4.8)^{\prime }
\end{equation*}

\section{Remarkable $\mathbf{Mod}$-endomorphisms}

Now, let us consider the following diagram:
\begin{equation*}
\begin{array}{c}
\xymatrix{\overset{\ast }{E}\ar[d]_{\overset{\ast }{\pi }}&\left( F,\left[
,\right] _{F,h},\left( \rho ,\eta \right) \right)\ar[d]^\nu\\ M\ar[r]^h&N}%
\end{array}%
\end{equation*}%
where $\left( E,\pi ,M\right) \in \left\vert \mathbf{B}^{\mathbf{v}%
}\right\vert $ and $\left( \left( F,\nu ,N\right) ,\left[ ,\right]
_{F,h},\left( \rho ,\eta \right) \right) $ is a generalized Lie algebroid.

Let $\left( \rho ,\eta \right) \Gamma $ be a $\left( \rho ,\eta \right) $%
-connection $\ $for the vector bundle $\left( \overset{\ast }{E},\overset{%
\ast }{\pi },M\right) $

\textbf{Definition 5.1 }For any $\mathbf{Mod}$-endomorphism $e$ of
\begin{equation*}
\left( \Gamma \left( \left( \rho ,\eta \right) T\overset{\ast }{E},\left(
\rho ,\eta \right) \tau _{\overset{\ast }{E}},\overset{\ast }{E}\right)
,+,\cdot \right)
\end{equation*}%
we define the application of Nijenhuis \ type
\begin{equation*}
\!\!\Gamma \left( \left( \rho ,\eta \right) T\overset{\ast }{E},\left( \rho
,\eta \right) \tau _{\overset{\ast }{E}},\overset{\ast }{E}\right) ^{2~\
\underrightarrow{~\ \ N_{e}~\ \ }}~\ \Gamma \!\left( \left( \rho ,\eta
\right) T\overset{\ast }{E},\left( \rho ,\eta \right) \tau _{\overset{\ast }{%
E}},\overset{\ast }{E}\right)
\end{equation*}%
defined by
\begin{equation*}
\begin{array}{c}
N_{e}\left( X,Y\right) =\left[ eX,eY\right] _{\left( \rho ,\eta \right) T%
\overset{\ast }{E}}+e^{2}\left[ X,Y\right] _{\left( \rho ,\eta \right) T%
\overset{\ast }{E}}-e\left[ eX,Y\right] _{\left( \rho ,\eta \right) T\overset%
{\ast }{E}}-e\left[ X,eY\right] _{\left( \rho ,\eta \right) T\overset{\ast }{%
E}},%
\end{array}%
\end{equation*}%
for any $X,Y\in \Gamma \!\left( \left( \rho ,\eta \right) T\overset{\ast }{E}%
,\left( \rho ,\eta \right) \tau _{\overset{\ast }{E}},\overset{\ast }{E}%
\right) .$\noindent

\subsection{Projectors}

\textbf{Definition 5.1.1 }Any $\mathbf{Mod}$-endomorphism $e$ of $\Gamma
\left( \left( \rho ,\eta \right) T\overset{\ast }{E},\left( \rho ,\eta
\right) \tau _{\overset{\ast }{E}},\overset{\ast }{E}\right) $ with the
property
\begin{equation*}
\begin{array}{l}
e^{2}=e%
\end{array}%
\leqno(5.1.1)
\end{equation*}%
will be called \emph{projector.}

\textbf{Example 5.1.1 }The $\mathbf{Mod}$-endomorphism
\begin{equation*}
\begin{array}{rcl}
\Gamma \left( \left( \rho ,\eta \right) T\overset{\ast }{E},\left( \rho
,\eta \right) \tau _{\overset{\ast }{E}},\overset{\ast }{E}\right) & ^{%
\underrightarrow{\ \ \overset{\ast }{\mathcal{V}}\ \ }} & \Gamma \left(
\left( \rho ,\eta \right) T\overset{\ast }{E},\left( \rho ,\eta \right) \tau
_{\overset{\ast }{E}},\overset{\ast }{E}\right) \vspace*{2mm} \\
\tilde{Z}^{\alpha }\overset{\ast }{\tilde{\delta}}_{\alpha }+Y_{a}\overset{%
\cdot }{\tilde{\partial}}^{a} & \longmapsto & Y_{a}\overset{\cdot }{\tilde{%
\partial}}^{a}%
\end{array}%
\end{equation*}%
is a projector which will be called the \emph{the vertical projector.}

\emph{\ Remark 5.1.1 } We have $\overset{\ast }{\mathcal{V}}\left( \overset{%
\ast }{\tilde{\delta}}_{\alpha }\right) =0$ and $\overset{\ast }{\mathcal{V}}%
\left( \overset{\cdot }{\tilde{\partial}}^{a}\right) =\overset{\cdot }{%
\tilde{\partial}}^{a}.$ Therefore, it follows\vspace*{-2mm}
\begin{equation*}
\overset{\ast }{\mathcal{V}}\left( \overset{\ast }{\tilde{\partial}}_{\alpha
}\right) =-\left( \rho ,\eta \right) \Gamma _{b\alpha }\overset{\cdot }{%
\tilde{\partial}}^{b}.
\end{equation*}

In addition, we obtain the equality%
\begin{equation*}
\begin{array}[b]{c}
\Gamma \left( \left( \rho ,\eta \right) \Gamma ,Id_{E}\right) \left(
Z^{\alpha }\overset{\ast }{\tilde{\partial}}_{\alpha }+Y_{a}\overset{\cdot }{%
\tilde{\partial}}^{a}\right) =\mathcal{V}\left( Z^{\alpha }\overset{\ast }{%
\tilde{\partial}}_{\alpha }+Y_{a}\overset{\cdot }{\tilde{\partial}}%
^{a}\right) ,%
\end{array}%
\leqno(5.1.2)
\end{equation*}%
for any $Z^{\alpha }\overset{\ast }{\tilde{\partial}}_{\alpha }+Y_{a}\overset%
{\cdot }{\tilde{\partial}}^{a}\in \Gamma \left( \left( \rho ,\eta \right) T%
\overset{\ast }{E},\left( \rho ,\eta \right) \tau _{\overset{\ast }{E}},%
\overset{\ast }{E}\right) .$

\textbf{Theorem 5.1.1 }\emph{A }$(\rho ,\eta )$\emph{-connection for the
vector bundle} $\left( \overset{\ast }{E},\overset{\ast }{\pi },M\right) $
\emph{is characterized by the existence of a} $\mathbf{Mod}$\emph{%
-endomorphism} $\overset{\ast }{\mathcal{V}}$ \emph{of }$\left( \Gamma
\left( \left( \rho ,\eta \right) T\overset{\ast }{E},\left( \rho ,\eta
\right) \tau _{\overset{\ast }{E}},\overset{\ast }{E}\right) ,+,\cdot
\right) $\emph{\ with the properties:}
\begin{equation*}
\begin{array}{c}
\overset{\ast }{\mathcal{V}}\left( \Gamma \left( \left( \rho ,\eta \right) T%
\overset{\ast }{E},\left( \rho ,\eta \right) \tau _{\overset{\ast }{E}},%
\overset{\ast }{E}\right) \right) \subset \Gamma \left( \left( V\left( \rho
,\eta \right) T\overset{\ast }{E},\left( \rho ,\eta \right) \tau _{\overset{%
\ast }{E}},\overset{\ast }{E}\right) \right) \vspace*{1mm} \\
\overset{\ast }{\mathcal{V}}\left( X\right) =X~\Longleftrightarrow ~X\in
\Gamma \left( \left( V\left( \rho ,\eta \right) T\overset{\ast }{E},\left(
\rho ,\eta \right) \tau _{\overset{\ast }{E}},\overset{\ast }{E}\right)
\right)%
\end{array}%
\leqno(5.1.3)
\end{equation*}

\textbf{Example 5.1.2 }The $\mathbf{Mod}$-endomorphism
\begin{equation*}
\begin{array}{rcl}
\Gamma \left( \left( \rho ,\eta \right) T\overset{\ast }{E},\left( \rho
,\eta \right) \tau _{\overset{\ast }{E}},\overset{\ast }{E}\right) & ^{%
\underrightarrow{\ \ \overset{\ast }{\mathcal{H}}\ \ }} & \Gamma \left(
\left( \rho ,\eta \right) T\overset{\ast }{E},\left( \rho ,\eta \right) \tau
_{\overset{\ast }{E}},\overset{\ast }{E}\right) \vspace*{1mm} \\
\tilde{Z}^{\alpha }\overset{\ast }{\tilde{\delta}}_{\alpha }+Y_{a}\overset{%
\cdot }{\tilde{\partial}}^{a} & \longmapsto & \tilde{Z}^{\alpha }\overset{%
\ast }{\tilde{\delta}}_{\alpha }%
\end{array}%
\end{equation*}%
is a projector which will be called the \emph{horizontal projector.}

\emph{Remark 5.1.2} We have $\ \overset{\ast }{\mathcal{H}}\left( \overset{%
\ast }{\tilde{\delta}}_{\alpha }\right) {=}\overset{\ast }{\tilde{\delta}}%
_{\alpha }$ and $\ \overset{\ast }{\mathcal{H}}\left( \overset{\cdot }{%
\tilde{\partial}}^{a}\right) {=}0.$ Therefore, we obtain $\ \overset{\ast }{%
\mathcal{H}}\left( \overset{\ast }{\tilde{\partial}}_{\alpha }\right) {=}%
\overset{\ast }{\tilde{\delta}}_{\alpha }.$

\textbf{Theorem 5.1.2 }\emph{A }$\left( \rho ,\eta \right) $\emph{%
-connection for the vector bundle} $\left( \overset{\ast }{E},\overset{\ast }%
{\pi },M\right) $ \emph{is characterized by the existence of a }$\mathbf{Mod}
$\emph{-endomorphism}$\ \overset{\ast }{\mathcal{H}}$\emph{\ of}
\begin{equation*}
\left( \Gamma \left( \left( \rho ,\eta \right) T\overset{\ast }{E},\left(
\rho ,\eta \right) \tau _{\overset{\ast }{E}},\overset{\ast }{E}\right)
,+,\cdot \right)
\end{equation*}%
\emph{\ with the properties:}
\begin{equation*}
\begin{array}{c}
\ \Gamma \left( \left( \rho ,\eta \right) T\overset{\ast }{E},\left( \rho
,\eta \right) \tau _{\overset{\ast }{E}},\overset{\ast }{E}\right) \subset
\Gamma \left( H\left( \rho ,\eta \right) T\overset{\ast }{E},\left( \rho
,\eta \right) \tau _{\overset{\ast }{E}},\overset{\ast }{E}\right) \vspace*{%
1mm} \\
\ \overset{\ast }{\mathcal{H}}\left( X\right) =X\Longleftrightarrow X\in
\Gamma \left( H\left( \rho ,\eta \right) T\overset{\ast }{E},\left( \rho
,\eta \right) \tau _{\overset{\ast }{E}},\overset{\ast }{E}\right) .%
\end{array}%
\leqno(5.1.4)
\end{equation*}

\textbf{Corollary 5.1.1} \emph{A }$\left( \rho ,\eta \right) $\emph{%
-connection for the vector bundle} $\left( \overset{\ast }{E},\overset{\ast }%
{\pi },M\right) $ \emph{is characterized by the existence of a }$\mathbf{Mod}
$\emph{-endomorphism }$\ \overset{\ast }{\mathcal{H}}$\emph{\ of}
\begin{equation*}
\left( \Gamma \left( \left( \rho ,\eta \right) T\overset{\ast }{E},\left(
\rho ,\eta \right) \tau _{\overset{\ast }{E}},\overset{\ast }{E}\right)
,+,\cdot \right)
\end{equation*}%
\emph{\ with the properties:}
\begin{equation*}
\begin{array}{c}
\ \overset{\ast }{\mathcal{H}}^{2}=\ \overset{\ast }{\mathcal{H}} \\
Ker\left( \overset{\ast }{\mathcal{H}}\right) =\left( \Gamma \left( V\left(
\rho ,\eta \right) T\overset{\ast }{E},\left( \rho ,\eta \right) \tau _{%
\overset{\ast }{E}},\overset{\ast }{E}\right) ,+,\cdot \right) .%
\end{array}%
\leqno(5.1.5)
\end{equation*}

\emph{Remark 5.1.3 }For any
\begin{equation*}
X\in \Gamma \left( \left( \rho ,\eta \right) T\overset{\ast }{E},\left( \rho
,\eta \right) \tau _{\overset{\ast }{E}},\overset{\ast }{E}\right)
\end{equation*}%
we obtain the following unique decomposition
\begin{equation*}
X=\ \overset{\ast }{\mathcal{H}}X+\overset{\ast }{\mathcal{V}}X.
\end{equation*}

\textbf{Proposition 5.1.1} \emph{After some calculations we obtain }%
\begin{equation*}
\begin{array}{c}
N_{\overset{\ast }{\mathcal{V}}}\left( X,Y\right) =\overset{\ast }{\mathcal{V%
}}\left[ \ \overset{\ast }{\mathcal{H}}X,\ \overset{\ast }{\mathcal{H}}Y%
\right] _{\left( \rho ,\eta \right) T\overset{\ast }{E}}=N_{\ \overset{\ast }%
{\mathcal{H}}}\left( X,Y\right) ,%
\end{array}%
\leqno(5.1.6)
\end{equation*}%
for any $X,Y\in \Gamma \left( \left( \rho ,\eta \right) T\overset{\ast }{E}%
,\left( \rho ,\eta \right) \tau _{\overset{\ast }{E}},\overset{\ast }{E}%
\right) .$

\textbf{Corollary 5.1.2} \emph{The horizontal interior differential system }%
\begin{equation*}
\left( H\left( \rho ,\eta \right) T\overset{\ast }{E},\left( \rho ,\eta
\right) \tau _{\overset{\ast }{E}},\overset{\ast }{E}\right)
\end{equation*}%
\emph{is involutive if and only if }$N_{\overset{\ast }{\mathcal{V}}}=0$%
\emph{\ or }$N_{\overset{\ast }{\mathcal{H}}}=0.$

\subsection{The almost product structure}

\textbf{Definition 5.2.1 }Any $\mathbf{Mod}$-endomorphism $e$ of
\begin{equation*}
\left( \Gamma \left( \left( \rho ,\eta \right) T\overset{\ast }{E},\left(
\rho ,\eta \right) \tau _{\overset{\ast }{E}},\overset{\ast }{E}\right)
,+,\cdot \right)
\end{equation*}
with the property%
\begin{equation*}
\begin{array}{c}
e^{2}=Id%
\end{array}%
\leqno(5.2.1)
\end{equation*}%
will be called the \emph{almost product structure}.

\textbf{Example 5.2.1 }The $\mathbf{Mod}$-endomorphism
\begin{equation*}
\begin{array}{rcl}
\Gamma \left( \left( \rho ,\eta \right) T\overset{\ast }{E},\left( \rho
,\eta \right) \tau _{\overset{\ast }{E}},\overset{\ast }{E}\right) & ^{%
\underrightarrow{\ \ \overset{\ast }{\mathcal{P}}\ \ }} & \Gamma \left(
\left( \rho ,\eta \right) T\overset{\ast }{E},\left( \rho ,\eta \right) \tau
_{\overset{\ast }{E}},\overset{\ast }{E}\right) \vspace*{1mm} \\
\tilde{Z}^{\alpha }\overset{\ast }{\tilde{\delta}}_{\alpha }+Y_{a}\overset{%
\cdot }{\tilde{\partial}}^{a} & \longmapsto & \tilde{Z}^{\alpha }\overset{%
\ast }{\tilde{\delta}}_{\alpha }-Y_{a}\overset{\cdot }{\tilde{\partial}}^{a}%
\end{array}%
\end{equation*}%
is an almost product structure.

\emph{\ Remark 5.2.1 }The previous almost product structure has the
properties:
\begin{equation*}
\begin{array}{l}
\overset{\ast }{\mathcal{P}}=2\overset{\ast }{\mathcal{H}}-Id; \\
\overset{\ast }{\mathcal{P}}=Id-2\overset{\ast }{\mathcal{V}}; \\
\overset{\ast }{\mathcal{P}}=\overset{\ast }{\mathcal{H}}-\overset{\ast }{%
\mathcal{V}}.%
\end{array}%
\leqno(5.2.2)
\end{equation*}

\emph{Remark 5.2.2 } We obtain that $\overset{\ast }{\mathcal{P}}\left(
\overset{\ast }{\tilde{\delta}}_{\alpha }\right) =\overset{\ast }{\tilde{%
\delta}}_{\alpha }$ and $\overset{\ast }{\mathcal{P}}\left( \overset{\cdot }{%
\tilde{\partial}}^{a}\right) =-\overset{\cdot }{\tilde{\partial}}^{a}.$
Therefore, it follows \vspace*{-2mm}
\begin{equation*}
\overset{\ast }{\mathcal{P}}\left( \overset{\ast }{\tilde{\partial}}_{\alpha
}\right) =\overset{\ast }{\tilde{\delta}}_{\alpha }-\rho \Gamma _{b\alpha }%
\overset{\cdot }{\tilde{\partial}}^{b}.
\end{equation*}

\textbf{Theorem 5.2.1 }\emph{A }$\left( \rho ,\eta \right) $\emph{%
-connection for the vector bundle }$\left( \overset{\ast }{E},\overset{\ast }%
{\pi },M\right) $ \emph{is characterized by the existence of a }$\mathbf{Mod}
$\emph{-endomorphism }$\overset{\ast }{\mathcal{P}}$\emph{\ of }%
\begin{equation*}
\left( \Gamma \left( \left( \rho ,\eta \right) T\overset{\ast }{E},\left(
\rho ,\eta \right) \tau _{\overset{\ast }{E}},\overset{\ast }{E}\right)
,+,\cdot \right)
\end{equation*}%
\emph{\ with the following property:}
\begin{equation*}
\begin{array}{c}
\overset{\ast }{\mathcal{P}}\left( X\right) =-X\Longleftrightarrow X\in
\Gamma \left( V\left( \rho ,\eta \right) T\overset{\ast }{E},\left( \rho
,\eta \right) \tau _{\overset{\ast }{E}},\overset{\ast }{E}\right) .%
\end{array}%
\leqno(5.2.3)
\end{equation*}

\textbf{Proposition 5.2.1} \emph{After some calculations, we obtain }%
\begin{equation*}
N_{\overset{\ast }{\mathcal{P}}}\left( X,Y\right) =4\overset{\ast }{\mathcal{%
V}}\left[ \overset{\ast }{\mathcal{H}}X,\overset{\ast }{\mathcal{H}}Y\right]
,
\end{equation*}%
\emph{for any }$X,Y\in \Gamma \left( \left( \rho ,\eta \right) T\overset{%
\ast }{E},\left( \rho ,\eta \right) \tau _{\overset{\ast }{E}},\overset{\ast
}{E}\right) .$

\textbf{Corollary 5.2.1} \emph{The horizontal interior differential system }$%
\left( H\left( \rho ,\eta \right) T\overset{\ast }{E},\left( \rho ,\eta
\right) \tau _{\overset{\ast }{E}},\overset{\ast }{E}\right) $ \emph{is
involutive if and only if }$N_{\overset{\ast }{\mathcal{P}}}=0.$

\subsection{\noindent The almost tangent structure}

\textbf{Definition 5.3.1 }Any $\mathbf{Mod}$-endomorphism $e$ of $\left(
\Gamma \!((\rho ,\eta )T\overset{\ast }{E},\left( \rho ,\eta \right) \tau _{%
\overset{\ast }{E}},\overset{\ast }{E}\right) $ with the property
\begin{equation*}
\begin{array}{c}
e^{2}=0%
\end{array}%
\leqno(5.3.1)
\end{equation*}%
will be called the \emph{almost tangent structure.}

\textbf{Example 5.3.1 }If $\left( E,\pi ,M\right) =\left( F,\nu ,N\right) $,
$g\in \mathbf{Man}\left( \overset{\ast }{E},E\right) $ such that $\left(
g,h\right) $ is a locally invertible $\mathbf{B}^{\mathbf{v}}$\textit{-}%
morphism, then the $\mathbf{Mod}$-endomorphism
\begin{equation*}
\begin{array}{rcl}
\Gamma \left( \left( \rho ,\eta \right) T\overset{\ast }{E},\left( \rho
,\eta \right) \tau _{\overset{\ast }{E}},\overset{\ast }{E}\right) & ^{%
\underrightarrow{\overset{\ast }{\mathcal{J}}_{\left( g,h\right) }}} &
\Gamma \left( \left( \rho ,\eta \right) T\overset{\ast }{E},\left( \rho
,\eta \right) \tau _{\overset{\ast }{E}},\overset{\ast }{E}\right) \\
\tilde{Z}^{a}\overset{\ast }{\tilde{\partial}}_{a}+Y_{b}\overset{\cdot }{%
\tilde{\partial}}^{b} & \longmapsto & \left( \tilde{g}_{ba}\circ h\circ
\overset{\ast }{\pi }\right) \tilde{Z}^{a}\overset{\cdot }{\tilde{\partial}}%
^{b}%
\end{array}%
\end{equation*}%
is an almost tangent structure which will be called the \emph{almost tangent
structure associated to the }$\mathbf{B}^{\mathbf{v}}$\emph{-morphism }$%
\left( g,h\right) $. (See: \emph{Definition 4.3}\textbf{)}

\emph{Remark 5.3.1 }We obtain that
\begin{equation*}
\begin{array}{c}
\overset{\ast }{\mathcal{J}}_{\left( g,h\right) }\left( \overset{\ast }{%
\tilde{\delta}}_{a}\right) =\overset{\ast }{\mathcal{J}}_{\left( g,h\right)
}\left( \overset{\ast }{\tilde{\partial}}_{a}\right) =\left( \tilde{g}%
_{ba}\circ h\circ \overset{\ast }{\pi }\right) \overset{\cdot }{\tilde{%
\partial}}^{b}%
\end{array}%
\end{equation*}%
and
\begin{equation*}
\begin{array}{c}
\overset{\ast }{\mathcal{J}}_{\left( g,h\right) }\left( \overset{\cdot }{%
\tilde{\partial}}^{b}\right) =0.%
\end{array}%
\end{equation*}%
and we have the following properties:
\begin{equation*}
\begin{array}{rcl}
\overset{\ast }{\mathcal{J}}_{\left( g,h\right) }\circ \overset{\ast }{%
\mathcal{P}} & = & \overset{\ast }{\mathcal{J}}_{\left( g,h\right) };%
\vspace*{1mm} \\
\overset{\ast }{\mathcal{P}}\circ \overset{\ast }{\mathcal{J}}_{\left(
g,h\right) } & = & -\overset{\ast }{\mathcal{J}}_{\left( g,h\right) };%
\vspace*{1mm} \\
\overset{\ast }{\mathcal{J}}_{\left( g,h\right) }\circ \overset{\ast }{%
\mathcal{H}} & = & \overset{\ast }{\mathcal{J}}_{\left( g,h\right) };%
\vspace*{1mm} \\
\overset{\ast }{\mathcal{H}}\circ \overset{\ast }{\mathcal{J}}_{\left(
g,h\right) } & = & 0;\vspace*{1mm} \\
\overset{\ast }{\mathcal{J}}_{\left( g,h\right) }\circ \overset{\ast }{%
\mathcal{V}} & = & 0;\vspace*{1mm} \\
\overset{\ast }{\mathcal{V}}\circ \overset{\ast }{\mathcal{J}}_{\left(
g,h\right) } & = & \overset{\ast }{\mathcal{J}}_{\left( g,h\right) };%
\vspace*{1mm} \\
N_{\overset{\ast }{\mathcal{J}}_{\left( g,h\right) }} & = & 0.%
\end{array}%
\leqno(5.3.2)
\end{equation*}

\section{Tensor $d$-fields. Distinguished linear $\left( \protect\rho ,%
\protect\eta \right) $-connections}

We consider the following diagram:
\begin{equation*}
\begin{array}{c}
\xymatrix{\overset{\ast }{E}\ar[d]_{\overset{\ast }{\pi }}&\left( F,\left[
,\right] _{F,h},\left( \rho ,\eta \right) \right)\ar[d]^\nu\\ M\ar[r]^h&N}%
\end{array}%
\end{equation*}%
where $\left( E,\pi ,M\right) \in \left\vert \mathbf{B}^{\mathbf{v}%
}\right\vert $ and $\left( \left( F,\nu ,N\right) ,\left[ ,\right]
_{F,h},\left( \rho ,\eta \right) \right) $ is a generalized Lie algebroid.

Let $\left( \rho ,\eta \right) \Gamma $ be a $\left( \rho ,\eta \right) $%
-connection $\ $for the vector bundle $\left( \overset{\ast }{E},\overset{%
\ast }{\pi },M\right) .$

Let
\begin{equation*}
\left( \mathcal{T}~_{q,s}^{p,r}\left( \left( \rho ,\eta \right) T\overset{%
\ast }{E},\left( \rho ,\eta \right) \tau _{\overset{\ast }{E}},\overset{\ast
}{E}\right) ,+,\cdot \right)
\end{equation*}%
be the $\mathcal{F}\left( \overset{\ast }{E}\right) $-module of tensor
fields by $\left( _{q,s}^{p,r}\right) $-type from the generalized tangent
bundle
\begin{equation*}
\left( H\left( \rho ,\eta \right) T\overset{\ast }{E}\oplus V\left( \rho
,\eta \right) T\overset{\ast }{E},\left( \rho ,\eta \right) \tau _{\overset{%
\ast }{E}},\overset{\ast }{E}\right) .
\end{equation*}

An arbitrarily tensor field $T$ is written by the form:
\begin{equation*}
T=T_{\beta _{1}...\beta _{q}b_{1}...b_{s}}^{\alpha _{1}...\alpha
_{p}a_{1}...a_{r}}\overset{\ast }{\tilde{\delta}}_{\alpha _{1}}\otimes
...\otimes \overset{\ast }{\tilde{\delta}}_{\alpha _{p}}\otimes d\tilde{z}%
^{\beta _{1}}\otimes ...\otimes d\tilde{z}^{\beta _{q}}\otimes \overset{%
\cdot }{\tilde{\partial}}^{b_{1}}\otimes ...\otimes \overset{\cdot }{\tilde{%
\partial}}^{b_{s}}\otimes \delta \tilde{p}_{a_{1}}\otimes ...\otimes \delta
\tilde{p}_{a_{r}}.
\end{equation*}

Let
\begin{equation*}
\left( ~\mathcal{T}\left( \left( \rho ,\eta \right) T\overset{\ast }{E}%
,\left( \rho ,\eta \right) \tau _{\overset{\ast }{E}},\overset{\ast }{E}%
\right) ,+,\cdot ,\otimes \right)
\end{equation*}%
be the tensor fields algebra of generalized tangent bundle $\left( \left(
\rho ,\eta \right) T\overset{\ast }{E},\left( \rho ,\eta \right) \tau _{%
\overset{\ast }{E}},\overset{\ast }{E}\right) $.

If $T_{1}{\in }\mathcal{T}_{q_{1},s_{1}}^{p_{1},r_{1}}\left( \left( \rho
,\eta \right) T\overset{\ast }{E},\left( \rho ,\eta \right) \tau _{\overset{%
\ast }{E}},\overset{\ast }{E}\right) $ and $T_{2}{\in }\mathcal{T}%
_{q_{2},s_{2}}^{p_{2},r_{2}}\left( \left( \rho ,\eta \right) T\overset{\ast }%
{E},\left( \rho ,\eta \right) \tau _{\overset{\ast }{E}},\overset{\ast }{E}%
\right) $, then the components of product tensor field $T_{1}\otimes T_{2}$
are the products of local components of $T_{1}$ and $T_{2}.$

Therefore, we obtain $T_{1}\otimes T_{2}\in \mathcal{T}%
_{q_{1}+q_{2},s_{1}+s_{2}}^{p_{1}+p_{2},r_{1}+r_{2}}\left( \left( \rho ,\eta
\right) T\overset{\ast }{E},\left( \rho ,\eta \right) \tau _{\overset{\ast }{%
E}},\overset{\ast }{E}\right) .$

Let $\mathcal{DT}\left( \left( \rho ,\eta \right) T\overset{\ast }{E},\left(
\rho ,\eta \right) \tau _{\overset{\ast }{E}},\overset{\ast }{E}\right) $ be
the family of tensor fields
\begin{equation*}
T\in \mathcal{T}\left( \left( \rho ,\eta \right) T\overset{\ast }{E},\left(
\rho ,\eta \right) \tau _{\overset{\ast }{E}},\overset{\ast }{E}\right)
\end{equation*}%
for which there exists%
\begin{equation*}
T_{1}{\in }\mathcal{T}_{q,0}^{p,0}\left( \left( \rho ,\eta \right) T\overset{%
\ast }{E},\left( \rho ,\eta \right) \tau _{\overset{\ast }{E}},\overset{\ast
}{E}\right)
\end{equation*}%
and%
\begin{equation*}
T_{2}{\in }\mathcal{T}_{0,s}^{0,r}\left( \left( \rho ,\eta \right) T\overset{%
\ast }{E},\left( \rho ,\eta \right) \tau _{\overset{\ast }{E}},\overset{\ast
}{E}\right)
\end{equation*}
such that $T=T_{1}+T_{2}.$

The $\mathcal{F}\left( \overset{\ast }{E}\right) $-module $\left( \mathcal{DT%
}\left( \left( \rho ,\eta \right) T\overset{\ast }{E},\left( \rho ,\eta
\right) \tau _{\overset{\ast }{E}},\overset{\ast }{E}\right) ,+,\cdot
\right) $ will be called the \emph{module of distinguished tensor fields} or
the \emph{module of tensor }$d$-\emph{fields.}

\emph{\ Remark 5.1 }The elements of
\begin{equation*}
\Gamma \left( \left( \rho ,\eta \right) T\overset{\ast }{E},\left( \rho
,\eta \right) \tau _{\overset{\ast }{E}},\overset{\ast }{E}\right)
\end{equation*}%
respectively
\begin{equation*}
\Gamma (((\rho ,\eta )T\overset{\ast }{E})^{\ast },\break ((\rho ,\eta )\tau
_{\overset{\ast }{E}})^{\ast },\overset{\ast }{E})
\end{equation*}%
are tensor $d$-fields.

\textbf{Definition 6.1 }Let $\left( \rho ,\eta \right) \Gamma $ be a $\left(
\rho ,\eta \right) $-connection $\ $for the vector bundle $\left( \overset{%
\ast }{E},\overset{\ast }{\pi },M\right) $ and let
\begin{equation*}
\begin{array}{l}
\left( X,T\right) ^{\underrightarrow{\left( \rho ,\eta \right) \overset{\ast
}{D}}\,}\vspace*{1mm}\left( \rho ,\eta \right) \overset{\ast }{D}_{X}T%
\end{array}%
\leqno(6.4.1)
\end{equation*}%
be a covariant $\left( \rho ,\eta \right) $-derivative for the tensor
algebra of generalized tangent bundle
\begin{equation*}
\left( \left( \rho ,\eta \right) T\overset{\ast }{E},\left( \rho ,\eta
\right) \tau _{\overset{\ast }{E}},\overset{\ast }{E}\right)
\end{equation*}%
which \ preserves \ the \ horizontal and vertical distributions by
parallelism.

If $\left( U,\overset{\ast }{s}_{U}\right) $ is a vector local $\left(
m+r\right) $-chart for $\left( \overset{\ast }{E},\overset{\ast }{\pi }%
,M\right) ,$ then the real local functions
\begin{equation*}
\left( \left( \rho ,\eta \right) \overset{\ast }{H}_{\beta \gamma }^{\alpha
},\left( \rho ,\eta \right) \overset{\ast }{H}_{b\gamma }^{a},\left( \rho
,\eta \right) \overset{\ast }{V}_{\beta }^{\alpha c},\left( \rho ,\eta
\right) \overset{\ast }{V}_{a}^{bc}\right)
\end{equation*}%
defined on $\overset{\ast }{\pi }^{-1}\left( U\right) $ and determined by
the following equalities:
\begin{equation*}
\begin{array}{ll}
\left( \rho ,\eta \right) \overset{\ast }{D}_{\overset{\ast }{\tilde{\delta}}%
_{\gamma }}\overset{\ast }{\tilde{\delta}}_{\beta }=\left( \rho ,\eta
\right) \overset{\ast }{H}_{\beta \gamma }^{\alpha }\overset{\ast }{\tilde{%
\delta}}_{\alpha }, & \left( \rho ,\eta \right) \overset{\ast }{D}_{\overset{%
\ast }{\tilde{\delta}}_{\gamma }}\overset{\cdot }{\tilde{\partial}}%
^{a}=\left( \rho ,\eta \right) \overset{\ast }{H}_{b\gamma }^{a}\overset{%
\cdot }{\tilde{\partial}}^{b} \\
\left( \rho ,\eta \right) \overset{\ast }{D}_{\overset{\cdot }{\tilde{%
\partial}}^{c}}\overset{\ast }{\tilde{\delta}}_{\beta }=\left( \rho ,\eta
\right) \overset{\ast }{V}_{\beta }^{\alpha c}\overset{\ast }{\tilde{\delta}}%
_{\alpha }, & \left( \rho ,\eta \right) \overset{\ast }{D}_{\overset{\cdot }{%
\tilde{\partial}}^{c}}\overset{\cdot }{\tilde{\partial}}^{b}=\left( \rho
,\eta \right) \overset{\ast }{V}_{a}^{bc}\overset{\cdot }{\tilde{\partial}}%
^{a}%
\end{array}%
\leqno(6.2)
\end{equation*}%
are the components of a linear $\left( \rho ,\eta \right) $-connection
\begin{equation*}
\left( \left( \rho ,\eta \right) \overset{\ast }{H},\left( \rho ,\eta
\right) \overset{\ast }{V}\right)
\end{equation*}%
for the generalized tangent bundle $\left( \left( \rho ,\eta \right) T%
\overset{\ast }{E},\left( \rho ,\eta \right) \tau _{\overset{\ast }{E}},%
\overset{\ast }{E}\right) $ which will be called the \emph{distinguished
linear }$\left( \rho ,\eta \right) $\emph{-connection.}

If $h=Id_{M},$ then the distinguished linear $\left( Id_{TM},Id_{M}\right) $%
-connection will be called the \emph{distinguished linear connection.}

The components of a distinguished linear connection $\left( \overset{\ast }{H%
},\overset{\ast }{V}\right) $ will be denoted
\begin{equation*}
\left( \overset{\ast }{H}_{jk}^{i},\overset{\ast }{H}_{bk}^{a},\overset{\ast
}{V}_{j}^{ic},\overset{\ast }{V}_{a}^{bc}\right) .
\end{equation*}

\textbf{Theorem 6.1 }\emph{If }$\left( \left( \rho ,\eta \right) \overset{%
\ast }{H},\left( \rho ,\eta \right) \overset{\ast }{V}\right) $ \emph{is a
distinguished linear} $(\rho ,\eta )$-\emph{connection for the generalized
tangent bundle }$\left( \left( \rho ,\eta \right) T\overset{\ast }{E},\left(
\rho ,\eta \right) \tau _{\overset{\ast }{E}},\overset{\ast }{E}\right) $%
\emph{, then its components satisfy the change relations: }

\begin{equation*}
\begin{array}{ll}
\left( \rho ,\eta \right) \overset{\ast }{H}_{\beta
%TCIMACRO{\U{b4}}%
%BeginExpansion
{\acute{}}%
%EndExpansion
\gamma
%TCIMACRO{\U{b4}}%
%BeginExpansion
{\acute{}}%
%EndExpansion
}^{\alpha
%TCIMACRO{\U{b4}}%
%BeginExpansion
{\acute{}}%
%EndExpansion
}\!\! & =\Lambda _{\alpha }^{\alpha
%TCIMACRO{\U{b4}}%
%BeginExpansion
{\acute{}}%
%EndExpansion
}\circ h\circ \overset{\ast }{\pi }\left[ \Gamma \left( \overset{\ast }{%
\tilde{\rho}},Id_{\overset{\ast }{E}}\right) \left( \overset{\ast }{\tilde{%
\delta}}_{\gamma }\right) \left( \Lambda _{\beta
%TCIMACRO{\U{b4}}%
%BeginExpansion
{\acute{}}%
%EndExpansion
}^{\alpha }\circ h\circ \overset{\ast }{\pi }\right) +\right. \vspace*{1mm}
\\
& +\left. \left( \rho ,\eta \right) \overset{\ast }{H}_{\beta \gamma
}^{\alpha }\cdot \Lambda _{\beta
%TCIMACRO{\U{b4}}%
%BeginExpansion
{\acute{}}%
%EndExpansion
}^{\beta }\circ h\circ \overset{\ast }{\pi }\right] \cdot \Lambda _{\gamma
%TCIMACRO{\U{b4}}%
%BeginExpansion
{\acute{}}%
%EndExpansion
}^{\gamma }\circ h\circ \overset{\ast }{\pi },\vspace*{2mm} \\
\left( \rho ,\eta \right) \overset{\ast }{H}_{b%
%TCIMACRO{\U{b4}}%
%BeginExpansion
{\acute{}}%
%EndExpansion
\gamma
%TCIMACRO{\U{b4}}%
%BeginExpansion
{\acute{}}%
%EndExpansion
}^{a%
%TCIMACRO{\U{b4}}%
%BeginExpansion
{\acute{}}%
%EndExpansion
}\!\! & =M_{a}^{a%
%TCIMACRO{\U{b4}}%
%BeginExpansion
{\acute{}}%
%EndExpansion
}\circ \overset{\ast }{\pi }\left[ \Gamma \left( \overset{\ast }{\tilde{\rho}%
},Id_{\overset{\ast }{E}}\right) \left( \overset{\ast }{\tilde{\delta}}%
_{\gamma }\right) \left( M_{b%
%TCIMACRO{\U{b4}}%
%BeginExpansion
{\acute{}}%
%EndExpansion
}^{a}\circ \overset{\ast }{\pi }\right) +\right. \vspace*{1mm} \\
& \left. +\left( \rho ,\eta \right) \overset{\ast }{H}_{b\gamma }^{a}\cdot
M_{b%
%TCIMACRO{\U{b4}}%
%BeginExpansion
{\acute{}}%
%EndExpansion
}^{b}\circ \overset{\ast }{\pi }\right] \cdot \Lambda _{\gamma
%TCIMACRO{\U{b4}}%
%BeginExpansion
{\acute{}}%
%EndExpansion
}^{\gamma }\circ h\circ \overset{\ast }{\pi },\vspace*{2mm} \\
\left( \rho ,\eta \right) \overset{\ast }{V}_{\beta
%TCIMACRO{\U{b4}}%
%BeginExpansion
{\acute{}}%
%EndExpansion
}^{\alpha
%TCIMACRO{\U{b4}}%
%BeginExpansion
{\acute{}}%
%EndExpansion
c%
%TCIMACRO{\U{b4}}%
%BeginExpansion
{\acute{}}%
%EndExpansion
}\!\! & =\Lambda _{\alpha }^{\alpha
%TCIMACRO{\U{b4}}%
%BeginExpansion
{\acute{}}%
%EndExpansion
}\circ h\circ \overset{\ast }{\pi }\cdot \left( \rho ,\eta \right) \overset{%
\ast }{V}_{\beta }^{\alpha c}\cdot \Lambda _{\beta
%TCIMACRO{\U{b4}}%
%BeginExpansion
{\acute{}}%
%EndExpansion
}^{\beta }\circ h\circ \overset{\ast }{\pi }\cdot M_{c}^{c%
%TCIMACRO{\U{b4}}%
%BeginExpansion
{\acute{}}%
%EndExpansion
}\circ \overset{\ast }{\pi },\vspace*{2mm} \\
\left( \rho ,\eta \right) \overset{\ast }{V}_{b%
%TCIMACRO{\U{b4}}%
%BeginExpansion
{\acute{}}%
%EndExpansion
}^{a%
%TCIMACRO{\U{b4}}%
%BeginExpansion
{\acute{}}%
%EndExpansion
c%
%TCIMACRO{\U{b4}}%
%BeginExpansion
{\acute{}}%
%EndExpansion
}\!\! & =M_{a}^{a%
%TCIMACRO{\U{b4}}%
%BeginExpansion
{\acute{}}%
%EndExpansion
}\circ \overset{\ast }{\pi }\cdot \left( \rho ,\eta \right) \overset{\ast }{V%
}_{b}^{ac}\cdot M_{b%
%TCIMACRO{\U{b4}}%
%BeginExpansion
{\acute{}}%
%EndExpansion
}^{b}\circ \overset{\ast }{\pi }\cdot M_{c}^{c%
%TCIMACRO{\U{b4}}%
%BeginExpansion
{\acute{}}%
%EndExpansion
}\circ \overset{\ast }{\pi }.%
\end{array}%
\leqno(6.3)
\end{equation*}

\emph{The components of a distinguished linear connection }$\left( \overset{%
\ast }{H},\overset{\ast }{V}\right) $\emph{\ verify the change relations:}
\begin{equation*}
\begin{array}{cl}
\overset{\ast }{H}_{j%
%TCIMACRO{\U{b4}}%
%BeginExpansion
{\acute{}}%
%EndExpansion
k%
%TCIMACRO{\U{b4}}%
%BeginExpansion
{\acute{}}%
%EndExpansion
}^{i%
%TCIMACRO{\U{b4}}%
%BeginExpansion
{\acute{}}%
%EndExpansion
} & =\displaystyle\frac{\partial x^{i%
%TCIMACRO{\U{b4}}%
%BeginExpansion
{\acute{}}%
%EndExpansion
}}{\partial x^{i}}\circ \overset{\ast }{\pi }\cdot \left[ \displaystyle\frac{%
\delta }{\delta x^{k}}\left( \displaystyle\frac{\partial x^{i}}{\partial x^{j%
%TCIMACRO{\U{b4}}%
%BeginExpansion
{\acute{}}%
%EndExpansion
}}\circ \overset{\ast }{\pi }\right) +\overset{\ast }{H}_{jk}^{i}\cdot %
\displaystyle\frac{\partial x^{j}}{\partial x^{j%
%TCIMACRO{\U{b4}}%
%BeginExpansion
{\acute{}}%
%EndExpansion
}}\circ \overset{\ast }{\pi }\right] \cdot \displaystyle\frac{\partial x^{k}%
}{\partial x^{k%
%TCIMACRO{\U{b4}}%
%BeginExpansion
{\acute{}}%
%EndExpansion
}}\circ \overset{\ast }{\pi }\vspace*{1mm}, \\
\overset{\ast }{H}_{b%
%TCIMACRO{\U{b4}}%
%BeginExpansion
{\acute{}}%
%EndExpansion
k%
%TCIMACRO{\U{b4}}%
%BeginExpansion
{\acute{}}%
%EndExpansion
}^{a%
%TCIMACRO{\U{b4}}%
%BeginExpansion
{\acute{}}%
%EndExpansion
} & =M_{a}^{a%
%TCIMACRO{\U{b4}}%
%BeginExpansion
{\acute{}}%
%EndExpansion
}\circ \overset{\ast }{\pi }\cdot \left[ \displaystyle\frac{\delta }{\delta
x^{k}}\left( M_{b%
%TCIMACRO{\U{b4}}%
%BeginExpansion
{\acute{}}%
%EndExpansion
}^{a}\circ \overset{\ast }{\pi }\right) +\overset{\ast }{H}_{bk}^{a}\cdot
M_{b%
%TCIMACRO{\U{b4}}%
%BeginExpansion
{\acute{}}%
%EndExpansion
}^{b}\circ \overset{\ast }{\pi }\right] \cdot \displaystyle\frac{\partial
x^{k}}{\partial x^{k%
%TCIMACRO{\U{b4}}%
%BeginExpansion
{\acute{}}%
%EndExpansion
}}\circ \overset{\ast }{\pi },\vspace*{1mm} \\
\overset{\ast }{V}_{j%
%TCIMACRO{\U{b4}}%
%BeginExpansion
{\acute{}}%
%EndExpansion
}^{i%
%TCIMACRO{\U{b4}}%
%BeginExpansion
{\acute{}}%
%EndExpansion
c%
%TCIMACRO{\U{b4}}%
%BeginExpansion
{\acute{}}%
%EndExpansion
} & =\displaystyle\frac{\partial x^{i%
%TCIMACRO{\U{b4}}%
%BeginExpansion
{\acute{}}%
%EndExpansion
}}{\partial x^{i}}\circ \overset{\ast }{\pi }\cdot \overset{\ast }{V}%
_{j}^{ic}\displaystyle\frac{\partial x^{j}}{\partial x^{j%
%TCIMACRO{\U{b4}}%
%BeginExpansion
{\acute{}}%
%EndExpansion
}}\circ \overset{\ast }{\pi }\cdot M_{c%
%TCIMACRO{\U{b4}}%
%BeginExpansion
{\acute{}}%
%EndExpansion
}^{c}\circ \overset{\ast }{\pi }\vspace*{2mm}, \\
\overset{\ast }{V}_{b%
%TCIMACRO{\U{b4}}%
%BeginExpansion
{\acute{}}%
%EndExpansion
}^{a%
%TCIMACRO{\U{b4}}%
%BeginExpansion
{\acute{}}%
%EndExpansion
c%
%TCIMACRO{\U{b4}}%
%BeginExpansion
{\acute{}}%
%EndExpansion
} & =M_{a}^{a%
%TCIMACRO{\U{b4}}%
%BeginExpansion
{\acute{}}%
%EndExpansion
}\circ \overset{\ast }{\pi }\cdot \overset{\ast }{V}_{b}^{ac}\cdot M_{b%
%TCIMACRO{\U{b4}}%
%BeginExpansion
{\acute{}}%
%EndExpansion
}^{b}\circ \overset{\ast }{\pi }M_{c%
%TCIMACRO{\U{b4}}%
%BeginExpansion
{\acute{}}%
%EndExpansion
}^{c}\circ \overset{\ast }{\pi }.%
\end{array}%
\leqno(6.3^{\prime })
\end{equation*}

\textbf{Example 6.1 }If $\left( \overset{\ast }{E},\overset{\ast }{\pi }%
,M\right) $ is endowed with the $\left( \rho ,\eta \right) $-connection $%
\left( \rho ,\eta \right) \Gamma $, then the local real functions
\begin{equation*}
\begin{array}[b]{c}
\left( \frac{\partial \left( \rho ,\eta \right) \Gamma _{b\gamma }}{\partial
p_{a}},\frac{\partial \left( \rho ,\eta \right) \Gamma _{b\gamma }}{\partial
p_{a}},0,0\right)%
\end{array}%
\leqno(6.4)
\end{equation*}%
are the components of a distinguished linear $\left( \rho ,\eta \right) $%
\textit{-}connection for the generalized tangent bundle
\begin{equation*}
\left( \left( \rho ,\eta \right) T\overset{\ast }{E},\left( \rho ,\eta
\right) \tau _{\overset{\ast }{E}},\overset{\ast }{E}\right) ,
\end{equation*}%
which will by called the \emph{Berwald linear }$\left( \rho ,\eta \right) $%
\emph{-connection.}

\textbf{Theorem 6.2} \emph{If the generalized tangent bundle} $\!\left(
\left( \rho ,\eta \right) T\overset{\ast }{E},\left( \rho ,\eta \right) \tau
_{\overset{\ast }{E}},\overset{\ast }{E}\right) $ \emph{is endowed with a
distinguished linear} $\!(\rho ,\!\eta )$\emph{-connection} $((\rho ,\eta )%
\overset{\ast }{H},(\rho ,\eta )\overset{\ast }{V})$, \emph{then, for any}
\begin{equation*}
X=\tilde{Z}^{\gamma }\overset{\ast }{\tilde{\delta}}_{\gamma }+Y_{a}\overset{%
\cdot }{\tilde{\partial}}^{a}\in \Gamma \left( \left( \rho ,\eta \right) T%
\overset{\ast }{E},\left( \rho ,\eta \right) \tau _{\overset{\ast }{E}},%
\overset{\ast }{E}\right)
\end{equation*}%
\emph{and for any}
\begin{equation*}
T\in \mathcal{T}_{qs}^{pr}\!\left( \left( \rho ,\eta \right) T\overset{\ast }%
{E},\left( \rho ,\eta \right) \tau _{\overset{\ast }{E}},\overset{\ast }{E}%
\right) ,
\end{equation*}%
\emph{we obtain the formula:}
\begin{equation*}
\begin{array}{l}
\left( \rho ,\eta \right) D_{X}\left( T_{\beta _{1}...\beta
_{q}b_{1}...b_{s}}^{\alpha _{1}...\alpha _{p}a_{1}...a_{r}}\overset{\ast }{%
\tilde{\delta}}_{\alpha _{1}}\otimes ...\otimes \overset{\ast }{\tilde{\delta%
}}_{\alpha _{p}}\otimes d\tilde{z}^{\beta _{1}}\otimes ...\otimes \right.
\vspace*{1mm} \\
\hspace*{9mm}\left. \otimes d\tilde{z}^{\beta _{q}}\otimes \overset{\cdot }{%
\tilde{\partial}}^{b_{1}}\otimes ...\otimes \overset{\cdot }{\tilde{\partial}%
}^{b_{s}}\otimes \delta \tilde{p}_{a_{1}}\otimes ...\otimes \delta \tilde{p}%
_{a_{r}}\right) =\vspace*{1mm} \\
\hspace*{9mm}=\tilde{Z}^{\gamma }T_{\beta _{1}...\beta _{q}b_{1}...b_{s}\mid
\gamma }^{\alpha _{1}...\alpha _{p}a_{1}...a_{r}}\overset{\ast }{\tilde{%
\delta}}_{\alpha _{1}}\otimes ...\otimes \overset{\ast }{\tilde{\delta}}%
_{\alpha _{p}}\otimes d\tilde{z}^{\beta _{1}}\otimes ...\otimes d\tilde{z}%
^{\beta _{q}}\otimes \overset{\cdot }{\tilde{\partial}}^{b_{1}}\otimes
...\otimes \vspace*{1mm} \\
\hspace*{9mm}\otimes \overset{\cdot }{\tilde{\partial}}^{b_{s}}\otimes
\delta \tilde{p}_{a_{1}}\otimes ...\otimes \delta \tilde{p}%
_{a_{r}}+Y_{c}T_{\beta _{1}...\beta _{q}b_{1}...b_{s}}^{\alpha _{1}...\alpha
_{p}a_{1}...a_{r}}\mid ^{c}\overset{\ast }{\tilde{\delta}}_{\alpha
_{1}}\otimes ...\otimes \vspace*{1mm} \\
\hspace*{9mm}\otimes \overset{\ast }{\tilde{\delta}}_{\alpha _{p}}\otimes d%
\tilde{z}^{\beta _{1}}\otimes ...\otimes d\tilde{z}^{\beta _{q}}\otimes
\overset{\cdot }{\tilde{\partial}}^{b_{1}}\otimes ...\otimes \overset{\cdot }%
{\tilde{\partial}}^{b_{s}}\otimes \delta \tilde{p}_{a_{1}}\otimes ...\otimes
\delta \tilde{p}_{a_{r}},%
\end{array}%
\end{equation*}%
\emph{where\ }%
\begin{equation*}
\begin{array}{l}
T_{\beta _{1}...\beta _{q}b_{1}...b_{s}\mid \gamma }^{\alpha _{1}...\alpha
_{p}a_{1}...a_{r}}=\Gamma \left( \overset{\ast }{\tilde{\rho}},Id_{\overset{%
\ast }{E}}\right) \left( \overset{\ast }{\tilde{\delta}}_{\gamma }\right)
T_{\beta _{1}...\beta _{q}b_{1}...b_{s}}^{\alpha _{1}...\alpha
_{p}a_{1}...a_{r}} \\
\hspace*{8mm}+\left( \rho ,\eta \right) \overset{\ast }{H}_{\alpha \gamma
}^{\alpha _{1}}T_{\beta _{1}...\beta _{q}b_{1}...b_{s}}^{\alpha \alpha
_{2}...\alpha _{p}a_{1}...a_{r}}+...+\vspace*{2mm}\left( \rho ,\eta \right)
\overset{\ast }{H}_{\alpha \gamma }^{\alpha _{p}}T_{\beta _{1}...\beta
_{q}b_{1}...b_{s}}^{\alpha _{1}...\alpha _{p-1}\alpha a_{1}...a_{r}} \\
\hspace*{8mm}-\left( \rho ,\eta \right) \overset{\ast }{H}_{\beta _{1}\gamma
}^{\beta }T_{\beta \beta _{2}...\beta _{q}b_{1}...b_{s}}^{\alpha
_{1}...\alpha _{p}a_{1}...a_{r}}-...-\vspace*{2mm}\left( \rho ,\eta \right)
\overset{\ast }{H}_{\beta _{q}\gamma }^{\beta }T_{\beta _{1}...\beta
_{q-1}\beta b_{1}...b_{s}}^{\alpha _{1}...\alpha _{p}a_{1}...a_{r}} \\
\hspace*{8mm}-\left( \rho ,\eta \right) \overset{\ast }{H}_{a\gamma
}^{a_{1}}T_{\beta _{1}...\beta _{q}b_{1}...b_{s}}^{\alpha _{1}...\alpha
_{p}aa_{2}...a_{r}}-...-\vspace*{2mm}\left( \rho ,\eta \right) \overset{\ast
}{H}_{a\gamma }^{a_{r}}T_{\beta _{1}...\beta _{q}b_{1}...b_{s}}^{\alpha
_{1}...\alpha _{p}a_{1}...a_{r-1}a} \\
\hspace*{8mm}+\left( \rho ,\eta \right) \overset{\ast }{H}_{b_{1}\gamma
}^{b}T_{\beta _{1}...\beta _{q}bb_{2}...b_{s}}^{\alpha _{1}...\alpha
_{p}a_{1}...a_{r}}+\vspace*{2mm}...+\left( \rho ,\eta \right) \overset{\ast }%
{H}_{b_{s}\gamma }^{b}T_{\beta _{1}...\beta _{q}b_{1}...b_{s-1}b}^{\alpha
_{1}...\alpha _{p}a_{1}...a_{r}}%
\end{array}%
\end{equation*}%
\emph{and }%
\begin{equation*}
\begin{array}{l}
T_{\beta _{1}...\beta _{q}b_{1}...b_{s}}^{\alpha _{1}...\alpha
_{p}a_{1}...a_{r}}\mid ^{c}=\Gamma \left( \overset{\ast }{\tilde{\rho}},Id_{%
\overset{\ast }{E}}\right) \left( \overset{\cdot }{\tilde{\partial}}%
^{c}\right) T_{\beta _{1}...\beta _{q}b_{1}...b_{s}}^{\alpha _{1}...\alpha
_{p}a_{1}...a_{r}}+\vspace*{2mm} \\
\hspace*{8mm}+\left( \rho ,\eta \right) \overset{\ast }{V}_{\alpha }^{\alpha
_{1}c}T_{\beta _{1}...\beta _{q}b_{1}...b_{s}}^{\alpha \alpha _{2}...\alpha
_{p}a_{1}...a_{r}}+...+\left( \rho ,\eta \right) \overset{\ast }{V}_{\alpha
}^{\alpha _{p}c}T_{\beta _{1}...\beta _{q}b_{1}...b_{s}}^{\alpha
_{1}...\alpha _{p-1}\alpha a_{1}...a_{r}}\vspace*{2mm} \\
\hspace*{8mm}-\left( \rho ,\eta \right) \overset{\ast }{V}_{\beta
_{1}}^{\beta c}T_{\beta \beta _{2}...\beta _{q}b_{1}...b_{s}}^{\alpha
_{1}...\alpha _{p}a_{1}...a_{r}}-...-\left( \rho ,\eta \right) \overset{\ast
}{V}_{\beta _{q}}^{\beta c}T_{\beta _{1}...\beta _{q-1}\beta
b_{1}...b_{s}}^{\alpha _{1}...\alpha _{p}a_{1}...a_{r}} \\
\hspace*{8mm}-\left( \rho ,\eta \right) \overset{\ast }{V}%
_{a}^{a_{1}c}T_{\beta _{1}...\beta _{q}b_{1}...b_{s}}^{\alpha _{1}...\alpha
_{p}aa_{2}...a_{r}}-...-\left( \rho ,\eta \right) \overset{\ast }{V}%
_{a}^{a_{r}c}T_{\beta _{1}...\beta _{q}b_{1}...b_{s}}^{\alpha _{1}...\alpha
_{p}a_{1}...a_{r-1}a}\vspace*{2mm} \\
\hspace*{8mm}+\left( \rho ,\eta \right) \overset{\ast }{V}%
_{b_{1}}^{bc}T_{\beta _{1}...\beta _{q}bb_{2}...b_{s}}^{\alpha _{1}...\alpha
_{p}a_{1}...a_{r}}...+\left( \rho ,\eta \right) \overset{\ast }{V}%
_{b_{s}}^{bc}T_{\beta _{1}...\beta _{q}b_{1}...b_{s-1}b}^{\alpha
_{1}...\alpha _{p}a_{1}...a_{r}}.%
\end{array}%
\end{equation*}

\textbf{Definition 6.2 }We assume that $\left( E,\pi ,M\right) =\left( F,\nu
,N\right) .$

If $\left( \rho ,\eta \right) \Gamma $ is a $\left( \rho ,\eta \right) $%
-connection for the vector bundle $\left( \overset{\ast }{E},\overset{\ast }{%
\pi },M\right) $ and
\begin{equation*}
\left( \left( \rho ,\eta \right) \overset{\ast }{H}_{bc}^{a},\left( \rho
,\eta \right) \overset{\ast }{\tilde{H}}_{bc}^{a},\left( \rho ,\eta \right)
\overset{\ast }{V}_{b}^{ac},\left( \rho ,\eta \right) \overset{\ast }{\tilde{%
V}}_{b}^{ac}\right)
\end{equation*}%
are the components of a distinguished linear $\left( \rho ,\eta \right) $%
\textit{-}connection for the generalized tangent bundle $\left( \left( \rho
,\eta \right) T\overset{\ast }{E},\left( \rho ,\eta \right) \tau _{\overset{%
\ast }{E}},\overset{\ast }{E}\right) $ such that
\begin{equation*}
\left( \rho ,\eta \right) \overset{\ast }{H}_{bc}^{a}=\left( \rho ,\eta
\right) \overset{\ast }{\tilde{H}}_{bc}^{a}\mbox{ and }\left( \rho ,\eta
\right) \overset{\ast }{V}_{b}^{ac}=\left( \rho ,\eta \right) \overset{\ast }%
{\tilde{V}}_{b}^{ac},
\end{equation*}%
then we will say that \emph{the generalized tangent bundle }$\!\left( \left(
\rho ,\eta \right) T\overset{\ast }{E},\left( \rho ,\eta \right) \tau _{%
\overset{\ast }{E}},\overset{\ast }{E}\right) $ \emph{is endowed with a
normal distinguished linear }$\left( \rho ,\eta \right) $\emph{-connection
on components }%
\begin{equation*}
\left( \left( \rho ,\eta \right) \overset{\ast }{H}_{bc}^{a},\left( \rho
,\eta \right) \overset{\ast }{V}_{b}^{ac}\right) .
\end{equation*}

The components of a normal distinguished linear $\left(
Id_{TM},Id_{M}\right) $-con\-nec\-tion $\left( \overset{\ast }{H},\overset{%
\ast }{V}\right) $ will be denoted $\left( \overset{\ast }{H}_{jk}^{i},%
\overset{\ast }{V}_{jk}^{i}\right) $.

\section{Dual mechanical systems}

Using the diagram:
\begin{equation*}
\begin{array}{c}
\xymatrix{\overset{\ast }{E}\ar[d]_{\overset{\ast }{\pi }}&\left( E,\left[
,\right] _{E,h},\left( \rho ,\eta \right) \right)\ar[d]^\pi\\ M\ar[r]^h&M}%
\end{array}%
\leqno(7.1)
\end{equation*}%
where $\left( \left( E,\pi ,M\right) ,\left[ ,\right] _{E,h},\left( \rho
,\eta \right) \right) $ is a generalized Lie algebroid, we build the
generalized tangent bundle
\begin{equation*}
\begin{array}{c}
(((\rho ,\eta )T\overset{\ast }{E},(\rho ,\eta )\tau _{\overset{\ast }{E}},%
\overset{\ast }{E}),[,]_{(\rho ,\eta )T\overset{\ast }{E}},(\overset{\ast }{%
\tilde{\rho}},Id_{\overset{\ast }{E}})).%
\end{array}%
\end{equation*}

\textbf{Definition 7.1 }A triple
\begin{equation*}
\begin{array}{c}
\left( \left( \overset{\ast }{E},\overset{\ast }{\pi },M\right) ,\overset{%
\ast }{F}_{e},\left( \rho ,\eta \right) \overset{\ast }{\Gamma }\right) ,%
\end{array}%
\leqno(7.2)
\end{equation*}%
where
\begin{equation*}
\begin{array}{c}
\overset{\ast }{F}_{e}=F_{a}\overset{\cdot }{\tilde{\partial}}^{a}\in \Gamma
\left( V\left( \rho ,\eta \right) T\overset{\ast }{E},(\rho ,\eta )\tau _{%
\overset{\ast }{E}},\overset{\ast }{E}\right)%
\end{array}%
\leqno(7.3)
\end{equation*}%
is an external force and $\left( \rho ,\eta \right) \Gamma $ is a $\left(
\rho ,\eta \right) $-connection for $\left( \overset{\ast }{E},\overset{\ast
}{\pi },M\right) $, will be called \emph{dual mechanical }$\left( \rho ,\eta
\right) $\emph{-system.}

\textbf{Definition 7.2 }A smooth \emph{Hamilton fundamental function} on the
dual vector bundle $\left( \overset{\ast }{E},\overset{\ast }{\pi },M\right)
$ is a mapping $%
\begin{array}[b]{ccc}
\overset{\ast }{E} & ^{\underrightarrow{~H\ }} & \mathbb{R}%
\end{array}%
$ which satisfies the following conditions:\medskip

1. $H\circ \overset{\ast }{u}\in C^{\infty }\left( M\right) $, for any $%
\overset{\ast }{u}\in \Gamma \left( \overset{\ast }{E},\overset{\ast }{\pi }%
,M\right) \setminus \left\{ 0\right\} $;\smallskip

2. $H\circ 0\in C^{0}\left( M\right) $, where $0$ means the null section of $%
\left( \overset{\ast }{E},\overset{\ast }{\pi },M\right) .$\medskip

Let $H$ be a differentiable Hamiltonian on the total space of the dual
vector bundle $\left( \overset{\ast }{E},\overset{\ast }{\pi },M\right) .$

If $\left( U,\overset{\ast }{s}_{U}\right) $ is a local vector $\left(
m+r\right) $-chart for $\left( \overset{\ast }{E},\overset{\ast }{\pi }%
,M\right) $, then we obtain the following real functions defined on $\overset%
{\ast }{\pi }^{-1}\left( U\right) $:%
\begin{equation*}
\begin{array}{cc}
H_{i}\overset{put}{=}\displaystyle\frac{\partial H}{\partial x^{i}}\overset{%
put}{=}\frac{\partial }{\partial x^{i}}\left( H\right) & H_{i}^{b}\overset{%
put}{=}\displaystyle\frac{\partial ^{2}H}{\partial x^{i}\partial p_{b}}%
\vspace*{2mm}\overset{put}{=}\frac{\partial }{\partial x^{i}}\left( \frac{%
\partial }{\partial p_{b}}\left( H\right) \right) \\
H^{a}\overset{put}{=}\displaystyle\frac{\partial H}{\partial p_{a}}\overset{%
put}{=}\frac{\partial }{\partial p_{a}}\left( H\right) & H^{ab}\overset{put}{%
=}\displaystyle\frac{\partial ^{2}H}{\partial p_{a}\partial p_{b}}\overset{%
put}{=}\frac{\partial }{\partial p_{a}}\left( \frac{\partial }{\partial p_{b}%
}\left( H\right) \right)%
\end{array}%
.\leqno(7.4)
\end{equation*}

\textbf{Definition 7.3 }If for any local vector $m+r$-chart $\left( U,%
\overset{\ast }{s}_{U}\right) $ of $\left( \overset{\ast }{E},\overset{\ast }%
{\pi },M\right) ,$ we have:
\begin{equation*}
\begin{array}{c}
rank\left\Vert H^{ab}\left( \overset{\ast }{u}_{x}\right) \right\Vert =r,%
\end{array}%
\leqno(7.5)
\end{equation*}%
for any $\overset{\ast }{u}_{x}\in \overset{\ast }{\pi }^{-1}\left( U\right)
\backslash \left\{ 0_{x}\right\} $, then we say that \emph{the Hamiltonian }$%
H$\emph{\ is regular.}

\textbf{Proposition 7.1} If the Hamiltonian $H$ is regular, then for any
local vector $m+r$-chart $\left( U,\overset{\ast }{s}_{U}\right) $ of $%
\left( \overset{\ast }{E},\overset{\ast }{\pi },M\right) ,$ we obtain the
real functions $\tilde{H}_{ba}$ locally defined by%
\begin{equation*}
\begin{array}{ccc}
\overset{\ast }{\pi }^{-1}\left( U\right) & ^{\underrightarrow{\ \tilde{H}%
_{ba}\ }} & \mathbb{R}\vspace*{1mm} \\
\overset{\ast }{u}_{x} & \longmapsto & \tilde{H}_{ba}\left( \overset{\ast }{u%
}_{x}\right)%
\end{array}%
\leqno(7.6)
\end{equation*}%
where $\left\Vert \tilde{H}_{ba}\left( \overset{\ast }{u}_{x}\right)
\right\Vert =\left\Vert H^{ab}\left( \overset{\ast }{u}_{x}\right)
\right\Vert ^{-1}$, for any $\overset{\ast }{u}_{x}\in \overset{\ast }{\pi }%
^{-1}\left( U\right) \backslash \left\{ 0_{x}\right\} $.

\textbf{Definition 7.4 }A smooth \emph{Cartan fundamental function} on the
vector bundle $\left( \overset{\ast }{E},\overset{\ast }{\pi },M\right) $ is
a mapping $%
\begin{array}[b]{ccc}
\overset{\ast }{E} & ^{\underrightarrow{\ K\ }} & \mathbb{R}_{+}%
\end{array}%
$ which satisfies the following conditions:\medskip

1. $K\circ \overset{\ast }{u}\in C^{\infty }\left( M\right) $, for any $%
\overset{\ast }{u}\in \Gamma \left( \overset{\ast }{E},\overset{\ast }{\pi }%
,M\right) \setminus \left\{ 0\right\} $;\smallskip

2. $K\circ 0\in C^{0}\left( M\right) $, where $0$ means the null section of $%
\left( \overset{\ast }{E},\overset{\ast }{\pi },M\right) $;

3. $K$ is positively $1$-homogenous on the fibres of vector bundle $\left(
\overset{\ast }{E},\overset{\ast }{\pi },M\right) ;$

4. For any local vector $m+r$-chart $\left( U,\overset{\ast }{s}_{U}\right) $
of $\left( \overset{\ast }{E},\overset{\ast }{\pi },M\right) ,$ the hessian:%
\begin{equation*}
\left\Vert K^{2~ab}\left( \overset{\ast }{u}_{x}\right) \right\Vert \leqno%
(7.7)
\end{equation*}%
is positively define for any $\overset{\ast }{u}_{x}\in \overset{\ast }{\pi }%
^{-1}\left( U\right) \backslash \left\{ 0_{x}\right\} $.

\medskip \noindent \textbf{Definition 7.5 }If $H$ respectively $K$ is a
smooth Hamilton respectively Cartan function, then we put the triple
\begin{equation*}
\begin{array}{c}
\left( \left( \overset{\ast }{E},\overset{\ast }{\pi },M\right) ,\overset{%
\ast }{F}_{e},H\right) ,%
\end{array}%
\end{equation*}%
respectively
\begin{equation*}
\begin{array}{c}
\left( \left( \overset{\ast }{E},\overset{\ast }{\pi },M\right) ,\overset{%
\ast }{F}_{e},K\right) ,%
\end{array}%
\end{equation*}%
where
\begin{equation*}
\begin{array}{c}
\overset{\ast }{F}_{e}=F_{a}\overset{\cdot }{\tilde{\partial}}^{a}\in \Gamma
\left( V\left( \rho ,\eta \right) T\overset{\ast }{E},(\rho ,\eta )\tau _{%
\overset{\ast }{E}},\overset{\ast }{E}\right)%
\end{array}%
\end{equation*}%
is an external force. These are called \emph{Hamilton mechanical }$\left(
\rho ,\eta \right) $\emph{-system }and \emph{Cartan mechanical }$\left( \rho
,\eta \right) $\emph{-system }respectively.

Any Hamilton mechanical $\left( Id_{TM},Id_{M}\right) $-system and any
Cartan mechanical\break $\ \left( Id_{TM},Id_{M}\right) $-system will be
called \emph{Hamilton mechanical system }and \emph{Cartan mechanical system,
respectively.}

\section{ $(\protect\rho ,\protect\eta )$-semisprays and $(\protect\rho ,%
\protect\eta )$-sprays for dual mechanical $(\protect\rho ,\protect\eta )$%
-systems}

Let $\left( \left( \overset{\ast }{E},\overset{\ast }{\pi },M\right) ,%
\overset{\ast }{F}_{e},\left( \rho ,\eta \right) \Gamma \right) $ be an
arbitrary dual mechanical $\left( \rho ,\eta \right) $-system.

\textbf{Definition 8.1 }The\textit{\ }vertical section $\overset{\ast }{%
\mathbb{C}}\mathbf{=}p_{a}\overset{\cdot }{\tilde{\partial}}^{a}$will be
called the\textit{\ }\emph{Liouville section.}

A section $\overset{\ast }{S}\in \Gamma \left( \left( \rho ,\eta \right) T%
\overset{\ast }{E},\left( \rho ,\eta \right) \tau _{\overset{\ast }{E}},%
\overset{\ast }{E}\right) $\ will be called $\left( \rho ,\eta \right) $%
\emph{-semispray}\ if there exists an almost tangent structure $e$ such that
$e\left( \overset{\ast }{S}\right) =\overset{\ast }{\mathbb{C}}.$

Let $g\in \mathbf{Man}\left( \overset{\ast }{E},E\right) $ be such that $%
\left( g,h\right) $ is a locally invertible $\mathbf{B}^{\mathbf{v}}$%
-morphism of $\left( \overset{\ast }{E},\overset{\ast }{\pi },M\right) $\
source and $\left( E,\pi ,M\right) $\ target.

\textbf{Theorem 8.1 }\emph{The section }%
\begin{equation*}
\begin{array}{l}
\overset{\ast }{S}=\left( g^{ab}\circ h\circ \overset{\ast }{\pi }\right)
p_{b}\overset{\ast }{\tilde{\partial}}_{a}-2\left( G_{a}-\frac{1}{4}%
F_{a}\right) \overset{\cdot }{\tilde{\partial}}^{a}%
\end{array}%
\leqno(8.1)
\end{equation*}%
\emph{is a }$\left( \rho ,\eta \right) $\emph{-semispray such that the real
local functions }$G_{a},\ a\in \overline{1,n},$\emph{\ satisfy the following
conditions}%
\begin{equation*}
\begin{array}{cl}
\left( \rho ,\eta \right) \Gamma _{bc} & =\left( \tilde{g}_{ca}\circ h\circ
\overset{\ast }{\pi }\right) \frac{\partial \left( G_{b}-\frac{1}{4}%
F_{b}\right) }{\partial p_{a}} \\
& -\frac{1}{2}\left( g^{de}\circ h\circ \overset{\ast }{\pi }\right)
p_{e}\left( L_{dc}^{f}\circ h\circ \overset{\ast }{\pi }\right) \left(
\tilde{g}_{fb}\circ h\circ \overset{\ast }{\pi }\right) \\
& +\frac{1}{2}\left( \rho _{c}^{j}\circ h\circ \overset{\ast }{\pi }\right)
\frac{\partial \left( g^{ae}\circ h\circ \overset{\ast }{\pi }\right) }{%
\partial x^{j}}p_{e}\left( \tilde{g}_{ab}\circ h\circ \overset{\ast }{\pi }%
\right) \\
& -\frac{1}{2}\left( g^{ae}\circ h\circ \overset{\ast }{\pi }\right)
p_{e}\left( \rho _{b}^{i}\circ h\circ \overset{\ast }{\pi }\right) \frac{%
\partial \left( \tilde{g}_{ac}\circ h\circ \overset{\ast }{\pi }\right) }{%
\partial x^{i}}%
\end{array}%
\leqno(8.2)
\end{equation*}%
\emph{\ }

\emph{In addition, we remark that the local real functions}%
\begin{equation*}
\begin{array}{cl}
\left( \rho ,\eta \right) \mathring{\Gamma}_{bc} & =\left( \tilde{g}%
_{ca}\circ h\circ \overset{\ast }{\pi }\right) \frac{\partial G_{b}}{%
\partial p_{a}} \\
& -\frac{1}{2}\left( g^{de}\circ h\circ \overset{\ast }{\pi }\right)
p_{e}\left( L_{dc}^{f}\circ h\circ \overset{\ast }{\pi }\right) \left(
\tilde{g}_{fb}\circ h\circ \overset{\ast }{\pi }\right) \\
& +\frac{1}{2}\left( \rho _{c}^{j}\circ h\circ \overset{\ast }{\pi }\right)
\frac{\partial \left( g^{ae}\circ h\circ \overset{\ast }{\pi }\right) }{%
\partial x^{j}}p_{e}\left( \tilde{g}_{ab}\circ h\circ \overset{\ast }{\pi }%
\right) \\
& -\frac{1}{2}\left( g^{ae}\circ h\circ \overset{\ast }{\pi }\right)
p_{e}\left( \rho _{b}^{i}\circ h\circ \overset{\ast }{\pi }\right) \frac{%
\partial \left( \tilde{g}_{ac}\circ h\circ \overset{\ast }{\pi }\right) }{%
\partial x^{i}}%
\end{array}%
\leqno(8.3)
\end{equation*}%
\emph{are the components of a }$\left( \rho ,\eta \right) $\emph{-connection
}$\left( \rho ,\eta \right) \mathring{\Gamma}$\emph{\ for the vector bundle }%
$\left( \overset{\ast }{E},\overset{\ast }{\pi },M\right) .$

The $\left( \rho ,\eta \right) $-semispray $\overset{\ast }{S}$\ will be
called \emph{the\ canonical }$\left( \rho ,\eta \right) $\emph{-semispray
associated to mechanical }$\left( \rho ,\eta \right) $\emph{-system }$\left(
\left( \overset{\ast }{E},\overset{\ast }{\pi },M\right) ,\overset{\ast }{F}%
_{e},\left( \rho ,\eta \right) \Gamma \right) $\emph{\ and from locally
invertible }$\mathbf{B}^{\mathbf{v}}$\emph{-morphism }$\left( g,h\right) .$

\emph{Proof.} We consider the $\mathbf{Mod}$-endomorphism%
\begin{equation*}
\begin{array}{rcl}
\Gamma \left( \left( \rho ,\eta \right) T\overset{\ast }{E},\left( \rho
,\eta \right) \tau _{\overset{\ast }{E}},\overset{\ast }{E}\right) & ^{%
\underrightarrow{\ \ \mathbb{P}\ \ }} & \Gamma \left( \left( \rho ,\eta
\right) T\overset{\ast }{E},\left( \rho ,\eta \right) \tau _{\overset{\ast }{%
E}},\overset{\ast }{E}\right) \vspace*{1mm} \\
X & \longmapsto & \overset{\ast }{\mathcal{J}}_{\left( g,h\right) }\left[
\overset{\ast }{S},X\right] _{\left( \rho ,\eta \right) T\overset{\ast }{E}}-%
\left[ \overset{\ast }{S},\overset{\ast }{\mathcal{J}}_{\left( g,h\right) }X%
\right] _{\left( \rho ,\eta \right) T\overset{\ast }{E}}.%
\end{array}%
\end{equation*}

Let $X=Z^{a}\overset{\ast }{\tilde{\partial}}_{a}+Y_{a}\overset{\cdot }{%
\tilde{\partial}}^{a}$ be an arbitrary section. Since
\begin{equation*}
\begin{array}{cl}
\left[ \overset{\ast }{S},X\right] _{\left( \rho ,\eta \right) T\overset{%
\ast }{E}} & =\displaystyle\left[ \left( g^{ae}\circ h\circ \overset{\ast }{%
\pi }\right) p_{e}\overset{\ast }{\tilde{\partial}}_{a},Z^{b}\overset{\ast }{%
\tilde{\partial}}_{b}\right] _{\left( \rho ,\eta \right) T\overset{\ast }{E}%
}+\left[ \left( g^{ae}\circ h\circ \overset{\ast }{\pi }\right) p_{e}\overset%
{\ast }{\tilde{\partial}}_{a},Y_{b}\overset{\cdot }{\tilde{\partial}}^{b}%
\right] _{\left( \rho ,\eta \right) T\overset{\ast }{E}}\vspace*{2mm} \\
& \displaystyle-\left[ 2\left( G_{a}-\frac{1}{4}F_{a}\right) \overset{\cdot }%
{\tilde{\partial}}^{a},Z^{b}\overset{\ast }{\tilde{\partial}}_{b}\right]
_{\left( \rho ,\eta \right) T\overset{\ast }{E}}-\left[ 2\left( G_{a}-\frac{1%
}{4}F_{a}\right) \overset{\cdot }{\tilde{\partial}}^{a},Y_{b}\overset{\cdot }%
{\tilde{\partial}}^{b}\right] _{\left( \rho ,\eta \right) T\overset{\ast }{E}%
}%
\end{array}%
\end{equation*}%
and
\begin{equation*}
\begin{array}{cl}
\left[ \left( g^{ae}\circ h\circ \overset{\ast }{\pi }\right) p_{e}\overset{%
\ast }{\tilde{\partial}}_{a},Z^{b}\overset{\ast }{\tilde{\partial}}_{b}%
\right] _{\left( \rho ,\eta \right) T\overset{\ast }{E}} & =\displaystyle%
\left( g^{ae}\circ h\circ \overset{\ast }{\pi }\right) p_{e}\left( \rho
_{a}^{i}\circ h\circ \overset{\ast }{\pi }\right) \frac{\partial Z^{c}}{%
\partial x^{i}}\overset{\ast }{\tilde{\partial}}_{c}\vspace*{2mm} \\
& -\displaystyle Z^{b}\left( \rho _{b}^{j}\circ h\circ \overset{\ast }{\pi }%
\right) \frac{\partial \left( g^{ce}\circ h\circ \overset{\ast }{\pi }%
\right) }{\partial x^{j}}p_{e}\overset{\ast }{\tilde{\partial}}_{c}\vspace*{%
2mm} \\
& \displaystyle+\left( g^{ae}\circ h\circ \overset{\ast }{\pi }\right)
p_{e}Z^{b}\left( L_{ab}^{c}\circ h\circ \overset{\ast }{\pi }\right) \overset%
{\ast }{\tilde{\partial}}_{c}\vspace*{2mm},%
\end{array}%
\end{equation*}%
\begin{equation*}
\begin{array}{cl}
\left[ \left( g^{ae}\circ h\circ \overset{\ast }{\pi }\right) p_{e}\overset{%
\ast }{\tilde{\partial}}_{a},Y_{b}\overset{\cdot }{\tilde{\partial}}^{b}%
\right] _{\left( \rho ,\eta \right) T\overset{\ast }{E}} & =\displaystyle%
\left( g^{ae}\circ h\circ \overset{\ast }{\pi }\right) p_{e}\left( \rho
_{a}^{i}\circ h\circ \overset{\ast }{\pi }\right) \frac{\partial Y_{c}}{%
\partial x^{i}}\overset{\cdot }{\tilde{\partial}}^{c} \\
& \displaystyle-Y_{b}\left( g^{bc}\circ h\circ \overset{\ast }{\pi }\right)
\overset{\ast }{\tilde{\partial}}_{c}\vspace*{2mm},%
\end{array}%
\end{equation*}%
\begin{equation*}
\begin{array}{cl}
\displaystyle\left[ 2\left( G_{a}-\frac{1}{4}F_{a}\right) \overset{\cdot }{%
\tilde{\partial}}^{a},Z^{b}\overset{\ast }{\tilde{\partial}}_{b}\vspace*{2mm}%
\right] _{\left( \rho ,\eta \right) T\overset{\ast }{E}} & \displaystyle%
=2\left( G_{a}-\frac{1}{4}F_{a}\right) \frac{\partial Z^{c}}{\partial p_{a}}%
\overset{\ast }{\tilde{\partial}}_{c}\vspace*{2mm} \\
& \displaystyle-2Z^{b}\left( \rho _{b}^{j}\circ h\circ \overset{\ast }{\pi }%
\right) \frac{\partial \left( G_{c}-\frac{1}{4}F_{c}\right) }{\partial x^{j}}%
\overset{\cdot }{\tilde{\partial}}^{c},%
\end{array}%
\end{equation*}%
\begin{equation*}
\begin{array}{cl}
\left[ 2\left( G_{a}-\frac{1}{4}F_{a}\right) \overset{\cdot }{\tilde{\partial%
}}^{a},Y_{b}\overset{\cdot }{\tilde{\partial}}^{b}\right] _{\left( \rho
,\eta \right) T\overset{\ast }{E}} & =2\left( G_{a}-\frac{1}{4}F_{a}\right)
\frac{\partial Y_{c}}{\partial y^{a}}\overset{\cdot }{\tilde{\partial}}%
^{c}-2Y_{b}\displaystyle\frac{\partial \left( G_{c}-\displaystyle\frac{1}{4}%
F_{c}\right) }{\partial p_{b}}\overset{\cdot }{\tilde{\partial}}^{c},%
\end{array}%
\end{equation*}%
it results that
\begin{equation*}
\begin{array}{cl}
\overset{\ast }{\mathcal{J}}_{\left( g,h\right) }\left[ \overset{\ast }{S},X%
\right] _{\left( \rho ,\eta \right) T\overset{\ast }{E}} & \displaystyle%
=\left( g^{ae}\circ h\circ \overset{\ast }{\pi }\right) p_{e}\left( \rho
_{a}^{i}\circ h\circ \overset{\ast }{\pi }\right) \frac{\partial Z^{c}}{%
\partial x^{i}}\left( \tilde{g}_{cd}\circ h\circ \overset{\ast }{\pi }%
\right) \overset{\cdot }{\tilde{\partial}}^{d}\vspace*{2mm} \\
& \displaystyle-Z^{b}\left( \rho _{b}^{j}\circ h\circ \overset{\ast }{\pi }%
\right) \frac{\partial \left( g^{ce}\circ h\circ \overset{\ast }{\pi }%
\right) }{\partial x^{j}}p_{e}\left( \tilde{g}_{cd}\circ h\circ \overset{%
\ast }{\pi }\right) \overset{\cdot }{\tilde{\partial}}^{d}\vspace*{2mm} \\
& \displaystyle+\left( g^{ae}\circ h\circ \overset{\ast }{\pi }\right)
p_{e}Z^{b}\left( L_{ab}^{c}\circ h\circ \overset{\ast }{\pi }\right) \overset%
{\cdot }{\tilde{\partial}}^{d}\vspace*{2mm}\displaystyle-Y_{d}\overset{\cdot
}{\tilde{\partial}}^{d}\vspace*{2mm} \\
& \displaystyle-2\left( G_{a}-\frac{1}{4}F_{a}\right) \frac{\partial Z^{c}}{%
\partial p_{a}}\left( \tilde{g}_{cd}\circ h\circ \overset{\ast }{\pi }%
\right) \overset{\cdot }{\tilde{\partial}}^{d}.%
\end{array}%
\leqno\left( P_{1}\right)
\end{equation*}

Since
\begin{equation*}
\begin{array}{cl}
\left[ S,\overset{\ast }{\mathcal{J}}_{\left( g,h\right) }X\right] _{\left(
\rho ,\eta \right) T\overset{\ast }{E}} & \displaystyle=\left[ \left(
g^{ae}\circ h\circ \overset{\ast }{\pi }\right) p_{e}\overset{\ast }{\tilde{%
\partial}}_{a}\vspace*{2mm},Z^{b}\left( \tilde{g}_{bc}\circ h\circ \overset{%
\ast }{\pi }\right) \overset{\cdot }{\tilde{\partial}}^{c}\right] _{\left(
\rho ,\eta \right) T\overset{\ast }{E}}\vspace*{2mm} \\
& \displaystyle-\left[ 2\left( G_{a}-\frac{1}{4}F_{a}\right) \overset{\cdot }%
{\tilde{\partial}}^{a},Z^{b}\left( \tilde{g}_{bc}\circ h\circ \overset{\ast }%
{\pi }\right) \overset{\cdot }{\tilde{\partial}}^{c}\right] _{\left( \rho
,\eta \right) T\overset{\ast }{E}}%
\end{array}%
\end{equation*}%
and
\begin{equation*}
\begin{array}[b]{cl}
\left[ \left( g^{ae}\circ h\circ \overset{\ast }{\pi }\right) p_{e}\overset{%
\ast }{\tilde{\partial}}_{a}\vspace*{2mm},Z^{b}\left( \tilde{g}_{bc}\circ
h\circ \overset{\ast }{\pi }\right) \overset{\cdot }{\tilde{\partial}}^{c}%
\right] _{\left( \rho ,\eta \right) T\overset{\ast }{E}} & =-Z^{d}\overset{%
\ast }{\tilde{\partial}}_{d} \\
& +\left( g^{ae}\circ h\circ \overset{\ast }{\pi }\right) p_{e}\left( \rho
_{a}^{i}\circ h\circ \overset{\ast }{\pi }\right) \frac{\partial Z^{b}}{%
\partial x^{i}}\left( \tilde{g}_{bd}\circ h\circ \overset{\ast }{\pi }%
\right) \overset{\cdot }{\tilde{\partial}}^{d} \\
& -\left( g^{ae}\circ h\circ \overset{\ast }{\pi }\right) p_{e}\left( \rho
_{a}^{i}\circ h\circ \overset{\ast }{\pi }\right) Z^{b}\frac{\partial \left(
\tilde{g}_{bd}\circ h\circ \overset{\ast }{\pi }\right) }{\partial x^{i}}%
\overset{\cdot }{\tilde{\partial}}^{d}%
\end{array}%
\end{equation*}%
\begin{equation*}
\begin{array}{cl}
\displaystyle\left[ 2\left( G_{a}-\frac{1}{4}F_{a}\right) \overset{\cdot }{%
\tilde{\partial}}^{a},Z^{b}\left( \tilde{g}_{bc}\circ h\circ \overset{\ast }{%
\pi }\right) \overset{\cdot }{\tilde{\partial}}^{c}\right] _{\left( \rho
,\eta \right) T\overset{\ast }{E}} & \displaystyle=2\left( G_{a}-\frac{1}{4}%
F_{a}\right) \frac{\partial Z^{b}}{\partial p_{a}}\left( \tilde{g}_{bd}\circ
h\circ \overset{\ast }{\pi }\right) \overset{\cdot }{\tilde{\partial}}^{d}%
\vspace*{2mm} \\
& \displaystyle-Z^{b}\left( \tilde{g}_{bc}\circ h\circ \overset{\ast }{\pi }%
\right) \frac{\partial 2\left( G_{d}-\frac{1}{4}F_{d}\right) }{\partial p_{c}%
}\overset{\cdot }{\tilde{\partial}}^{d}%
\end{array}%
\end{equation*}%
it results that%
\begin{equation*}
\begin{array}{cl}
\left[ \overset{\ast }{S},\overset{\ast }{\mathcal{J}}_{\left( g,h\right) }X%
\right] _{\left( \rho ,\eta \right) T\overset{\ast }{E}} & \displaystyle%
=-Z^{d}\overset{\ast }{\tilde{\partial}}_{d}\vspace*{2mm}+\left( g^{ae}\circ
h\circ \overset{\ast }{\pi }\right) p_{e}\left( \rho _{a}^{i}\circ h\circ
\overset{\ast }{\pi }\right) \frac{\partial Z^{b}}{\partial x^{i}}\left(
\tilde{g}_{bd}\circ h\circ \overset{\ast }{\pi }\right) \overset{\cdot }{%
\tilde{\partial}}^{d} \\
& \displaystyle-\left( g^{ae}\circ h\circ \overset{\ast }{\pi }\right)
p_{e}\left( \rho _{a}^{i}\circ h\circ \overset{\ast }{\pi }\right) Z^{b}%
\frac{\partial \left( \tilde{g}_{bd}\circ h\circ \overset{\ast }{\pi }%
\right) }{\partial x^{i}}\overset{\cdot }{\tilde{\partial}}^{d} \\
& \displaystyle-2\left( G_{a}-\frac{1}{4}F_{a}\right) \frac{\partial Z^{b}}{%
\partial p_{a}}\left( \tilde{g}_{bd}\circ h\circ \overset{\ast }{\pi }%
\right) \overset{\cdot }{\tilde{\partial}}^{d}\vspace*{2mm} \\
& \displaystyle+Z^{b}\left( \tilde{g}_{bc}\circ h\circ \overset{\ast }{\pi }%
\right) \frac{\partial 2\left( G_{d}-\frac{1}{4}F_{d}\right) }{\partial p_{c}%
}\overset{\cdot }{\tilde{\partial}}^{d}.%
\end{array}%
\leqno\left( P_{2}\right)
\end{equation*}

Using equalities $\left( P_{1}\right) $ and $\left( P_{2}\right) $, we
obtain:%
\begin{equation*}
\begin{array}[b]{cl}
\mathbb{P}\left( Z^{a}\tilde{\partial}_{a}+Y\overset{\cdot }{\tilde{\partial}%
}^{a}\right) & =Z^{a}\overset{\ast }{\tilde{\partial}}_{a}\vspace*{2mm}-Y_{d}%
\overset{\cdot }{\tilde{\partial}}^{d}+\left( g^{ae}\circ h\circ \overset{%
\ast }{\pi }\right) p_{e}Z^{b}\left( L_{ab}^{c}\circ h\circ \overset{\ast }{%
\pi }\right) \left( \tilde{g}_{cd}\circ h\circ \overset{\ast }{\pi }\right)
\overset{\cdot }{\tilde{\partial}}^{d} \\
& -Z^{b}\left( \rho _{b}^{j}\circ h\circ \overset{\ast }{\pi }\right) \frac{%
\partial \left( g^{ce}\circ h\circ \overset{\ast }{\pi }\right) }{\partial
x^{j}}p_{e}\left( \tilde{g}_{cd}\circ h\circ \overset{\ast }{\pi }\right)
\overset{\cdot }{\tilde{\partial}}^{d} \\
& +\left( g^{ae}\circ h\circ \overset{\ast }{\pi }\right) p_{e}\left( \rho
_{a}^{i}\circ h\circ \overset{\ast }{\pi }\right) Z^{b}\frac{\partial \left(
\tilde{g}_{bd}\circ h\circ \overset{\ast }{\pi }\right) }{\partial x^{i}}%
\overset{\cdot }{\tilde{\partial}}^{d} \\
& -Z^{b}\left( \tilde{g}_{bc}\circ h\circ \overset{\ast }{\pi }\right) \frac{%
\partial 2\left( G_{d}-\frac{1}{4}F_{d}\right) }{\partial p_{c}}\overset{%
\cdot }{\tilde{\partial}}^{d}%
\end{array}%
\end{equation*}

After some calculations, it results that $\mathbb{P}$ is an almost product
structure.

Using the equalities $\left( 5.1.2\right) $ and $\left( 5.2.2\right) $ it
results that
\begin{equation*}
\mathbb{P}\left( Z^{a}\overset{\ast }{\tilde{\partial}}_{a}\vspace*{2mm}%
+Y_{a}\overset{\cdot }{\tilde{\partial}}^{a}\right) =\left( Id-2\left( \rho
,\eta \right) \Gamma \right) \left( Z^{a}\overset{\ast }{\tilde{\partial}}%
_{a}\vspace*{2mm}+Y_{a}\overset{\cdot }{\tilde{\partial}}^{a}\right) ,
\end{equation*}%
for any $Z^{a}\overset{\ast }{\tilde{\partial}}_{a}+Y\overset{\cdot }{\tilde{%
\partial}}^{a}\in \Gamma \left( \left( \rho ,\eta \right) T\overset{\ast }{E}%
,\left( \rho ,\eta \right) \tau _{\overset{\ast }{E}},\overset{\ast }{E}%
\right) $ and we obtain%
\begin{equation*}
\begin{array}[b]{cl}
\left( \rho ,\eta \right) \Gamma \left( Z^{a}\overset{\ast }{\tilde{\partial}%
}_{a}+Y\overset{\cdot }{\tilde{\partial}}^{a}\right) & =Y_{d}\overset{\cdot }%
{\tilde{\partial}}^{d}-\frac{1}{2}\left( g^{ae}\circ h\circ \overset{\ast }{%
\pi }\right) p_{e}Z^{b}\left( L_{ab}^{c}\circ h\circ \overset{\ast }{\pi }%
\right) \left( \tilde{g}_{cd}\circ h\circ \overset{\ast }{\pi }\right)
\overset{\cdot }{\tilde{\partial}}^{d} \\
& +\frac{1}{2}Z^{b}\left( \rho _{b}^{j}\circ h\circ \overset{\ast }{\pi }%
\right) \frac{\partial \left( g^{ce}\circ h\circ \overset{\ast }{\pi }%
\right) }{\partial x^{j}}p_{e}\left( \tilde{g}_{cd}\circ h\circ \overset{%
\ast }{\pi }\right) \overset{\cdot }{\tilde{\partial}}^{d} \\
& -\frac{1}{2}\left( g^{ae}\circ h\circ \overset{\ast }{\pi }\right)
p_{e}\left( \rho _{a}^{i}\circ h\circ \overset{\ast }{\pi }\right) Z^{b}%
\frac{\partial \left( \tilde{g}_{bd}\circ h\circ \overset{\ast }{\pi }%
\right) }{\partial x^{i}}\overset{\cdot }{\tilde{\partial}}^{d} \\
& +Z^{b}\left( \tilde{g}_{bc}\circ h\circ \overset{\ast }{\pi }\right) \frac{%
\partial \left( G_{d}-\frac{1}{4}F_{d}\right) }{\partial p_{c}}\overset{%
\cdot }{\tilde{\partial}}^{d}.%
\end{array}%
\end{equation*}

Since
\begin{equation*}
\begin{array}{c}
\left( \rho ,\eta \right) \Gamma \left( Z^{a}\overset{\ast }{\tilde{\partial}%
}_{a}+Y_{a}\overset{\cdot }{\tilde{\partial}}^{a}\right) =\left(
Y_{d}+\left( \rho ,\eta \right) \Gamma _{db}Z^{b}\right) \overset{\cdot }{%
\tilde{\partial}}^{d}%
\end{array}%
\end{equation*}%
it results the relations $\left( 8.3\right) $. In addition, since
\begin{equation*}
\begin{array}[b]{c}
\left( \rho ,\eta \right) \mathring{\Gamma}_{bc}=\left( \rho ,\eta \right)
\Gamma _{bc}+\frac{1}{4}\left( \tilde{g}_{cd}\circ h\circ \overset{\ast }{%
\pi }\right) \frac{\partial F_{b}}{\partial p_{d}}%
\end{array}%
\end{equation*}%
and
\begin{equation*}
\begin{array}{l}
\left( \rho ,\eta \right) \mathring{\Gamma}_{b%
%TCIMACRO{\U{b4}}%
%BeginExpansion
{\acute{}}%
%EndExpansion
c%
%TCIMACRO{\U{b4}}%
%BeginExpansion
{\acute{}}%
%EndExpansion
}=\left( \rho ,\eta \right) \Gamma _{b%
%TCIMACRO{\U{b4}}%
%BeginExpansion
{\acute{}}%
%EndExpansion
c%
%TCIMACRO{\U{b4}}%
%BeginExpansion
{\acute{}}%
%EndExpansion
}-\frac{1}{4}\left( \tilde{g}_{c%
%TCIMACRO{\U{b4}}%
%BeginExpansion
{\acute{}}%
%EndExpansion
e%
%TCIMACRO{\U{b4}}%
%BeginExpansion
{\acute{}}%
%EndExpansion
}\circ h\circ \overset{\ast }{\pi }\right) \frac{\partial F_{b%
%TCIMACRO{\U{b4}}%
%BeginExpansion
{\acute{}}%
%EndExpansion
}}{\partial p_{e%
%TCIMACRO{\U{b4}}%
%BeginExpansion
{\acute{}}%
%EndExpansion
}}\vspace*{1mm} \\
=M_{b%
%TCIMACRO{\U{b4}}%
%BeginExpansion
{\acute{}}%
%EndExpansion
}^{b}\circ \overset{\ast }{\pi }\left( -\left( \rho _{c}^{i}\circ h\circ
\overset{\ast }{\pi }\right) \frac{\partial M_{b}^{a%
%TCIMACRO{\U{b4}}%
%BeginExpansion
{\acute{}}%
%EndExpansion
}}{\partial x^{i}}p_{a%
%TCIMACRO{\U{b4}}%
%BeginExpansion
{\acute{}}%
%EndExpansion
}+\left( \rho ,\eta \right) \Gamma _{bc}\right) M_{c%
%TCIMACRO{\U{b4}}%
%BeginExpansion
{\acute{}}%
%EndExpansion
}^{c}{\circ }h\circ \overset{\ast }{\pi }\vspace*{1mm} \\
+M_{b%
%TCIMACRO{\U{b4}}%
%BeginExpansion
{\acute{}}%
%EndExpansion
}^{b}\circ \overset{\ast }{\pi }\left( \frac{1}{4}\left( \tilde{g}_{ce}\circ
h\circ \overset{\ast }{\pi }\right) \frac{\partial F_{b}}{\partial p_{e}}%
\right) M_{c%
%TCIMACRO{\U{b4}}%
%BeginExpansion
{\acute{}}%
%EndExpansion
}^{c}{\circ }h\circ \overset{\ast }{\pi }\vspace*{1mm} \\
=M_{b%
%TCIMACRO{\U{b4}}%
%BeginExpansion
{\acute{}}%
%EndExpansion
}^{b}\circ \overset{\ast }{\pi }\left( -\left( \rho _{c}^{i}\circ h\circ
\overset{\ast }{\pi }\right) \frac{\partial M_{b}^{a%
%TCIMACRO{\U{b4}}%
%BeginExpansion
{\acute{}}%
%EndExpansion
}}{\partial x^{i}}p_{a%
%TCIMACRO{\U{b4}}%
%BeginExpansion
{\acute{}}%
%EndExpansion
}+\left( \left( \rho ,\eta \right) \Gamma _{bc}-\frac{1}{4}\left( \tilde{g}%
_{ce}\circ h\circ \overset{\ast }{\pi }\right) \cdot \frac{\partial F_{b}}{%
\partial p_{e}}\right) \right) M_{c%
%TCIMACRO{\U{b4}}%
%BeginExpansion
{\acute{}}%
%EndExpansion
}^{c}{\circ }h{\circ }\overset{\ast }{\pi } \\
=M_{b%
%TCIMACRO{\U{b4}}%
%BeginExpansion
{\acute{}}%
%EndExpansion
}^{b}\circ \overset{\ast }{\pi }\left( -\left( \rho _{c}^{i}\circ h\circ
\overset{\ast }{\pi }\right) \frac{\partial M_{b}^{a%
%TCIMACRO{\U{b4}}%
%BeginExpansion
{\acute{}}%
%EndExpansion
}}{\partial x^{i}}p_{a%
%TCIMACRO{\U{b4}}%
%BeginExpansion
{\acute{}}%
%EndExpansion
}+\left( \rho ,\eta \right) \mathring{\Gamma}_{bc}\right) M_{c%
%TCIMACRO{\U{b4}}%
%BeginExpansion
{\acute{}}%
%EndExpansion
}^{c}{\circ }h{\circ }\overset{\ast }{\pi }%
\end{array}%
\end{equation*}%
it results the conclusion of the theorem. \hfill \emph{q.e.d.}

\emph{Remark 8.1 }In particular, if $\left( \rho ,\eta \right) =\left(
Id_{TM},Id_{M}\right) $, $\left( g,h\right) =\left( Id_{E},Id_{M}\right) $,
and $F_{e}=0$, then we obtain the classical canonical semispray associated
to connection $\Gamma $.

Using \emph{Theorem 8.1}, we obtain the following:

\textbf{Theorem 8.2 }\emph{The following properties hold good:}\medskip

$1^{\circ }$\ \emph{Since } $\overset{\ast }{\overset{\circ }{\tilde{\delta}}%
}_{c}=\overset{\ast }{\tilde{\partial}}_{c}+\left( \rho ,\eta \right)
\mathring{\Gamma}_{bc}\overset{\cdot }{\tilde{\partial}}^{b},~c\in \overline{%
1,r},$ \emph{it results that }%
\begin{equation*}
\begin{array}[t]{l}
\overset{\ast }{\overset{\circ }{\tilde{\delta}}}_{c}=\overset{\ast }{\tilde{%
\delta}}_{c}-\frac{1}{4}\left( \tilde{g}_{ce}\circ h\circ \overset{\ast }{%
\pi }\right) \frac{\partial F_{b}}{\partial p_{e}}\overset{\cdot }{\tilde{%
\partial}}^{b},~c\in \overline{1,r}.%
\end{array}%
\leqno(8.4)
\end{equation*}

$2^{\circ }\ $\emph{Since} $\mathring{\delta}\tilde{p}_{b}=-\left( \rho
,\eta \right) \overset{\ast }{\mathring{\Gamma}}_{bc}d\tilde{z}^{c}+d\tilde{p%
}_{b},~b\in \overline{1,r},$ \emph{it results that \ \ }%
\begin{equation*}
\begin{array}[t]{l}
\mathring{\delta}\tilde{p}_{b}=\delta \tilde{p}_{b}+\frac{1}{4}\left( \tilde{%
g}_{ec}\circ h\circ \overset{\ast }{\pi }\right) \frac{\partial F_{b}}{%
\partial \tilde{p}_{e}}d\tilde{z}^{c},~b\in \overline{1,r}.%
\end{array}%
\leqno(8.5)
\end{equation*}

\textbf{Theorem 8.3 }\emph{The real local functions}%
\begin{equation*}
\begin{array}{c}
\left( \frac{\partial \left( \rho ,\eta \right) \Gamma _{bc}}{\partial p_{a}}%
,\frac{\partial \left( \rho ,\eta \right) \Gamma _{bc}}{\partial p_{a}}%
,0,~0\right) ,~a,b,c\in \overline{1,r}%
\end{array}%
\leqno(8.6)
\end{equation*}%
\emph{and}
\begin{equation*}
\begin{array}{c}
\left( \frac{\partial \left( \rho ,\eta \right) \mathring{\Gamma}_{bc}}{%
\partial p_{a}},\frac{\partial \left( \rho ,\eta \right) \mathring{\Gamma}%
_{bc}}{\partial p_{a}},0,~0\right) ,~a,b,c\in \overline{1,r}%
\end{array}%
\leqno(8.6)^{\prime }
\end{equation*}%
\emph{respectively, are the coefficients to a normal Berwald linear }$\left(
\rho ,\eta \right) $\emph{-connection for the generalized tangent bundle }$%
\left( \left( \rho ,\eta \right) T\overset{\ast }{E},\left( \rho ,\eta
\right) \tau _{\overset{\ast }{E}},\overset{\ast }{E}\right) $.

\textbf{Theorem 8.4 }\emph{The tensor of integrability of the }$\left( \rho
,\eta \right) $\emph{-connection }$\left( \rho ,\eta \right) \mathring{\Gamma%
}$\emph{\ is as follows:}%
\begin{equation*}
\begin{array}{c}
\left( \rho ,\eta ,h\right) \mathbb{\mathring{R}}_{b~cd}=\left( \rho ,\eta
,h\right) \mathbb{R}_{b~cd}+\frac{1}{4}\left( \left( \tilde{g}_{de}\circ
h\circ \overset{\ast }{\pi }\right) \frac{\partial F_{b}}{\partial p_{e}}%
_{|c}-\left( \tilde{g}_{ce}\circ h\circ \overset{\ast }{\pi }\right) \frac{%
\partial F_{b}}{\partial p_{e}}_{|d}\right) +\vspace*{1mm} \\
+\frac{1}{16}\left( \left( \tilde{g}_{ed}\circ h\circ \overset{\ast }{\pi }%
\right) \frac{\partial F_{l}}{\partial p_{e}}\left( \tilde{g}_{cf}\circ
h\circ \overset{\ast }{\pi }\right) \frac{\partial ^{2}F_{b}}{\partial
p_{l}\partial p_{f}}-\left( \tilde{g}_{cf}\circ h\circ \overset{\ast }{\pi }%
\right) \frac{\partial F_{l}}{\partial p_{f}}\left( \tilde{g}_{de}\circ
h\circ \overset{\ast }{\pi }\right) \frac{\partial ^{2}F_{b}}{\partial
p_{l}\partial p_{e}}\right) +\vspace*{1mm} \\
+\frac{1}{4}\left( L_{cd}^{f}\circ h\circ \overset{\ast }{\pi }\right)
\left( \tilde{g}_{fe}\circ h\circ \overset{\ast }{\pi }\right) \frac{%
\partial F_{b}}{\partial p_{e}},%
\end{array}%
\leqno(8.7)
\end{equation*}%
\emph{where }$_{|c}$\emph{\ is the }$h$\emph{-covariant derivation with
respect to the normal Berwald linear }$\left( \rho ,\eta \right) $\emph{%
-connection }$(8.6)$\emph{.}

\emph{Proof. }Since
\begin{equation*}
\begin{array}{cl}
\left( \rho ,\eta ,h\right) \mathbb{\mathring{R}}_{b~cd}= & \Gamma \left(
\overset{\ast }{\tilde{\rho}},Id_{\overset{\ast }{E}}\right) \left( \overset{%
\ast }{\overset{\circ }{\tilde{\delta}}}_{c}\right) \left( \left( \rho ,\eta
\right) \mathring{\Gamma}_{bd}\right) -\Gamma \left( \overset{\ast }{\tilde{%
\rho}},Id_{\overset{\ast }{E}}\right) \left( \overset{\ast }{\overset{\circ }%
{\tilde{\delta}}}_{d}\right) \left( \left( \rho ,\eta \right) \mathring{%
\Gamma}_{bc}\right) \\
& -\left( L_{cd}^{e}\circ h\circ \overset{\ast }{\pi }\right) \left( \rho
,\eta \right) \mathring{\Gamma}_{be},%
\end{array}%
\end{equation*}%
and%
\begin{equation*}
\begin{array}{cl}
\Gamma \left( \overset{\ast }{\tilde{\rho}},Id_{\overset{\ast }{E}}\right)
\left( \overset{\ast }{\overset{\circ }{\tilde{\delta}}}_{c}\right) \left(
\left( \rho ,\eta \right) \mathring{\Gamma}_{bd}\right) & =\Gamma \left(
\overset{\ast }{\tilde{\rho}},Id_{\overset{\ast }{E}}\right) \left( \overset{%
\ast }{\tilde{\delta}}_{c}\right) \left( \left( \rho ,\eta \right) \Gamma
_{bd}\right) \\
& +\frac{1}{4}\Gamma \left( \overset{\ast }{\tilde{\rho}},Id_{\overset{\ast }%
{E}}\right) \left( \overset{\ast }{\tilde{\delta}}_{c}\right) \left( \left(
\tilde{g}_{de}\circ h\circ \overset{\ast }{\pi }\right) \frac{\partial F_{b}%
}{\partial p_{e}}\right) \\
& -\frac{1}{4}\left( \tilde{g}_{ce}\circ h\circ \overset{\ast }{\pi }\right)
\frac{\partial F_{f}}{\partial p_{e}}\frac{\partial }{\partial p_{f}}\left(
\left( \rho ,\eta \right) \Gamma _{bd}\right) \\
& -\frac{1}{16}\left( \tilde{g}_{ce}\circ h\circ \overset{\ast }{\pi }%
\right) \frac{\partial F_{f}}{\partial p_{e}}\frac{\partial }{\partial p_{f}}%
\left( \left( \tilde{g}_{de}\circ h\circ \overset{\ast }{\pi }\right) \frac{%
\partial F_{b}}{\partial p_{e}}\right) ,%
\end{array}%
\end{equation*}%
\begin{equation*}
\begin{array}{cl}
\Gamma \left( \overset{\ast }{\tilde{\rho}},Id_{\overset{\ast }{E}}\right)
\left( \overset{\ast }{\overset{\circ }{\tilde{\delta}}}_{d}\right) \left(
\left( \rho ,\eta \right) \mathring{\Gamma}_{bc}\right) & =\Gamma \left(
\overset{\ast }{\tilde{\rho}},Id_{\overset{\ast }{E}}\right) \left( \overset{%
\ast }{\tilde{\delta}}_{d}\right) \left( \left( \rho ,\eta \right) \Gamma
_{bc}\right) \\
& +\frac{1}{4}\Gamma \left( \overset{\ast }{\tilde{\rho}},Id_{\overset{\ast }%
{E}}\right) \left( \overset{\ast }{\tilde{\delta}}_{d}\right) \left( \left(
\tilde{g}_{ce}\circ h\circ \overset{\ast }{\pi }\right) \frac{\partial F_{b}%
}{\partial p_{e}}\right) \\
& -\frac{1}{4}\left( \tilde{g}_{de}\circ h\circ \overset{\ast }{\pi }\right)
\frac{\partial F_{f}}{\partial p_{e}}\frac{\partial }{\partial p_{f}}\left(
\left( \rho ,\eta \right) \Gamma _{bc}\right) \\
& -\frac{1}{16}\left( \tilde{g}_{de}\circ h\circ \overset{\ast }{\pi }%
\right) \frac{\partial F_{f}}{\partial p_{e}}\frac{\partial }{\partial p_{f}}%
\left( \left( \tilde{g}_{ce}\circ h\circ \overset{\ast }{\pi }\right) \frac{%
\partial F_{b}}{\partial p_{e}}\right) ,%
\end{array}%
\end{equation*}%
\begin{equation*}
\begin{array}{cl}
\left( L_{cd}^{e}\circ h\circ \overset{\ast }{\pi }\right) \left( \rho ,\eta
\right) \mathring{\Gamma}_{be} & =\left( L_{cd}^{e}\circ h\circ \overset{%
\ast }{\pi }\right) \left( \rho ,\eta \right) \Gamma _{be} \\
& +\left( L_{cd}^{e}\circ h\circ \overset{\ast }{\pi }\right) \left( \left(
\tilde{g}_{fe}\circ h\circ \overset{\ast }{\pi }\right) \frac{\partial F_{b}%
}{\partial p_{e}}\right)%
\end{array}%
\end{equation*}%
it results the conclusion of the theorem.\hfill \emph{q.e.d.}

\textbf{Proposition 8.1 }\emph{If }$\overset{\ast }{S}$\emph{\ is the
canonical }$\left( \rho ,\eta \right) $\emph{-semispray asso\-cia\-ted to
the mechanical }$\left( \rho ,\eta \right) $\emph{-system }$\left( \left(
\overset{\ast }{E},\overset{\ast }{\pi },M\right) ,\overset{\ast }{F}%
_{e},\left( \rho ,\eta \right) \overset{\ast }{\Gamma }\right) $\emph{\ and
from locally invertible }$\mathbf{B}^{\mathbf{v}}$\emph{-mor\-phism }$\left(
g,h\right) $\emph{,\ then }%
\begin{equation*}
\begin{array}{r}
2G_{b%
%TCIMACRO{\U{b4}}%
%BeginExpansion
{\acute{}}%
%EndExpansion
}=2G_{b}\cdot M_{b%
%TCIMACRO{\U{b4}}%
%BeginExpansion
{\acute{}}%
%EndExpansion
}^{b}\circ h\circ \overset{\ast }{\pi }-\left( g^{ae}\circ h\circ \overset{%
\ast }{\pi }\right) p_{e}\left( \rho _{a}^{i}\circ h\circ \overset{\ast }{%
\pi }\right) \frac{\partial p_{b%
%TCIMACRO{\U{b4}}%
%BeginExpansion
{\acute{}}%
%EndExpansion
}}{\partial x^{i}}.%
\end{array}%
\leqno(8.8)
\end{equation*}%
\emph{\ }

\emph{Proof.} Since the Jacobian matrix of coordinates transformation is
\begin{equation*}
\left\Vert
\begin{array}{ll}
\,\ \ \ \ \ \ \ M_{a}^{a%
%TCIMACRO{\U{b4}}%
%BeginExpansion
{\acute{}}%
%EndExpansion
}\circ h\circ \overset{\ast }{\pi } & \,\ 0\vspace*{1mm} \\
\left( \rho _{a}^{i}\circ h\circ \overset{\ast }{\pi }\right) \frac{\partial
M_{b%
%TCIMACRO{\U{b4}}%
%BeginExpansion
{\acute{}}%
%EndExpansion
}^{a}\circ \overset{\ast }{\pi }}{\partial x^{i}}p_{a} & M_{b%
%TCIMACRO{\U{b4}}%
%BeginExpansion
{\acute{}}%
%EndExpansion
}^{b}\circ \overset{\ast }{\pi }%
\end{array}%
\right\Vert =\left\Vert
\begin{array}{ll}
\,\ \ \ \ \ \ \ M_{a}^{a%
%TCIMACRO{\U{b4}}%
%BeginExpansion
{\acute{}}%
%EndExpansion
}\circ h\circ \overset{\ast }{\pi } & \,\ 0\vspace*{1mm} \\
\left( \rho _{a}^{i}\circ h\circ \overset{\ast }{\pi }\right) \frac{\partial
p_{b%
%TCIMACRO{\U{b4}}%
%BeginExpansion
{\acute{}}%
%EndExpansion
}}{\partial x^{i}} & M_{b%
%TCIMACRO{\U{b4}}%
%BeginExpansion
{\acute{}}%
%EndExpansion
}^{b}\circ \overset{\ast }{\pi }%
\end{array}%
\right\Vert
\end{equation*}%
and
\begin{equation*}
\left\Vert
\begin{array}{ll}
\,\ \ \ \ \ \ \ M_{a}^{a%
%TCIMACRO{\U{b4}}%
%BeginExpansion
{\acute{}}%
%EndExpansion
}\circ h\circ \overset{\ast }{\pi } & \,\ 0\vspace*{1mm} \\
\left( \rho _{a}^{i}\circ h\circ \overset{\ast }{\pi }\right) \frac{\partial
p_{b%
%TCIMACRO{\U{b4}}%
%BeginExpansion
{\acute{}}%
%EndExpansion
}}{\partial x^{i}} & M_{b%
%TCIMACRO{\U{b4}}%
%BeginExpansion
{\acute{}}%
%EndExpansion
}^{b}\circ \overset{\ast }{\pi }%
\end{array}%
\right\Vert \left(
\begin{array}{l}
\,\ \ \ \left( g^{ae}\circ h\circ \overset{\ast }{\pi }\right) p_{e} \\
-2\left( G_{b}-\frac{1}{4}F_{b}\right)%
\end{array}%
\right) =\left(
\begin{array}{l}
\,\ \ \left( g^{a%
%TCIMACRO{\U{b4}}%
%BeginExpansion
{\acute{}}%
%EndExpansion
e%
%TCIMACRO{\U{b4}}%
%BeginExpansion
{\acute{}}%
%EndExpansion
}\circ h\circ \overset{\ast }{\pi }\right) p_{e%
%TCIMACRO{\U{b4}}%
%BeginExpansion
{\acute{}}%
%EndExpansion
}\vspace*{1mm} \\
-2\left( G_{b%
%TCIMACRO{\U{b4}}%
%BeginExpansion
{\acute{}}%
%EndExpansion
}-\frac{1}{4}F_{b%
%TCIMACRO{\U{b4}}%
%BeginExpansion
{\acute{}}%
%EndExpansion
}\right)%
\end{array}%
\right) ,\vspace*{2mm}
\end{equation*}%
the conclusion results. \hfill \emph{q.e.d.}

In the following we consider a differentiable curve $I~\ ^{\underrightarrow{c%
}}~\ M$ and its $\left( g,h\right) $-lift $\dot{c}.$

\textbf{Definition 8.3 }If it is verifies the following equality:\textit{\ }%
\begin{equation*}
\begin{array}{l}
\frac{d\dot{c}\left( t\right) }{dt}=\Gamma \left( \overset{\ast }{\tilde{\rho%
}},Id_{\overset{\ast }{E}}\right) \overset{\ast }{S}\left( \dot{c}\left(
t\right) \right) ,%
\end{array}%
\leqno(8.9)
\end{equation*}%
then we say that \emph{the curve }$\dot{c}$\emph{\ is an integral curve of
the }$\left( \rho ,\eta \right) $\emph{-semispray }$\overset{\ast }{S}$
\emph{of the dual mechanical }$\left( \rho ,\eta \right) $\emph{-system} $%
\left( \left( \overset{\ast }{E},\overset{\ast }{\pi },M\right) ,\overset{%
\ast }{F}_{e},\left( \rho ,\eta \right) \overset{\ast }{\Gamma }\right) .$

\textbf{Theorem 8.5 }\emph{The integral curves of the canonical }$\left(
\rho ,\eta \right) $\emph{-semispray asso\-cia\-ted to the dual mechanical }$%
\left( \rho ,\eta \right) $\emph{-system }$\left( \left( \overset{\ast }{E},%
\overset{\ast }{\pi },M\right) ,\overset{\ast }{F}_{e},\left( \rho ,\eta
\right) \overset{\ast }{\Gamma }\right) $\emph{\ and from locally invertible
}$\mathbf{B}^{\mathbf{v}}$\emph{-mor\-phism }$\left( g,h\right) $\emph{, are
the }$\left( g,h\right) $\emph{-lifts solutions of the equations:\ }%
\begin{equation*}
\begin{array}{l}
\frac{dp_{b}\left( t\right) }{dt}+2G_{b}\!\circ \overset{\ast }{u}\left( c,%
\dot{c}\right) \left( x\left( t\right) \right) {=}\frac{1}{2}F_{b}\!\circ
\overset{\ast }{u}\left( c,\dot{c}\right) \left( x\left( t\right) \right)
\!,\,b{\in }\overline{1,\!r},%
\end{array}%
\leqno(8.10)
\end{equation*}%
\emph{where }$x\left( t\right) =\left( \eta \circ h\circ c\right) \left(
t\right) .$\medskip

\emph{Proof.} Since the equality
\begin{equation*}
\begin{array}{c}
\frac{d\dot{c}\left( t\right) }{dt}=\Gamma \left( \overset{\ast }{\tilde{\rho%
}},Id_{\overset{\ast }{E}}\right) \overset{\ast }{S}\left( \dot{c}\left(
t\right) \right)%
\end{array}%
\end{equation*}%
is equivalent with
\begin{equation*}
\begin{array}{c}
\frac{d}{dt}((\eta \circ h\circ c)^{i}(t),p_{b}(t))=\vspace*{1mm} \\
=\left( \rho _{a}^{i}\circ \eta \circ h\circ c(t)g^{ae}\circ h\circ
c(t)p_{e}(t),-2\left( G_{b}-\frac{1}{4}F_{b}\right) ((\eta \circ h\circ
c)(t),p\left( t\right) )\right) ,%
\end{array}%
\end{equation*}%
it results
\begin{equation*}
\begin{array}{c}
\frac{dp_{b}\left( t\right) }{dt}+2G_{b}\!\left( x\left( t\right) ,p\left(
t\right) \right) {=}\frac{1}{2}F_{b}\!\left( x\left( t\right) ,p\left(
t\right) \right) \!,\,b{\in }\overline{1,\!r},\vspace*{1mm} \\
\frac{dx^{i}\left( t\right) }{dt}=\rho _{a}^{i}\circ \eta \circ h\circ
c\left( t\right) g^{ae}\circ h\circ c\left( t\right) p_{e}\left( t\right) ,%
\end{array}%
\end{equation*}%
where $x^{i}\left( t\right) =\left( \eta \circ h\circ c\right) ^{i}\left(
t\right) $. \hfill \emph{q.e.d.}\medskip

\textbf{Definition 8.4 }If $\overset{\ast }{S}$\ is a $\left( \rho ,\eta
\right) $-semispray, then the vector field
\begin{equation*}
\begin{array}{l}
\left[ \overset{\ast }{\mathbb{C}},\overset{\ast }{S}\right] _{\left( \rho
,\eta \right) T\overset{\ast }{E}}-\overset{\ast }{S}%
\end{array}%
\leqno(8.11)
\end{equation*}%
will be called the \emph{derivation of }$\left( \rho ,\eta \right) $\emph{%
-semispray }$\overset{\ast }{S}.$

The $\left( \rho ,\eta \right) $-semispray $\overset{\ast }{S}$\ will be
called $\left( \rho ,\eta \right) $\emph{-spray} if there are verified the
following conditions:\medskip

1. $\overset{\ast }{S}\circ 0\in C^{1},$\ where $0$\ is the null
section;\smallskip

2. Its derivation is the null vector field.\medskip

The $\left( \rho ,\eta \right) $-semispray $\overset{\ast }{S}$\ will be
called \emph{quadratic }$\left( \rho ,\eta \right) $\emph{-spray }if there
are verified the following conditions:\medskip

1. $\overset{\ast }{S}\circ 0\in C^{2},$\ where $0$\ is the null
section;\smallskip

2. Its derivation is the null vector field.\medskip

In particular, \ if $\ \left( \rho ,\eta \right) =\left(
id_{TM},Id_{M}\right) $ and $\left( g,h\right) =\left( Id_{E},Id_{M}\right)
, $ \ then \ we \ obtain \ the \ \emph{spray} \ and the \emph{quadratic
spray }which is similar with the classical spray and quadratic spray.

\textbf{Theorem 8.6 }\emph{If }$S$\emph{\ is the canonical }$\left( \rho
,\eta \right) $\emph{-spray associated to the dual mechanical }$\left( \rho
,\eta \right) $\emph{-system }$\left( \left( \overset{\ast }{E},\overset{%
\ast }{\pi },M\right) ,\overset{\ast }{F}_{e},\left( \rho ,\eta \right)
\overset{\ast }{\Gamma }\right) $\emph{\ and from locally invertible }$%
\mathbf{B}^{\mathbf{v}}$\emph{-morphism }$\left( g,h\right) $\emph{, then}%
\begin{equation*}
\begin{array}{cl}
2\left( G_{b}-\frac{1}{4}F_{b}\right) & =\left( \rho ,\eta \right) \Gamma
_{bc}\left( g^{cf}\circ h\circ \overset{\ast }{\pi }\right) p_{f} \\
& +\frac{1}{2}\left( g^{de}\circ h\circ \overset{\ast }{\pi }\right)
p_{e}\left( L_{dc}^{a}\circ h\circ \overset{\ast }{\pi }\right) \left(
\tilde{g}_{ab}\circ h\circ \overset{\ast }{\pi }\right) \left( g^{cf}\circ
h\circ \overset{\ast }{\pi }\right) p_{f} \\
& -\frac{1}{2}\left( \rho _{c}^{j}\circ h\circ \overset{\ast }{\pi }\right)
\frac{\partial \left( g^{ae}\circ h\circ \overset{\ast }{\pi }\right) }{%
\partial x^{j}}p_{e}\left( \tilde{g}_{ab}\circ h\circ \overset{\ast }{\pi }%
\right) \left( g^{cf}\circ h\circ \overset{\ast }{\pi }\right) p_{f} \\
& +\frac{1}{2}\left( g^{ae}\circ h\circ \overset{\ast }{\pi }\right)
p_{e}\left( \rho _{a}^{i}\circ h\circ \overset{\ast }{\pi }\right) \frac{%
\partial \left( \tilde{g}_{bc}\circ h\circ \overset{\ast }{\pi }\right) }{%
\partial x^{i}}\left( g^{cf}\circ h\circ \overset{\ast }{\pi }\right) p_{f}%
\end{array}%
\leqno(8.12)
\end{equation*}

\emph{We obtain the spray}%
\begin{equation*}
\begin{array}{cl}
S & =\left( g^{ae}\circ h\circ \pi \right) p_{e}\overset{\ast }{\tilde{%
\partial}}_{a}-\left( \rho ,\eta \right) \Gamma _{bc}\left( g^{cf}\circ
h\circ \pi \right) p_{f}\overset{\cdot }{\tilde{\partial}}^{b} \\
& -\frac{1}{2}\left( g^{de}\circ h\circ \pi \right) p_{e}\left(
L_{dc}^{a}\circ h\circ \pi \right) \left( \tilde{g}_{ab}\circ h\circ \pi
\right) \left( g^{cf}\circ h\circ \pi \right) p_{f}\overset{\cdot }{\tilde{%
\partial}}^{b} \\
& +\frac{1}{2}\left( \rho _{c}^{j}\circ h\circ \pi \right) \frac{\partial
\left( g^{ae}\circ h\circ \pi \right) }{\partial x^{j}}p_{e}\left( \tilde{g}%
_{ab}\circ h\circ \pi \right) \left( g^{cf}\circ h\circ \pi \right) p_{f}%
\overset{\cdot }{\tilde{\partial}}^{b} \\
& -\frac{1}{2}\left( g^{ae}\circ h\circ \pi \right) p_{e}\left( \rho
_{a}^{i}\circ h\circ \pi \right) \frac{\partial \left( \tilde{g}_{bc}\circ
h\circ \pi \right) }{\partial x^{i}}\left( g^{cf}\circ h\circ \pi \right)
p_{f}\overset{\cdot }{\tilde{\partial}}^{b}%
\end{array}%
\leqno(8.13)
\end{equation*}

\emph{This spray will be called the canonical }$\left( \rho ,\eta \right) $%
\emph{-spray associated to the dual mechanical system }$\left( \left(
\overset{\ast }{E},\overset{\ast }{\pi },M\right) ,\overset{\ast }{F}%
_{e},\left( \rho ,\eta \right) \overset{\ast }{\Gamma }\right) $\emph{\ and
from locally invertible }$\mathbf{B}^{\mathbf{v}}$\emph{-morphism }$(g,h).$

\emph{In particular, if }$\left( \rho ,\eta \right) =\left(
id_{TM},Id_{M}\right) $\emph{\ and }$\left( g,h\right) =\left(
Id_{E},Id_{M}\right) ,$\emph{\ then we get the canonical spray associated to
connection }$\Gamma $\emph{\ which is similar with the classical canonical
spray associated to connection }$\Gamma $.

\emph{Proof.} Since
\begin{equation*}
\begin{array}[t]{l}
\left[ \overset{\ast }{\mathbb{C}},\overset{\ast }{S}\right] _{\left( \rho
,\eta \right) T\overset{\ast }{E}}=\left[ p_{a}\overset{\cdot }{\tilde{%
\partial}}^{a},\left( g^{be}\circ h\circ \overset{\ast }{\pi }\right) p_{e}%
\overset{\ast }{\tilde{\partial}}_{b}\right] _{\left( \rho ,\eta \right) T%
\overset{\ast }{E}}-2\left[ p_{a}\overset{\cdot }{\tilde{\partial}}%
^{a},\left( G_{b}-\frac{1}{4}F_{b}\right) \overset{\cdot }{\tilde{\partial}}%
^{b}\right] _{\left( \rho ,\eta \right) T\overset{\ast }{E}},%
\end{array}%
\end{equation*}

\begin{equation*}
\!\!%
\begin{array}{cl}
\left[ p_{a}\overset{\cdot }{\tilde{\partial}}^{a},\left( g^{be}\circ h\circ
\overset{\ast }{\pi }\right) p_{e}\overset{\ast }{\tilde{\partial}}_{b}%
\right] _{\left( \rho ,\eta \right) T\overset{\ast }{E}}\!\!\!\! & %
\displaystyle=\left( g^{be}\circ h\circ \overset{\ast }{\pi }\right) p_{e}%
\overset{\ast }{\tilde{\partial}}_{b}\vspace*{2mm}%
\end{array}%
\end{equation*}%
and
\begin{equation*}
\begin{array}{cl}
\left[ p_{a}\overset{\cdot }{\tilde{\partial}}^{a},\left( G_{b}-\frac{1}{4}%
F_{b}\right) \overset{\cdot }{\tilde{\partial}}^{b}\right] _{\left( \rho
,\eta \right) T\overset{\ast }{E}} & \displaystyle=p_{a}\frac{\partial
\left( G_{b}-\frac{1}{4}F_{b}\right) }{\partial p_{a}}\overset{\cdot }{%
\tilde{\partial}}^{b}-\left( G_{b}-\frac{1}{4}F_{b}\right) \overset{\cdot }{%
\tilde{\partial}}^{b}\vspace*{2mm}%
\end{array}%
\end{equation*}%
it results that
\begin{equation*}
\begin{array}{cc}
\left[ \overset{\ast }{\mathbb{C}},\overset{\ast }{S}\right] _{\left( \rho
,\eta \right) T\overset{\ast }{E}}-\overset{\ast }{S} & \displaystyle%
=2\left( -p_{f}\frac{\partial \left( G_{b}-\frac{1}{4}F_{b}\right) }{%
\partial p_{f}}+2\left( G_{b}-\frac{1}{4}F_{b}\right) \right) \overset{\cdot
}{\tilde{\partial}}^{b}%
\end{array}%
\leqno\left( S_{1}\right)
\end{equation*}

Using equality $(8.3)$, it results that%
\begin{equation*}
\begin{array}{cl}
\displaystyle\frac{\partial \left( G_{b}-\frac{1}{4}F_{b}\right) }{\partial
p_{f}} & =\left( \rho ,\eta \right) \Gamma _{bc}\left( g^{cf}\circ h\circ
\overset{\ast }{\pi }\right) \\
& +\frac{1}{2}\left( g^{de}\circ h\circ \overset{\ast }{\pi }\right)
p_{e}\left( L_{dc}^{a}\circ h\circ \overset{\ast }{\pi }\right) \left(
\tilde{g}_{ab}\circ h\circ \overset{\ast }{\pi }\right) \left( g^{cf}\circ
h\circ \overset{\ast }{\pi }\right) \\
& -\frac{1}{2}\left( \rho _{c}^{j}\circ h\circ \overset{\ast }{\pi }\right)
\frac{\partial \left( g^{ae}\circ h\circ \overset{\ast }{\pi }\right) }{%
\partial x^{j}}p_{e}\left( \tilde{g}_{ab}\circ h\circ \overset{\ast }{\pi }%
\right) \left( g^{cf}\circ h\circ \overset{\ast }{\pi }\right) \\
& +\frac{1}{2}\left( g^{ae}\circ h\circ \overset{\ast }{\pi }\right)
p_{e}\left( \rho _{a}^{i}\circ h\circ \overset{\ast }{\pi }\right) \frac{%
\partial \left( \tilde{g}_{bc}\circ h\circ \overset{\ast }{\pi }\right) }{%
\partial x^{i}}\left( g^{cf}\circ h\circ \overset{\ast }{\pi }\right)%
\end{array}%
\leqno\left( S_{2}\right)
\end{equation*}

Using equalities $\left( S_{1}\right) $ and $\left( S_{2}\right) $, it
results the conclusion of the theorem.\hfill \emph{q.e.d.}

\textbf{Theorem 8.7 }\emph{\ All }$\left( g,h\right) $\emph{-lifts solutions
of the following system of equations:\ }%
\begin{equation*}
\begin{array}{l}
\displaystyle\frac{dp_{b}}{dt}+\left( \rho ,\eta \right) \Gamma _{bc}\left(
g^{cf}\circ h\circ \overset{\ast }{\pi }\right) p_{f}\vspace*{2mm} \\
\displaystyle+\frac{1}{2}\left( g^{de}\circ h\circ \overset{\ast }{\pi }%
\right) p_{e}\left( L_{dc}^{b}\circ h\circ \overset{\ast }{\pi }\right)
\left( \tilde{g}_{ba}\circ h\circ \overset{\ast }{\pi }\right) \left(
g^{cf}\circ h\circ \overset{\ast }{\pi }\right) p_{f}\vspace*{2mm} \\
\displaystyle-\frac{1}{2}\left( \rho _{c}^{j}\circ h\circ \overset{\ast }{%
\pi }\right) \frac{\partial \left( g^{be}\circ h\circ \overset{\ast }{\pi }%
\right) }{\partial x^{j}}p_{e}\left( \tilde{g}_{ba}\circ h\circ \overset{%
\ast }{\pi }\right) \left( g^{cf}\circ h\circ \overset{\ast }{\pi }\right)
p_{f}\vspace*{2mm} \\
\displaystyle+\frac{1}{2}\left( g^{ae}\circ h\circ \overset{\ast }{\pi }%
\right) p_{e}\left( \rho _{a}^{i}\circ h\circ \overset{\ast }{\pi }\right)
\frac{\partial \left( \tilde{g}_{bc}\circ h\circ \overset{\ast }{\pi }%
\right) }{\partial x^{i}}\left( g^{cf}\circ h\circ \overset{\ast }{\pi }%
\right) p_{f}\vspace*{2mm}=0,%
\end{array}%
\leqno(8.14)
\end{equation*}%
\emph{are the integral curves of canonical }$\left( \rho ,\eta \right) $%
\emph{-spray associated to the dual mechanical }$\left( \rho ,\eta \right) $%
\emph{-system }$\left( \left( \overset{\ast }{E},\overset{\ast }{\pi }%
,M\right) ,\overset{\ast }{F}_{e},\left( \rho ,\eta \right) \overset{\ast }{%
\Gamma }\right) $\emph{\ and from locally invertible }$\mathbf{B}^{\mathbf{v}%
}$\emph{-morphism\ }$\left( g,h\right) .$

\section{A Hamiltonian formalism for Hamilton mechanical $\left( \protect%
\rho ,\protect\eta \right) $-systems}

Let $\left( \left( \overset{\ast }{E},\overset{\ast }{\pi },M\right) ,%
\overset{\ast }{F}_{e},H\right) $ be an arbitrarily Hamilton mechanical $%
\left( \rho ,\eta \right) $-system.

Let $\left( d\tilde{z}^{a},d\tilde{p}_{a}\right) $ be the natural dual $%
\left( \rho ,\eta \right) $\emph{-base} of the natural $\left( \rho ,\eta
\right) $-base $\left( \overset{\ast }{\tilde{\partial}}_{a},\overset{\cdot }%
{\tilde{\partial}}^{a}\right) .$

It is very important to remark that the $1$-forms $d\tilde{z}^{a},d\tilde{p}%
_{a},~a\in \overline{1,p}$ are not the differentials of coordinates
functions as in the classical case, but we will use the same notations. In
this case
\begin{equation*}
\left( d\tilde{z}^{a}\right) \neq d^{\left( \rho ,\eta \right) T\overset{%
\ast }{E}}\left( \tilde{z}^{a}\right) ,
\end{equation*}%
where $d^{\left( \rho ,\eta \right) T\overset{\ast }{E}}$ is the exterior
differentiation operator associated to exterior differential $\mathcal{F}%
\left( \overset{\ast }{E}\right) $-algebra
\begin{equation*}
\left( \Lambda \left( \left( \rho ,\eta \right) T\overset{\ast }{E},\left(
\rho ,\eta \right) \tau _{\overset{\ast }{E}},\overset{\ast }{E}\right)
,+,\cdot ,\wedge \right) .
\end{equation*}

Let $H$ be a regular Hamiltonian and let $\left( g,h\right) $\ be a locally
invertible $\mathbf{B}^{\mathbf{v}}$-morphism of $\left( \overset{\ast }{E},%
\overset{\ast }{\pi },M\right) $ source and $\left( E,\pi ,M\right) $ target.

\textbf{Definition 9.1 } The $1$-form
\begin{equation*}
\begin{array}{c}
\theta _{H}=\left( \tilde{g}_{ae}\circ h\circ \overset{\ast }{\pi }\right)
H^{e}d\tilde{z}^{a}%
\end{array}%
\leqno(9.1)
\end{equation*}%
will be called the $1$\emph{-form of Poincar\'{e}-Cartan type associated to
the regular Hamiltonian }$H$ \emph{and from locally invertible }$\mathbf{B}^{%
\mathbf{v}}$\emph{-morphism }$\left( g,h\right) $.\medskip

We obtain easily:
\begin{equation*}
\begin{array}[t]{l}
\theta _{H}\left( \overset{\ast }{\tilde{\partial}}_{b}\right) =\left(
\tilde{g}_{be}\circ h\circ \overset{\ast }{\pi }\right) \cdot H^{e},\,\,\
\theta _{H}\left( \overset{\cdot }{\tilde{\partial}}^{b}\right) =0.%
\end{array}%
\leqno(9.2)
\end{equation*}

\textbf{Definition 9.2 } The $2$-form
\begin{equation*}
\omega _{H}=d^{\left( \rho ,\eta \right) T\overset{\ast }{E}}\theta _{H}
\end{equation*}%
will be called the $2$\emph{-form of Poincar\'{e}-Cartan type associated to
the Hamiltonian }$H$\emph{\ and to the locally invertible }$\mathbf{B}^{%
\mathbf{v}}$\emph{-morphism }$\left( g,h\right) $.\medskip

By the definition of $d^{\left( \rho ,\eta \right) T\overset{\ast }{E}},$ we
obtain:
\begin{equation*}
\begin{array}{ll}
\omega _{H}\left( U,V\right) & \displaystyle=\Gamma \left( \overset{\ast }{%
\tilde{\rho}},Id_{\overset{\ast }{E}}\right) \left( U\right) \left( \theta
_{H}\left( V\right) \right) \vspace*{2mm} \\
& \displaystyle-\,\Gamma \left( \overset{\ast }{\tilde{\rho}},Id_{\overset{%
\ast }{E}}\right) \left( V\right) \left( \theta _{H}\left( U\right) \right)
-\theta _{H}\left( \left[ U,V\right] _{\left( \rho ,\eta \right) T\overset{%
\ast }{E}}\right) ,%
\end{array}%
\leqno(9.3)
\end{equation*}%
for any $U,V\in \Gamma \left( \left( \rho ,\eta \right) T\overset{\ast }{E}%
,\left( \rho ,\eta \right) \tau _{\overset{\ast }{E}},\overset{\ast }{E}%
\right) $.\smallskip

\textbf{Definition 9.3 } The real function
\begin{equation*}
\begin{array}{c}
\mathcal{E}_{H}=p_{a}H^{a}-H%
\end{array}%
\leqno(9.4)
\end{equation*}%
will be called the \emph{energy of regular Hamiltonian }$H.$

\textbf{Theorem 9.1 }\emph{The equation }%
\begin{equation*}
\begin{array}{c}
i_{S}\left( \omega _{H}\right) =-d^{\left( \rho ,\eta \right) T\overset{\ast
}{E}}\left( \mathcal{E}_{H}\right) ,\,S\in \Gamma \left( \left( \rho ,\eta
\right) T\overset{\ast }{E},\left( \rho ,\eta \right) \tau _{\overset{\ast }{%
E}},\overset{\ast }{E}\right) ,%
\end{array}%
\leqno\left( 9.5\right)
\end{equation*}%
\emph{has an unique solution }$\overset{\ast }{S}_{H}\left( g,h\right) $%
\emph{\ of the type: }%
\begin{equation*}
\begin{array}[t]{l}
\left( g^{ae}\circ h\circ \overset{\ast }{\pi }\right) p_{e}\overset{\ast }{%
\tilde{\partial}}_{a}-2\left( G_{a}-\frac{1}{4}F_{a}\right) \overset{\cdot }{%
\tilde{\partial}}^{a},%
\end{array}%
\leqno(9.6)
\end{equation*}%
\emph{where }%
\begin{equation*}
\begin{array}[t]{l}
-2\left( G_{a}-\frac{1}{4}F_{a}\right) =E_{b}\left( H,g,h\right) \tilde{H}%
_{ae}\left( g^{eb}\circ h\circ \overset{\ast }{\pi }\right)%
\end{array}%
\leqno(9.7)
\end{equation*}%
\emph{and}%
\begin{equation*}
\begin{array}{cl}
E_{b}\left( H,g,h\right) & =\left( \rho _{b}^{i}{\circ }h{\circ }\overset{%
\ast }{\pi }\right) H_{i}-\left( \rho _{b}^{i}{\circ }h{\circ }\overset{\ast
}{\pi }\right) p_{a}H_{i}^{a} \\
& -\left( g^{df}\circ h\circ \overset{\ast }{\pi }\right) p_{f}\left( \rho
_{d}^{i}{\circ }h{\circ }\overset{\ast }{\pi }\right) \frac{\partial \left(
\left( \tilde{g}_{be}\circ h\circ \overset{\ast }{\pi }\right) H^{e}\right)
}{\partial x^{i}} \\
& +\left( g^{df}\circ h\circ \overset{\ast }{\pi }\right) p_{f}\left( \rho
_{b}^{i}{\circ }h{\circ }\overset{\ast }{\pi }\right) \frac{\partial \left(
\left( \tilde{g}_{de}\circ h\circ \overset{\ast }{\pi }\right) H^{e}\right)
}{\partial x^{i}} \\
& +\left( g^{df}\circ h\circ \overset{\ast }{\pi }\right) p_{f}\left(
L_{db}^{c}{\circ }h{\circ }\overset{\ast }{\pi }\right) \left( \tilde{g}%
_{ce}\circ h\circ \overset{\ast }{\pi }\right) H^{e}%
\end{array}%
\hspace*{-4mm}\leqno(9.8)
\end{equation*}%
$\overset{\ast }{S}_{H}\left( g,h\right) $\textit{\ }will be called \emph{%
the\ canonical }$\left( \rho ,\eta \right) $\emph{-semispray associated to
the Hamilton mechanical }$\left( \rho ,\eta \right) $\emph{-system }$\left(
\left( \overset{\ast }{E},\overset{\ast }{\pi },M\right) ,\overset{\ast }{F}%
_{e},H\right) $\emph{\ and from locally invertible }$\mathbf{B}^{\mathbf{v}}$%
\emph{-morphism }$(g,h).$

\emph{Proof.} We obtain that
\begin{equation*}
i_{\overset{\ast }{S}}\left( \omega _{H}\right) =-d^{\left( \rho ,\eta
\right) T\overset{\ast }{E}}\left( \mathcal{E}_{H}\right)
\end{equation*}%
if and only if
\begin{equation*}
\omega _{H}\left( \overset{\ast }{S},X\right) =-\Gamma \left( \overset{\ast }%
{\tilde{\rho}},Id_{\overset{\ast }{E}}\right) \left( X\right) \left(
\mathcal{E}_{H}\right) ,
\end{equation*}%
for any $X\in \Gamma \left( \left( \rho ,\eta \right) T\overset{\ast }{E}%
,\left( \rho ,\eta \right) \tau _{\overset{\ast }{E}},\overset{\ast }{E}%
\right) .$

Particularly, we obtain:%
\begin{equation*}
\begin{array}[t]{l}
\omega _{L}\left( \overset{\ast }{S},\overset{\ast }{\tilde{\partial}}%
_{b}\right) =-\Gamma \left( \overset{\ast }{\tilde{\rho}},Id_{\overset{\ast }%
{E}}\right) \left( \overset{\ast }{\tilde{\partial}}_{b}\right) \left(
\mathcal{E}_{H}\right) .%
\end{array}%
\end{equation*}

If we expand this equality, we obtain%
\begin{equation*}
\begin{array}{l}
\left( g^{df}\circ h\circ \overset{\ast }{\pi }\right) p_{f}\left[ \left(
\rho _{d}^{i}{\circ }h{\circ }\overset{\ast }{\pi }\right) \frac{\partial
\left( \left( \tilde{g}_{be}\circ h\circ \overset{\ast }{\pi }\right)
H^{e}\right) }{\partial x^{i}}-\left( \rho _{b}^{i}{\circ }h{\circ }\overset{%
\ast }{\pi }\right) \frac{\partial \left( \left( \tilde{g}_{de}\circ h\circ
\overset{\ast }{\pi }\right) H^{e}\right) }{\partial x^{i}}\right. \\
\displaystyle\left. -\left( L_{db}^{c}{\circ }h{\circ }\overset{\ast }{\pi }%
\right) \left( \tilde{g}_{ce}\circ h\circ \overset{\ast }{\pi }\right) H^{e}%
\right] -2\left( G_{b}-\frac{1}{4}F_{b}\right) \left( \tilde{g}_{ae}\circ
h\circ \overset{\ast }{\pi }\right) \cdot H^{eb}\vspace*{2mm} \\
\qquad \displaystyle=\left( \rho _{b}^{i}{\circ }h{\circ }\overset{\ast }{%
\pi }\right) L_{i}-\left( \rho _{b}^{i}{\circ }h{\circ }\overset{\ast }{\pi }%
\right) \frac{\partial \left( p_{a}H^{a}\right) }{\partial x^{i}}.%
\end{array}%
\end{equation*}

After some calculations, we obtain the conclusion of the theorem.\hfill
\emph{q.e.d.}

\textbf{Theorem 9.2}\textit{\ }\emph{If }$\overset{\ast }{S}_{H}\left(
g,h\right) $\textit{\ }\emph{is the\ canonical }$\left( \rho ,\eta \right) $%
\emph{-semispray associated to the Hamilton mechanical }$\left( \rho ,\eta
\right) $\emph{-system }$\left( \left( \overset{\ast }{E},\overset{\ast }{%
\pi },M\right) ,\overset{\ast }{F}_{e},H\right) $\emph{\ and from locally
invertible }$\mathbf{B}^{\mathbf{v}}$\emph{-morphism }$(g,h),$ \emph{then
the\ real local functions }%
\begin{equation*}
\begin{array}{cl}
\left( \rho ,\eta \right) \Gamma _{bc} & =-\frac{1}{2}\left( \tilde{g}%
_{cd}\circ h\circ \overset{\ast }{\pi }\right) \frac{\partial \left(
E_{b}\left( H,g,h\right) \tilde{H}_{ae}\left( g^{eb}\circ h\circ \overset{%
\ast }{\pi }\right) \right) }{\partial p_{d}} \\
& -\frac{1}{2}\left( g^{de}\circ h\circ \overset{\ast }{\pi }\right)
p_{e}\left( L_{dc}^{f}\circ h\circ \overset{\ast }{\pi }\right) \left(
\tilde{g}_{fb}\circ h\circ \overset{\ast }{\pi }\right) \\
& +\frac{1}{2}\left( \rho _{c}^{j}\circ h\circ \overset{\ast }{\pi }\right)
\frac{\partial \left( g^{be}\circ h\circ \overset{\ast }{\pi }\right) }{%
\partial x^{j}}p_{e}\left( \tilde{g}_{ba}\circ h\circ \overset{\ast }{\pi }%
\right) \\
& -\frac{1}{2}\left( g^{de}\circ h\circ \overset{\ast }{\pi }\right)
p_{e}\left( \rho _{d}^{i}\circ h\circ \overset{\ast }{\pi }\right) \frac{%
\partial \left( \tilde{g}_{bc}\circ h\circ \overset{\ast }{\pi }\right) }{%
\partial x^{i}}%
\end{array}%
\leqno(9.9)
\end{equation*}%
\emph{are the components of a }$\left( \rho ,\eta \right) $\emph{-connection
}$\left( \rho ,\eta \right) \Gamma $\emph{\ for the vector bundle }$\left(
\overset{\ast }{E},\overset{\ast }{\pi },M\right) $\emph{\ which will be
called the }$\left( \rho ,\eta \right) $\emph{-connection associated to the
Hamilton mechanical }$\left( \rho ,\eta \right) $\emph{-system}\break $%
\left( \left( \overset{\ast }{E},\overset{\ast }{\pi },M\right) ,\overset{%
\ast }{F}_{e},H\right) $\emph{\ and from locally invertible }$\mathbf{B}^{%
\mathbf{v}}$\emph{-morphism} $(g,h).$

\textbf{Theorem 9.3 }\emph{The parallel }$\left( g,h\right) $\emph{-lifts
with respect to }$\left( \rho ,\eta \right) $\emph{-connection }$\left( \rho
,\eta \right) \Gamma $ \emph{are the\ integral curves of the canonical }$%
\left( \rho ,\eta \right) $\emph{-semispray associated to the Hamilton\
mechanical }$\left( \rho ,\eta \right) $\emph{-system }$\left( \left(
\overset{\ast }{E},\overset{\ast }{\pi },M\right) ,\overset{\ast }{F}%
_{e},H\right) $ \emph{and from locally invertible }$\mathbf{B}^{\mathbf{v}}$%
\emph{-morphism }$\left( g,h\right) .$

\textbf{Definition 9.4 }The equations
\begin{equation*}
\begin{array}{c}
\,\dfrac{dp_{a}\left( t\right) }{dt}-E_{b}\left( H,g,h\right) \tilde{H}%
_{ae}\left( g^{eb}\circ h\circ \overset{\ast }{\pi }\right) \circ u\left( c,%
\dot{c}\right) \left( x\left( t\right) \right) =0,%
\end{array}%
\leqno(9.10)
\end{equation*}%
where $x\left( t\right) =\eta \circ h\circ c\left( t\right) $, will be
called the \emph{equations of Hamilton-Jacobi type associated to the
Hamilton mechanical }$\left( \rho ,\eta \right) $\emph{-system }$\left(
\left( \overset{\ast }{E},\overset{\ast }{\pi },M\right) ,\overset{\ast }{F}%
_{e},H\right) $\emph{\ and from locally invertible }$\mathbf{B}^{\mathbf{v}}$%
\emph{-morphism }$\left( g,h\right) .$

\emph{Remark 9.1 }The\ integral curves of the canonical $\left( \rho ,\eta
\right) $-semispray associated to the Hamilton mechanical $\left( \rho ,\eta
\right) $-system $\left( \left( \overset{\ast }{E},\overset{\ast }{\pi }%
,M\right) ,\overset{\ast }{F}_{e},H\right) $\ and from locally invertible%
\emph{\ }$\mathbf{B}^{\mathbf{v}}$-morphism $\left( g,h\right) $\ are the $%
\left( g,h\right) $-lifts solutions for the equations of Hamilton-Jacobi
type $\left( 9.10\right) $.

Using our theory, we obtain the following

\textbf{Theorem 9.4 }\emph{If }$K$\emph{\ is a Cartan fundamental function,
then the\ geodesics on the manifold }$M$\emph{\ are the curves such that the
components of their }$\left( g,h\right) $\emph{-lifts are\ solutions for the
equations of Hamilton-Jacobi type }$\left( 9.10\right) .$\bigskip

Therefore, it is natural to propose to extend the study of Cartan geometry
from the dual of the Lie algebroid $\left( \left( TM,\tau _{M},M\right) ,%
\left[ ,\right] ,\left( Id_{TM},Id_{M}\right) \right) ,$ to the dual of an
arbitrary (generalized) Lie algebroid $\left( \left( E,\pi ,M\right) ,\left[
,\right] _{E,h},\left( \rho ,\eta \right) \right) .$

\hfill
\begin{tabular}{c}
SECONDARY SCHOOL \textquotedblleft CORNELIUS RADU\textquotedblright , \\
RADINESTI VILLAGE, 217196, GORJ COUNTY, ROMANIA \\
e-mail: c\_arcus@yahoo.com, c\_arcus@radinesti.ro%
\end{tabular}

\end{document}